\pdfoutput=1

\documentclass[iop,revtex4]{emulateapj}
\usepackage{color}
\usepackage{amsmath}
\usepackage{xcolor}
\usepackage{mathrsfs}
\usepackage{color}
\usepackage{natbib}

\newcommand{\sval}{\ensuremath{S_{\mbox{\scriptsize HK}}}}

\shortauthors{Bryan et al.}
\shorttitle{Giant Planet Statistics}
\slugcomment{{\sc Accepted to ApJ:} January 22, 2016}

\begin{document}

\title{Statistics of Long Period Gas Giant Planets in Known Planetary Systems}


\author{
Marta~L.~Bryan\altaffilmark{1},
Heather A. Knutson\altaffilmark{2},
Andrew W. Howard\altaffilmark{3},
Henry Ngo\altaffilmark{2},
Konstantin Batygin\altaffilmark{2},
Justin R. Crepp\altaffilmark{4},
B. J. Fulton\altaffilmark{3},
Sasha Hinkley\altaffilmark{5},
Howard Isaacson\altaffilmark{7}
John A. Johnson\altaffilmark{6},
Geoffry W. Marcy\altaffilmark{7}
Jason T. Wright\altaffilmark{8}
}

\altaffiltext{1}{Cahill Center for Astronomy and Astrophysics, California Institute of Technology,
1200 East California Boulevard, MC 249-17, Pasadena, CA 91125, USA}

\altaffiltext{2}{Division of Geological and Planetary Sciences, California Institute of Technology, Pasadena, CA 91125 USA}

\altaffiltext{3}{Institute for Astronomy, University of Hawaii at Manoa, Honolulu, HI, USA}

\altaffiltext{4}{Department of Physics, University of Notre Dame, Notre Dame, IN, USA}

\altaffiltext{5}{Department of Astronomy, University of Exeter, Exeter, Devon UK}

\altaffiltext{6}{Harvard-Smithsonian Center for Astrophysics, Cambridge, MA, USA}
\altaffiltext{7}{University of California Berkeley Astronomy Department, Berkeley, Ca, USA}
\altaffiltext{8}{Penn State University Department of Astronomy $\&$ Astrophysics, University Park, PA, USA}

\begin{abstract}

We conducted a Doppler survey at Keck combined with NIRC2 K-band AO imaging to search for massive, long-period companions to 123 known exoplanet systems with one or two planets detected using the radial velocity (RV) method.  Our survey is sensitive to Jupiter mass planets out to 20 AU for a majority of stars in our sample, and we report the discovery of eight new long-period planets, in addition to 20 systems with statistically significant RV trends indicating the presence of an outer companion beyond 5 AU.  We combine our RV observations with AO imaging to determine the range of allowed masses and orbital separations for these companions, and account for variations in our sensitivity to companions among stars in our sample.  We estimate the total occurrence rate of companions in our sample to be $52 \pm 5\%$ over the range 1 - 20 $M_{\rm Jup}$ and 5 - 20 AU.  Our data also suggest a declining frequency for gas giant planets in these systems beyond 3-10 AU, in contrast to earlier studies that found a rising frequency for giant planets in the range 0.01-3 AU.  This suggests either that the frequency of gas giant planets peaks between 3-10 AU, or that outer companions in these systems have a different semi-major axis distribution than the overall gas giant planet population.  Our results also suggest that hot gas giants may be more likely to have an outer companion than cold gas giants. We find that planets with an outer companion have higher average eccentricities than their single counterparts, suggesting that dynamical interactions between planets may play an important role in these systems.


\end{abstract}

\section{Introduction}

The presence of a substantial population of gas giant planets on orbits interior to 1 AU poses a challenge to models of planet formation and migration.  Standard core accretion models favor giant planet formation beyond the ice line, where core-nucleated accretion may proceed on a timescale substantially shorter than the lifetime of the disk \citep{Pollack1996, Alibert2005, Rafikov2006}.  In this scenario, gas giant planets on short period orbits most likely migrated in from their original formation locations \citep{Lin1996}.  Migration models for these planets can be divided into two broad categories.  The first is smooth disk migration, in which exchanges of angular momentum with the disk causes the planet's orbit to gradually decay.  This mechanism would be expected to produce close to, if not completely, circular orbits that are well aligned with the spin axis of the host star \citep{Goldreich1980, Lin1986, Tanaka2002}.  The second migration channel is three-body interactions.  These include the Kozai mechanism, in which the presence of a stellar or planetary companion causes the argument of periastron to undergo resonant librations, allowing the planet's orbit to exchange between mutual inclination and eccentricity.  Alternatively, planet-planet scattering or long term secular interactions between planets could impart a large orbital eccentricity to the inner planet \citep{Chatterjee2008, Nagasawa2008, Wu2010}.  This highly eccentric orbit can then shrink and circularize at short periods via tidal dissipation.  

High eccentricity migration channels and dynamical interactions between planets are thought to frequently produce planets whose orbits are misaligned with the rotation axes of their host stars\footnote{This assessment is however sensitive to the dynamical evolution of the stellar spin-axis itself, as spin-orbit misalignments may be suppressed by adiabatic coupling \citep{Storch2014}}.  Over the past decade, Rossiter-McLaughlin measurements of spin-orbit alignment have found a number of hot Jupiter systems that are misaligned \citep{Winn2010, Hebrard2011, Albrecht2012}.  However, previous studies demonstrated that there is no correlation between the presence of an outer planetary or stellar companion and the spin-orbit angle of hot Jupiters \citep{Knutson2014, Ngo2014}.  Furthermore, Batygin (2012) and Batygin $\&$ Adams (2013) have suggested that a distant stellar companion could tilt the protoplanetary disk with respect to the star's spin axis, in which case disk migration could lead to a misaligned orbit \citep{Spalding2014}.  This scenario is supported by the discovery of apparently coplanar multi-planet systems with spin-orbit misalignments \citep{Huber2013, Bourrier2014}, although other surveys have suggested that such systems may be relatively rare \citep{Albrecht2013, Morton2014}.  In either case, it appears that the cause of hot Jupiter misalignment is more complicated than the simple picture presented above.

Measurements of orbital eccentricities for a large sample of single and multi-planet systems provide a more direct diagnostic of the importance of dynamical interactions in shaping the observed architectures of planetary systems.  We expect dynamical interactions between planets to pump up the eccentricities of their orbits, a process that could result in migration if the periapse of an orbit gets close enough to the star for tidal forces to become significant \citep{Rasio1996, Juric2008}.  However, previous radial velocity studies of gas giants indicate that high eccentricities are more common in apparently single systems \citep{Howard2013}.  It has been suggested that this enhanced eccentricity may be due to planet-planet scattering, where one planet was ejected from the system \citep{Chatterjee2008}.  This is consistent with the results of \citet{Dawson2013}, which suggest that higher eccentricities are more common when the star has a high metallicity, and infer that this is because higher metallicity stars are more likely to form multiple giant planets, which then interact and pump up planet eccentricities.  \citet{Limbach2014} also find a positive correlation between lower eccentricity and higher system multiplicity.  Conversely, \citet{Dong2014} finds that warm Jupiters with outer companions are more likely to have higher eccentricities than single warm Jupiters, albeit with a relatively small sample size of just 26 systems.  We can test these trends by directly searching for outer companions at wide orbital separations in a large sample of known planetary systems, and checking to see if these companions are associated with a larger orbital eccentricity for the inner planet.  

In order to understand whether or not dynamical interactions between planets are responsible for the inward migration of a subset of these planets, it is useful to study systems where we can obtain a complete census of gas giant planets across a broad range of orbital separations.  While large surveys have made it possible to understand the statistical properties of exoplanet populations, recent studies have focused on determining mass distributions and occurrence rates of short period, low mass planets around apparently single main sequence FGK stars \citep[e.g.][]{Howard2012, Fressin2013, Howard2013, Petigura2013}. Many of these surveys are primarily sensitive to short-period planets, making it difficult to evaluate the role that a massive distant planetary companion might have on the formation and orbital evolution of the inner planets.  Early studies of hot Jupiters, which are among the best-studied exoplanet populations, indicated that they rarely contain nearby companions (Steffen et al. 2012, but see Becker et al. 2015 for a recent exception). In contrast, recent work by Knutson et al (2014) looked at 51 hot Jupiter systems and found that they are not lonely - the occurrence rate of massive, outer companions was $51 \pm10 \%$ for companions with masses of 1-13 $M_{\rm Jup}$ and separations of 1-20 AU.  This implies that long period companions to hot Jupiters are common, and thus might play an important role in the orbital evolution of these systems.  

In this study we combine Keck HIRES radial velocity measurements with NIRC2 K band adaptive optics (AO) imaging to search for massive, long period companions to a sample of 123 known exoplanet systems detected using the radial velocity (RV) method.  Unlike our previous survey, which focused exclusively on transiting hot Jupiter systems, our new sample includes planets with a wide range of masses and orbital separations (Fig. 1).  We present results from this survey in two papers.  In this paper, we focus on long-term RV monitoring of the confirmed exoplanet systems, probing planetary and brown dwarf mass companions out to $\sim$100 AU.  We test whether close-in gas giant planets are more likely to have outer companions than their long period counterparts, and whether planets in two-planet systems are more likely to have higher eccentricities than single planet systems. 
In the second paper, we will use our complementary K-band AO images to find and confirm low mass stellar companions in these systems in order to determine how stellar companions might influence the formation and evolution of the inner planets.

In section 2 we describe the selected sample of systems, as well as the methods for obtaining the RV and K-band AO imaging data.  In section 3 we describe fits to the RV data, generation of contrast curves from the AO data, identification of significant RV accelerations, calculation of two-dimensional companion probability distributions, and the completeness analysis that was performed for each individual system.  Finally, in section 4 we discuss our occurrence rate calculations and analysis of eccentricity distributions.  
 
 \begin{figure}
\includegraphics[width=0.5\textwidth]{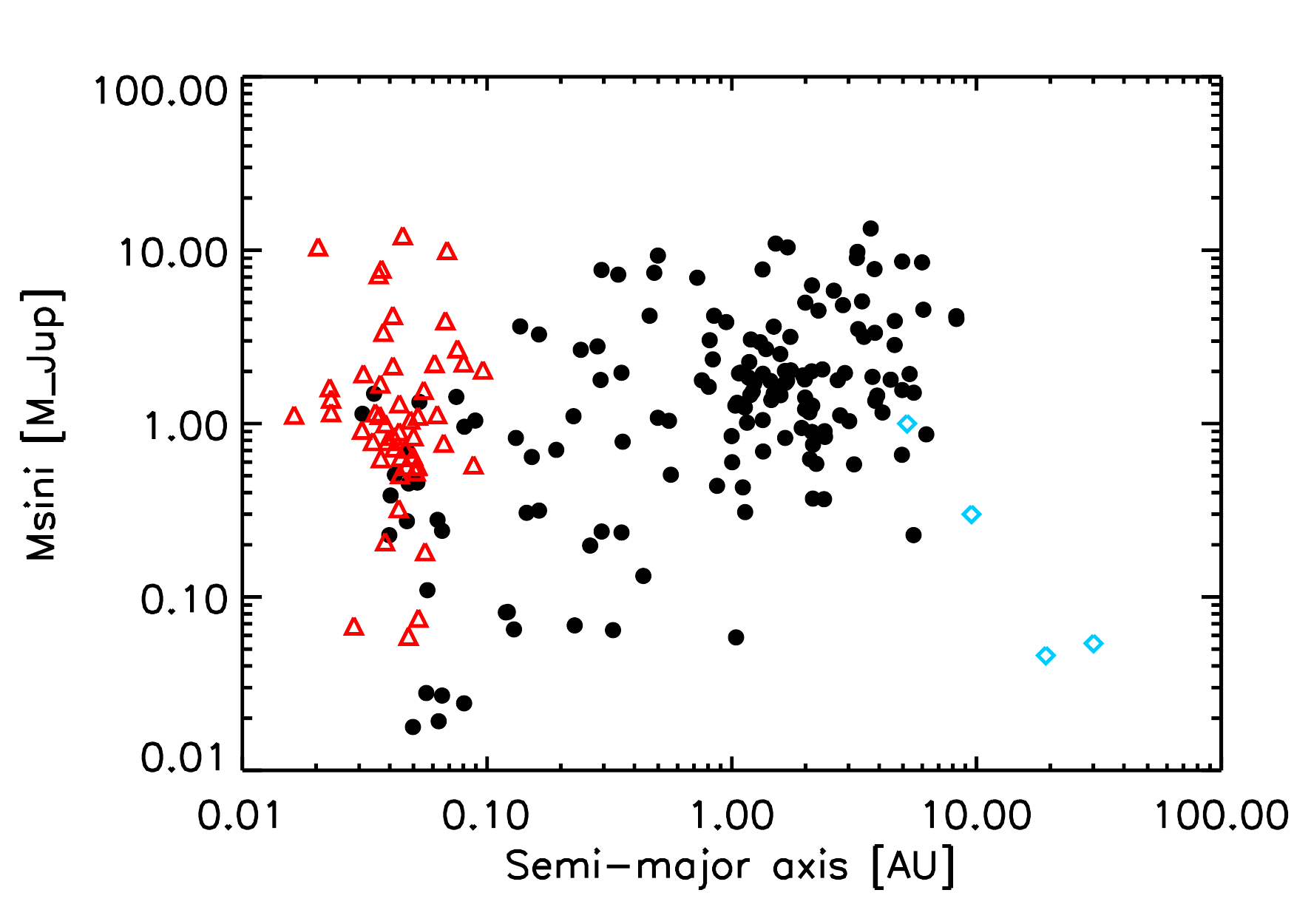}
\caption{Transiting hot Jupiters from our previous radial velocity study \citep{Knutson2014} are shown as red triangles, and the new sample of gas giant planets in this study are shown as black circles.  The blue diamonds represent the gas and ice giant planets in the solar system for comparison.}
\end{figure}

\begin{figure}
\includegraphics[width=0.5\textwidth]{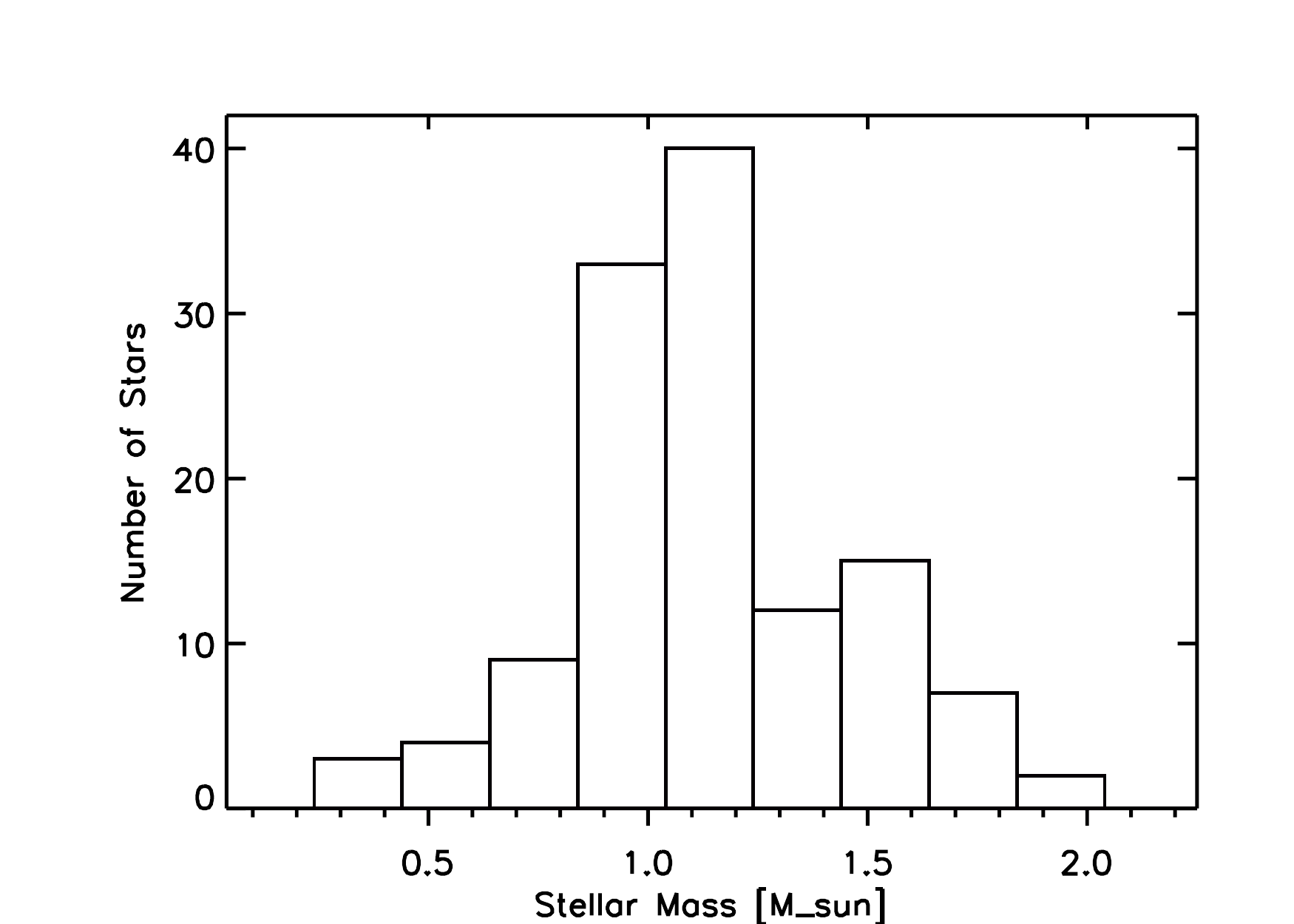}
\caption{Distribution of masses for the stars in our sample.  }
\end{figure}

\section{Observations}

Radial velocity measurements were made at Keck Observatory as part of more than a dozen PI-led programs falling under the umbrella of the California Planet Survey \citep[CPS;][]{Howard2010}.  We observed each target star using the High Resolution Echelle Spectrometer (HIRES) \citep{Vogt94} following standard practices of CPS.  Our selected sample includes all known one- and two-planet systems discovered via the radial velocity method with at least ten RV observations obtained using HIRES.  We also excluded systems with a Keck baseline shorter than the published orbital period.  The published planets in our resulting sample of 123 systems span a range of masses and semi-major axes, as shown in Figure 1.  RV baselines for these targets range from 5.02 to 18.18 years, making it possible to detect gas giant planets spanning a broad range of orbital semi-major axes.  Properties of the target stars are described in Table 1.  Figure 2 shows the distribution of stellar masses in our sample.  While most stars are F and G stars, there are significant numbers of M, K, and A stars.  The A stars in this sample are all moderately evolved, which facilitates precise radial velocity measurements \citep{Johnson2010, Johnson2011}.

\subsection{Keck HIRES Radial Velocities}

All of the target stars were observed using the High Resolution Echelle Spectrometer (HIRES) on Keck I \citep{Vogt94}.  While the majority of the RV data used in this study was published in previous papers, we also obtained new observations that extend these published baselines by up to 12 years.  To reduce the RV data, the standard CPS HIRES configuration and reduction pipeline were used (Wright et al 2004; Howard et al 2009; Johnson et al 2010).  We measured Doppler shifts from the echelle spectra using an iodine absorption spectrum and a modeling procedure descended from \citet{Butler96} and described in \citet{Howard2011}.  The set of observations for each star comprise a ``template spectrum" taken without iodine and de-convolved using a reference point spread function (PSF) inferred from near-in-time observations of B-stars through iodine, and a set of dozens to hundreds of observations through iodine that each yield an RV.  We used one of the 0\farcs{86}-wide slits (`B5' or `C2') for the observations taken through iodine and a 0\farcs{57} (`B1' or `B3') or 0\farcs{86}-wide slit for the template observations.  Using a real-time exposure meter, integration times of 1--8 minutes were chosen to achieve (in most cases) a signal-to-noise ratio of $\sim$220 in the reduced spectrum at the peak of the blaze function near 550 nm.  All Doppler observations were made with an iodine cell mounted directly in front of the spectrometer entrance slit.  The dense set of molecular absorption lines imprinted on the stellar spectra provide a robust wavelength fiducial against which Doppler shifts are measured, as well as strong constraints on the shape of the spectrometer instrumental profile at the time of each observations \citep{Marcy92,Valenti95}.  The velocity and corresponding uncertainty for each observation is based on separate measurements for $\sim$700 spectral chunks each 2\AA\ wide.  The RVs are corrected for motion of Keck Observatory through the solar System (barycentric corrections).  The measurements span 1996--2015 (see Table 2).  Measurements made after the HIRES CCD upgrade in 2004 August have a different (arbitrary) velocity zero point (not the star's systemic velocity) and suffer from somewhat smaller systematic errors.  A summary of the radial velocity data used in this work is provided in Table 2.  We include best-fit stellar jitter and RV acceleration ``trend" values from our orbital solution fitting described in section 3.  

\subsection{NIRC2 AO Imaging}

We observed $K$ band images for all targets using the NIRC2 instrument (Instrument PI: Keith Matthews) on Keck II. We used natural guide star AO imaging and the narrow camera setting (10$\,$mas$\,$pixel$^{-1}$) to achieve better contrast and spatial resolution. For most targets, we imaged using the full NIRC2 array (1024$\times$1024 pixels) and used a 3-point dither pattern that avoids NIRC2's noisier quadrant. Because NIRC2 does not have neutral density filters, we used the subarray mode (2.5" or 5" field of view) to decrease readout time when it was necessary to avoid saturation.  We typically obtained two minutes of on-target integration time per system in position angle mode.

We use dome flat fields and dark frames to calibrate the images. We identify image artifacts by searching for pixels that are $8\sigma$ outliers compared to the counts in the surrounding 5$\times$5 box. We replace these pixels by the median value of the same 5$\times$5 box. To compute contrast curves, we register all frames with the target star and then combine using a median stack.  Table 3 summarizes the NIRC2 AO observations taken during this survey that were used in subsequent analysis.  

\clearpage

\LongTables
\begin{deluxetable*}{lccccc}
\tabletypesize{\scriptsize}
\tablewidth{0pc}
\tablecaption{Stellar Parameters}
\tablehead{
\colhead{Star}&
\colhead{Mass [$M_\odot$]}   &
\colhead{[Fe/H]} &
\colhead{$B - V$}&
\colhead{$S_{HK}$}&
\colhead{References}
}

\startdata
$\rho$ CrB\footnotemark[1]&0.97&-0.20&0.61&0.15& \citet{Takeda2007}\\%
16 Cyg B\footnotemark[2]&0.96&0.04&0.66&0.15& \citet{Takeda2007}\\%
24 Sex\footnotemark[3]&1.54&-0.0&0.91&0.14& \citet{Mortier2013}\\%
51 Peg\footnotemark[4]&1.05&0.20&0.67&0.15& \citet{Takeda2007}\\%
70 Vir\footnotemark[5]&1.10&-0.012&0.69&0.14& \citet{Takeda2007}\\%
GJ 176\footnotemark[6]&0.49&-0.10&1.5&1.5& \citet{Endl2007}\\%
GJ 179\footnotemark[7]&0.36&0.30&1.6&1.1& \citet{Howard2009} \\
GJ 317&0.24&-0.23&1.6&1.2 & \citet{Johnson2007} \\
GJ 649\footnotemark[8]&0.54&0.08&1.6&1.6& \citet{Johnson20102} \\
GJ 849\footnotemark[9]&0.49&0.16&1.5&1.0& \citet{Butler2006}\\%
HD 1461&1.03&0.18&0.68&0.16& \citet{Takeda2007}\\%
HD 1502&1.61&-0.04&0.92&0.10&\citet{Mortier2013} \\
HD 3651&0.88&0.16&0.92&0.17& \citet{Takeda2007} \\
HD 4203&1.13&0.45&0.73&0.14 & \citet{Takeda2007} \\
HD 4208&0.88&-0.28&0.67&0.15 & \citet{Takeda2007} \\
HD 4313&1.72&0.05&0.96&0.12&  \citet{Mortier2013} \\
HD 5319&1.56&0.02&0.98&12.& \citet{Mortier2013}\\%
HD 5891&1.61&-0.38&0.99&0.12&\citet{Mortier2013} \\
HD 8574&1.12&-0.01&0.57&0.15& \citet{Takeda2007} \\
HD 10697&1.11&0.19&0.66&0.15& \citet{Takeda2007} \\
HD 11506&1.19&0.31&0.60&0.15& \citet{Fischer2007} \\
HD 11964A&1.11&0.14&0.83&0.13& \citet{Mortier2013}\\%
HD 12661&1.14&0.36&0.72&0.15& \citet{Takeda2007}\\%
HD 13931&1.02&0.03&0.64&0.15& \citet{Takeda2007} \\
HD 16141&1.05&0.17&0.71&0.15& \citet{Takeda2007} \\
HD 17156&1.29&0.24&0.64&0.15& \citet{Gilliland2011} \\
HD 24040&1.09&0.21&0.66&0.15& \citet{Boisse2012} \\
HD 28678&1.74&-0.21&1.0&0.13& \citet{Mortier2013} \\
HD 30856&1.35&-0.14&0.96&0.13& \citet{Mortier2013} \\
HD 33142&1.48&0.03&0.95&0.14& \citet{Mortier2013} \\
HD 33283&1.24&0.37&0.61&0.13& \citet{Johnson2006} \\
HD 33636&1.02&-0.13&0.58&0.17 & \citet{Takeda2007} \\
HD 34445&1.07&0.14&0.62&0.16& \citet{Takeda2007} \\
HD 37605&1.00&0.34&0.82&0.16& \citet{Wang2012}\\%
HD 38529&1.48&0.40&0.77&0.16& \citet{Mortier2013}\\%
HD 38801&1.36&0.25&0.87&0.16& \citet{Mortier2013} \\
HD 40979&1.15&0.17&0.52&0.22& \citet{Takeda2007} \\
HD 43691&1.38&0.25&0.59&0.15& \citet{Silva2007} \\
HD 45350&1.05&0.29&0.74&0.14& \citet{Takeda2007} \\
HD 46375&0.93&0.24&0.86&0.18& \citet{Takeda2007} \\
HD 49674&1.02&0.31&0.71&0.19& \citet{Takeda2007} \\
HD 50499&1.28&0.34&0.57&0.14& \citet{Takeda2007}\\%
HD 50554&1.03&-0.07&0.53&0.16& \citet{Takeda2007} \\
HD 52265&1.17&0.19&0.53&0.14& \citet{Takeda2007} \\
HIP 57050&0.34&0.32&1.6&0.76& \citet{Haghighipour2010} \\
HD 66428&1.06&0.31&0.71&0.14& \citet{Takeda2007} \\
HD 68988&1.12&0.32&0.62&0.16 & \citet{Takeda2007} \\
HD 72659&1.07&-0.0&0.57&0.15& \citet{Takeda2007}\\%
HD 73534&1.23&0.16&0.95&0.12& \citet{Mortier2013} \\
HD 74156&1.24&0.13&0.54&0.14& \citet{Takeda2007}\\%
HD 75898&1.28&0.27&0.59&0.14& \citet{Robinson2007} \\
HIP 79431&0.49&0.40&1.5&0.90& \citet{Delfosse2000} \\
HD 80606&1.06&0.34&0.71&0.15& \citet{Takeda2007} \\
HD 82886&1.06&-0.31&0.86&0.14& \citet{Johnson2011} \\
HD 83443&0.99&0.36&0.79&0.19& \citet{Takeda2007} \\
HD 86081&1.21&0.26&0.66&0.16& \citet{Johnson2006} \\
HD 88133&1.20&0.33&0.82&0.13& \citet{Mortier2013} \\
HD 92788&1.08&0.32&0.69&0.16& \citet{Takeda2007} \\
HD 96063&1.02&-0.20&0.85&0.14& \citet{Mortier2013} \\
HD 96167&1.31&0.34&0.68&0.13& \citet{Peek2009} \\
HD 97658&0.78&-0.30&0.80&0.17& \citet{Dragomir2013} \\
HD 99109&0.94&0.32&0.87&0.15& \citet{Takeda2007}\\
HD 99492&0.83&0.36&1.0&0.25& \citet{Takeda2007}\\
HD 99706&1.72&0.14&0.99&0.12& \citet{Johnson2011}\\
HD 102195&0.87&0.05&0.90&0.35& \citet{Melo2007}\\  
HD 102329&1.95&0.05&1.0&0.12& \citet{Mortier2013}\\%
HD 102956&1.68&0.19&0.97&0.15& \citet{Johnson20102}\\%
HD 104067&0.79&-0.06&0.99&0.33& \citet{Segransan2011}\\%
HD 106270&1.32&0.06&0.74&0.21& \citet{Mortier2013}\\%
HD 107148&1.14&0.31&0.66&0.15& \citet{Takeda2007}\\%
HD 108863&1.85&0.20&0.99&0.13& \citet{Mortier2013}\\%
HD 108874&0.95&0.18&0.71&0.15& \citet{Takeda2007}\\%
HD 109749&1.21&0.25&0.70&0.16& \citet{Fischer2006}\\%
HD 114729&1.00&-0.26&0.62&0.15& \citet{Takeda2007}\\%
HD 114783&0.85&0.12&0.90&0.18 & \citet{Takeda2007}\\%
HD 116029&1.58&0.08&1.0&0.12& \citet{Mortier2013}\\%
HD 117207&1.03&0.27&0.72&0.15& \citet{Takeda2007}\\%
HD 126614&1.15&0.56&1.2&0.14& \citet{Takeda2007}\\%
HD 128311&0.83&0.21&0.99&0.57& \citet{Takeda2007}\\%
HD 130322&0.84&0.01&0.75&0.23& \citet{Takeda2007}\\%
HD 131496&1.61&0.25&1.0&0.13& \citet{Johnson2011}\\%
HD 134987&1.05&0.28&0.70&0.14& \citet{Takeda2007}\\%
HD 141937&1.05&0.13&0.60&0.20& \citet{Takeda2007}\\%
HD 142245&1.69&0.23&1.0&0.14& \citet{Johnson2011}\\%
HD 149143&1.20&0.26&0.68&0.16& \citet{Fischer2006}\\%
HD 152581&0.93&-0.46&0.90&0.14& \citet{Johnson2011}\\%
HD 154345&0.89&-0.11&0.76&0.20& \citet{Takeda2007}\\%
HD 156279&0.93&0.14&0.80&0.16& \citet{Diaz2012}\\%
HD 156668&0.77&0.05&1.0&0.23& \citet{Howard20112}\\%
HD 158038&1.65&0.28&1.0&0.13& \citet{Johnson2011}\\%
HD 163607&1.09&0.21&0.77&0.16& \citet{Giguere2012}\\%
HD 164509&1.13&0.21&0.66&0.18& \citet{Giguere2012}\\%
HD 164922&0.93&0.17&0.80&0.15& \citet{Takeda2007}\\%
HD 168443&1.00&0.04&0.70&0.14& \citet{Pilyavsky2011}\\%
HD 168746&0.92&-0.08&0.69&0.15& \citet{Takeda2007}\\%
HD 169830&1.41&0.15&0.47&0.14& \citet{Takeda2007}\\%
HD 170469&1.14&0.30&0.62&0.15& \citet{Takeda2007}\\%
HD 175541&1.52&-0.11&0.89&0.13& \citet{Mortier2013}\\%
HD 177830&1.46&0.30&1.1&0.12& \citet{Mortier2013}\\%
HD 178911B&1.06&0.29&0.73&0.18& \citet{Valenti2005}\\%
HD 179079&1.09&0.29&0.74&0.16& \citet{Valenti2009}\\%
HD 180902&1.52&0.0&0.93&0.15& \citet{Mortier2013}\\%
HD 181342&1.58&0.15&1.0&0.12& \citet{Mortier2013}\\%
HD 183263&1.12&0.30&0.63&0.15& \citet{Takeda2007}\\%
HD 187123&1.04&0.12&0.61&0.15& \citet{Takeda2007}\\%
HD 188015&1.06&0.29&0.70&0.16& \citet{Takeda2007}\\%
HD 189733&0.81&-0.03&0.93&0.50& \citet{Torres2008}\\%
HD 190228&1.82&-0.18&0.75&0.17& \citet{Takeda2007}\\%
HD 190360&0.98&0.21&0.73&0.14& \citet{Takeda2007}\\%
HD 192263&0.80&0.05&0.93&0.48& \citet{Takeda2007}\\%
HD 192310&0.85&-0.04&0.87&0.19& \citet{Pepe2011}\\%
HD 195019&1.03&0.07&0.64&0.16& \citet{Takeda2007}\\%
HD 200964&1.44&-0.15&0.88&0.14& \citet{Mortier2013}\\%
HD 206610&1.56&0.10&1.0&0.14& \citet{Mortier2013}\\%
HD 207832&0.94&0.06&0.69&0.24& \citet{Haghighipour2012}\\%
HD 209458&1.13&0.0&0.53&0.16& \citet{Takeda2007}\\%
HD 210277&0.99&0.21&0.71&0.15& \citet{Takeda2007}\\%
HD 212771&1.15&-0.14&0.88&0.14& \citet{Mortier2013}\\%
HD 217107&1.11&0.39&0.72&0.14& \citet{Takeda2007}\\%
HD 222582&0.97&-0.03&0.60&0.16& \citet{Takeda2007}\\%
HD 224693&1.33&0.34&0.63&0.14& \citet{Johnson2006}\\%
HD 231701&1.14&0.07&0.53&0.17& \citet{Fischer2007}%
\enddata
\end{deluxetable*}
\footnotetext[1]{Alternate name HD 143761}
\footnotetext[2]{Alternate name HD 186427}
\footnotetext[3]{Alternate name HD 90043}
\footnotetext[4]{Alternate name HD 217014}
\footnotetext[5]{Alternate name HD 117176}
\footnotetext[6]{Alternate name HD 285968}
\footnotetext[7]{Alternate name HIP 22627}
\footnotetext[8]{Alternate name HIP 83043}
\footnotetext[9]{Alternate name HIP 109388}

\pagebreak

\clearpage
\begin{deluxetable*}{lccccccc}
\tabletypesize{\scriptsize}
\tablecaption{Radial Velocity Observations}
\tablewidth{0pc}
\tablehead{
\colhead{System}         &
\colhead{$N_{obs}$}   &
\colhead{Start Date}   &
\colhead{End Date} &
\colhead{Duration [days]}  &
\colhead{Trend [m s$^{-1}$ yr$^{-1}$]}  &
\colhead{Jitter [m s$^{-1}$]}  & 
\colhead{Orbital Solution Reference}
}

\startdata
$\rho$ CrB&210&1997 Jun 2&2015 Feb 8&6460&$0.16^{+0.13}_{-0.12}$&$1.1^{+0.0036}_{-0.0037}$&  \citet{Vogt2006}\\
16 Cyg B&135&2006 Jul 11&2014 Dec 9&3073&$0.099^{+0.13}_{-0.13}$&$2.4^{+0.19}_{-0.18}$&  \citet{Vogt2006}\\
24 Sex&44&2008 Dec 5&2013 Dec 12&1833&$-0.062^{+1.4}_{-1.5}$&$7.3^{+1.3}_{-0.96}$&  \citet{Johnson2011}\\
51 Peg&43&2006 Jul 10&2014 Sep 13&2987&$-0.42^{+0.20}_{-0.20}$&$2.4^{+0.37}_{-0.31}$& \citet{Vogt2006}\\
70 Vir&56&2006 Jul 17&2015 Feb 4&3124&$0.14^{+0.25}_{-0.25}$&$3.5^{+0.44}_{-0.37}$&  \citet{Kane2015}\\
GJ 176&71&1998 Jan 26&2014 Sep 6&6067&$0.33^{+0.35}_{-0.34}$&$4.9^{+0.61}_{-0.48}$&  \citet{Forveille2009}\\
GJ 179&43&2000 Feb 6&2014 Aug 24&5313&$-0.62^{+0.55}_{-0.57}$&$5.8^{+1.1}_{-0.93}$& \citet{Howard2010} \\
GJ 317&48&2000 Jan 7&2013 Dec 10&2535&$ = 0\pm0 \hspace{0.1cm}\tablenotemark{10}$&$8.6^{+1.2}_{-1.0}$&  \citet{Anglada2012}\\
GJ 649&52&1999 Aug 19&2014 Feb 20&5299&$0.58^{+0.49}_{-0.48}$&$4.5^{+0.63}_{-0.51}$& \citet{Johnson2010}\\
GJ 849&87&1997 Jun 6&2014 Aug 14&6278&$0.32^{+2.5}_{-2.6}$&$3.5^{+0.41}_{-0.37}$&  \citet{Bonfils2013}\\
HD 1461&218&1996 Oct 10&2015 Feb 7&6694&$-0.0064^{+0.87}_{-0.65}$&$3.8^{+0.14}_{-0.13}$&  \citet{Rivera2010}\\
HD 1502&61&2007 Aug 27&2013 Dec 12&2299&$-0.46^{+1.1}_{-1.1}$&$11.^{+1.2}_{-1.0}$& \citet{Johnson2011}\\
$\bf{HD 3651}$&91&1996 Oct 10&2015 Feb 7&6694&$\bf{0.50^{+0.14}_{-0.14}}$&$3.1^{+0.30}_{-0.26}$& \citet{Wittenmyer2009}\\
HD 4203&46&2000 Jul 31&2014 Dec 11&5246&$ = 0\pm0 \hspace{0.1cm}\tablenotemark{10}$&$3.4^{+0.54}_{-0.45}$ &  \citet{Vogt2006}\\
$\bf{HD 4208}$&12&2005 Aug 21&2014 Sep 6&3303&$\bf{-1.2^{+0.30}_{-0.30}}$&$3.8^{+0.51}_{-0.44}$&  \citet{Vogt2006}\\
HD 4313&43&2007 Aug 27&2014 Aug 4&2534&$-1.1^{+0.42}_{-0.42}$&$4.2^{+0.65}_{-0.54}$&  \citet{Johnson2010}\\
HD 5319&87&2004 Jan 10&2014 Dec 11&3988&$0.50^{+0.31}_{-0.31}$&$6.7^{+0.61}_{-0.55}$&  \citet{Robinson2007}\\
HD 5891&63&2007 Aug 27&2013 Dec 14&2301&$0.86^{+4.2}_{-4.2}$&$33.^{+3.4}_{-2.9}$&\citet{Johnson2011} \\
HD 8574&25&1999 Feb 17&2014 Aug 12&5655&$0.31^{+0.96}_{-1.0}$&$-7.2^{+15.}_{-2.1}$ &  \citet{Wittenmyer2009}\\
HD 10697&77&1996 Oct 10&2014 Jul 8&6480&$0.17^{+0.36}_{-0.35}$&$6.0^{+0.59}_{-0.51}$&  \citet{Wittenmyer2009}\\
$\bf{HD 11506}$&125&2004 Jan 10&2015 Feb 7&4046&$\bf{-7.4^{+0.47}_{-0.47}}$&$9.9^{+0.71}_{-0.63}$&  \citet{Fischer2007}\\
HD 11964A&149&1996 Oct 9&2014 Aug 4&6508&$-0.22^{+0.13}_{-0.13}$&$3.2^{+0.23}_{-0.21}$&  \citet{Wright2009}\\
HD 12661&98&1998 Dec 23&2014 Aug 12&5711&$-0.11^{+0.19}_{-0.18}$&$2.7^{+0.28}_{-0.25}$&  \citet{Wright2009}\\
HD 13931&57&1998 Jan 24&2014 Jul 27&6028&$-0.14^{+0.37}_{-0.39}$&$2.8^{+0.39}_{-0.32}$&  \citet{Howard2010}\\
HD 16141&90&1996 Oct 9&2014 Aug 11&6515&$-0.37^{+0.19}_{-0.19}$&$3.3^{+0.33}_{-0.30}$& \citet{Vogt2006} \\
HD 17156&48&2006 Jan 11&2014 Sep 10&3164&$-0.13^{+0.41}_{-0.41}$&$3.2^{+0.85}_{-0.96}$& \citet{Barbieri2009} \\
$\bf{HD 24040}$&60&1998 Jan 25&2014 Aug 5&6036&$\bf{2.0^{+0.34}_{-0.35}}$&$4.7^{+0.54}_{-0.47}$& \citet{Boisse2012}\\
$\bf{HD 28678}$&39&2007 Aug 27&2014 Aug 25&2555&$\bf{3.9^{+0.99}_{-1.0}}$&$6.4^{+0.99}_{-0.82}$&\citet{Johnson2011}\\
HD 30856&22&2007 Aug 27&2013 Dec 14&2301&$-2.4^{+1.4}_{-1.5}$&$6.1^{+1.5}_{-1.1}$&\citet{Johnson2011}\\
HD 33142&40&2007 Aug 27&2014 Sep 12&2573&$-1.3^{+0.97}_{-1.0}$&$1.4^{+0.079}_{-0.074}$& \citet{Johnson2011}\\
HD 33283&42&2004 Jan 10&2014 Sep 7&3893&$-0.18^{+0.27}_{-0.26}$&$3.3^{+0.55}_{-0.46}$ &  \citet{Johnson2006}\\
HD 33636&48&1998 Jan 25&2014 Sep 7&6069&$-0.56^{+0.35}_{-0.34}$&$4.2^{+0.59}_{-0.50}$& \citet{Vogt2006}\\
HD 34445&117&1998 Jan 25&2015 Feb 4&6219&$-0.93^{+0.32}_{-0.32}$&$6.7^{+0.51}_{-0.45}$& \citet{Howard2010} \\
HD 37605&41&2006 Sep 3&2014 Sep 7&2926&$3.8^{+1.7}_{-5.3}$&$2.3^{+0.41}_{-0.35}$&  \citet{Wang2012}\\
HD 38529&96&1996 Dec 1&2014 Aug 19&6470&$0.65^{+0.57}_{-0.55}$&$8.9^{+0.76}_{-0.67}$&  \citet{Wright2009}\\
$\bf{HD 38801}$&17&2006 Sep 3&2014 Sep 7&2926&$\bf{4.1^{+1.2}_{-1.3}}$&$10.^{+3.3}_{-2.2}$&  \citet{Harakawa2010}\\
HD 40979&35&2001 Nov 6&2014 Sep 8&4689&$-0.99^{+1.6}_{-1.6}$&$19.^{+3.1}_{-2.4}$& \citet{Wittenmyer2009}\\
HD 43691&19&2004 Jan 10&2014 Sep 6&3892&$-0.51^{+0.63}_{-0.60}$&$4.8^{+1.7}_{-1.1}$&  \citet{Silva2007}\\
HD 45350&58&1999 Dec 31&2014 Sep 10&5367&$-0.29^{+0.21}_{-0.20}$&$3.7^{+0.46}_{-0.39}$ & \citet{Endl2006} \\
HD 46375&57&1998 Sep 13&2014 Sep 10&5841&$-0.29^{+0.31}_{-0.30}$&$3.8^{+0.55}_{-0.46}$& \citet{Vogt2006}\\
HD 49674&79&2000 Dec 4&2014 Sep 8&5026&$-0.21^{+0.34}_{-0.33}$&$5.2^{+0.50}_{-0.44}$& \citet{Vogt2006}\\
$\bf{HD 50499}$&61&1996 Dec 1&2013 Dec 14&6222&$\bf{ = 0\pm 0 \hspace{0.1cm}\tablenotemark{11}}$&$4.6^{+0.60}_{-0.53}$&  \citet{Vogt2005}\\
$\bf{HD 50554}$&41&1998 Dec 23&2015 Feb 4&5887&$\bf{-1.2^{+0.39}_{-0.37}}$&$4.8^{+0.77}_{-0.64}$& \citet{Vogt2006} \\
HD 52265&65&1998 Jan 25&2014 Sep 7&6069&$0.63^{+0.25}_{-0.24}$&$4.4^{+0.51}_{-0.43}$& \citet{Vogt2006} \\
HIP 57050&43&2000 Feb 6&2013 Dec 14&5060&$0.88^{+0.85}_{-0.85}$&$8.1^{+1.3}_{-1.0}$& \citet{Haghighipour2010} \\
$\bf{HD 66428}$&57&2000 Dec 4&2015 Feb 4&5175&$\bf{-3.1^{+0.23}_{-0.23}}$&$3.5^{+0.45}_{-0.38}$& \citet{Vogt2006} \\
$\bf{HD 68988}$&48&2000 Jan 8&2013 Dec 13&5088&$\bf{ = 0\pm 0 \hspace{0.1cm}\tablenotemark{11} }$&$1.8^{+0.036}_{-0.042}$& \citet{Vogt2006} \\
$\bf{HD 72659}$&61&1998 Jan 25&2015 Feb 4&6219&$\bf{ = 0\pm 0 \hspace{0.1cm}\tablenotemark{11} }$&$3.5^{+0.46}_{-0.40}$&  \citet{Moutou2011}\\
HD 73534&46&2004 Jan 10&2015 Feb 8&4047&$0.62^{+0.29}_{-0.29}$&$3.8^{+0.53}_{-0.44}$&  \citet{Valenti2009}\\
HD 74156&53&2001 Apr 8&2013 Dec 12&4631&$1.9^{+0.73}_{-0.74}$&$6.9^{+0.99}_{-0.85}$&  \citet{Meschiari2011}\\
$\bf{HD 75898}$&54&2004 Jan 10&2015 Feb 5&4044&$ \bf{= 0\pm 0 \hspace{0.1cm}\tablenotemark{11}} $&$2.7^{+0.074}_{-0.076}$& \citet{Robinson2007} \\
HIP 79431&31&2009 Apr 6&2014 Aug 23&1965&$1.8^{+1.9}_{-2.0}$&$6.0^{+1.1}_{-0.88}$ &  \citet{Apps2010}\\
HD 80606&79&2001 Apr 8&2013 Dec 13&4632&$0.23^{+0.27}_{-0.28}$&$3.8^{+0.40}_{-0.35}$& \citet{Moutou2009}\\
HD 82886&35&2007 Apr 26&2013 Dec 12&2422&$-1.2^{+1.4}_{-1.5}$&$9.6^{+1.6}_{-1.3}$& \citet{Johnson2011}\\
HD 83443&37&2000 Dec 19&2015 Feb 8&5164&$-0.081^{+0.64}_{-0.64}$&$5.8^{+1.0}_{-0.82}$& \citet{Vogt2006}\\
$\bf{HD 86081}$&41&2005 Nov 19&2013 Dec 14&2947&$\bf{-1.3^{+0.25}_{-0.25}}$&$4.2^{+0.66}_{-0.55}$&  \citet{Johnson2006}\\
HD 88133&53&2004 Jan 10&2013 Dec 11&3623&$-0.48^{+0.36}_{-0.35}$&$4.7^{+0.61}_{-0.51}$&  \citet{Vogt2006}\\
$\bf{HD 92788}$&37&2000 Jan 8&2014 Feb 20&5157&$\bf{ = 0\pm0 \hspace{0.1cm}\tablenotemark{11}} $&$3.7^{+0.069}_{-0.065}$&  \citet{Vogt2006}\\
HD 95089&37&2007 Apr 26&2013 Dec 12&2422&$ = 0\pm0 \hspace{0.1cm}\tablenotemark{10} $&$7.6^{+1.3}_{-1.1}$&  \citet{Johnson2010}\\
HD 96063&22&2007 Apr 26&2013 Dec 11&2421&$-0.69^{+0.10}_{-1.0}$&$6.0^{+1.5}_{-1.1}$&  \citet{Johnson2011}\\
HD 96167&59&2004 Jan 10&2013 Dec 14&3626&$-0.047^{+0.29}_{-0.29}$&$4.3^{+0.51}_{-0.44}$&  \citet{Peek2009}\\
HD 97658&209&1997 Jan 14&2015 Feb 11&6602&$0.39^{+0.11}_{-0.12}$&$2.9^{+0.16}_{-0.15}$&  \citet{Dragomir2013}\\
HD 99109&54&1998 Dec 24&2013 Dec 11&5466&$-0.73^{+0.56}_{-0.53}$&$7.0^{+0.10}_{-0.84}$ & \citet{Vogt2006}\\
HD 99492&104&1997 Jan 13&2015 Feb 11&6603&$0.42^{+0.19}_{-0.19}$&$4.1^{+0.35}_{-0.31}$ & \citet{Vogt2006}\\
HD 99706&33&2007 Nov 23&2014 Jul 7&2418&$-2.5^{+1.2}_{-1.1}$&$1.7^{+2.5}_{-0.57}$&  \citet{Johnson2011}\\
HD 102195&31&2006 Jan 11&2013 Dec 11&2891&$1.3^{+0.69}_{-0.69}$&$10.^{+1.8}_{-1.4}$&  \citet{Melo2007}\\
HD 102329&27&2007 Apr 26&2013 Dec 11&2421&$3.5^{+1.7}_{-1.8}$&$3.7^{+0.31}_{-0.35}$&  \citet{Johnson2011}\\
HD 102956&31&2007 Apr 26&2013 Aug 9&2297&$0.39^{+1.3}_{-1.3}$&$7.3^{+1.2}_{-1.0}$&  \citet{Johnson2010}\\
HD 104067&61&1997 Jan 13&2013 Dec 14&6179&$-0.16^{+0.43}_{-0.42}$&$6.0^{+0.71}_{-0.60}$&  \citet{Segransan2011}\\
HD 106270&27&2007 Apr 26&2014 Jul 13&2635&$1.9^{+1.7}_{-1.6}$&$12.^{+2.4}_{-1.8}$&  \citet{Johnson2011}\\
HD 107148&57&2000 Jan 9&2013 Dec 11&5085&$0.20^{+0.44}_{-0.50}$&$5.0^{+0.64}_{-0.54}$ &  \citet{Vogt2006}\\
HD 108863&41&2007 Apr 26&2013 Dec 10&2420&$-1.2^{+0.98}_{-0.93}$&$6.5^{+0.91}_{-0.75}$&  \citet{Johnson2011}\\
HD 108874&89&1999 Jun 11&2014 Aug 19&5548&$-0.30^{+0.23}_{-0.23}$&$3.4^{+0.36}_{-0.32}$&  \citet{Wright2009}\\
$\bf{HD 109749}$&28&2004 Jan 10&2013 Dec 14&3626&$\bf{0.75^{+0.19}_{-0.20}}$&$1.9^{+0.48}_{-0.39}$&  \citet{Fischer2006}\\
HD 114729&48&1997 Jan 14&2013 Dec 12&6176&$0.16^{+0.35}_{-0.37}$&$4.2^{+0.61}_{-0.52}$&  \citet{Vogt2006}\\
HD 114783&119&1998 Jun 19&2015 Feb 4&6074&$-0.18^{+0.34}_{-0.34}$&$3.8^{+0.30}_{-0.28}$&  \citet{Wittenmyer2009}\\
HD 116029&28&2007 Apr 26&2014 Aug 25&2678&$0.83^{+1.0}_{-1.0}$&$6.2^{+1.5}_{-1.1}$&  \citet{Johnson2011}\\
HD 117207&52&1997 Jan 14&2014 Jun 18&6364&$-0.074^{+0.33}_{-0.32}$&$3.2^{+0.47}_{-0.41}$&  \citet{Vogt2006}\\
HD 126614&81&1999 Jan 21&2015 Feb 7&5861&$87.^{+1.2}_{-1.6}$&$3.1^{+0.33}_{-0.29}$&  \citet{Howard2010}\\
HD 128311&118&1998 Jun 19&2015 Feb 11&6081&$-0.18^{+0.70}_{-0.69}$&$16.^{+1.2}_{-1.1}$&  \citet{Wittenmyer2009}\\
HD 130322&25&2000 Jul 30&2014 Jun 18&5071&$0.36^{+1.1}_{-1.1}$&$7.0^{+1.5}_{-1.1}$&  \citet{Wittenmyer2009}\\
HD 131496&48&2007 Jun 6&2014 Jul 7&2588&$-1.5^{+1.1}_{-0.99}$&$7.6^{+1.0}_{-0.82}$&  \citet{Johnson2011}\\
HD 134987&103&1996 Jul 12&2015 Feb 11&6788&$-0.32^{+0.68}_{-0.69}$&$3.1^{+0.29}_{-0.26}$ &  \citet{Jones2010}\\
HD 141937&33&2002 Aug 29&2014 Jul 9&4332&$-0.61^{+0.52}_{-0.53}$&$6.3^{+1.1}_{-0.86}$&  \citet{Udry2002}\\
HD 142245&26&2007 Jun 6&2014 Jul 7&2588&$0.82^{+0.76}_{-0.74}$&$6.0^{+1.2}_{-0.91}$&  \citet{Johnson2011}\\
HD 149143&48&2004 Jul 11&2014 Aug 13&3685&$0.12^{+0.40}_{-0.40}$&$6.7^{+0.88}_{-0.75}$&  \citet{Fischer2006}\\
HD 152581&30&2007 Jun 6&2014 Jul 24&2605&$0.22^{+0.71}_{-0.71}$&$5.1^{+0.93}_{-0.75}$&  \citet{Johnson2011}\\
HD 154345&113&1997 Apr 8&2015 Feb 4&6511&$0.053^{+0.19}_{-0.19}$&$2.8^{+0.25}_{-0.22}$&  \citet{Wright2008}\\
HD 156279&73&2003 Jul 12&2015 Feb 4&4225&$-5.1^{+6.0}_{-2.8}$&$2.2^{+0.27}_{-0.23}$&  \citet{Diaz2012}\\
HD 156668&219&2003 Jul 12&2015 Feb 5&4226&$-0.24^{+0.080}_{-0.085}$&$2.060^{+0.092}_{-0.085}$&  \citet{Howard2011}\\
$\bf{HD 158038}$&33&2007 Jun 6&2015 Feb 4&2800&$ \bf{= 0\pm0 \hspace{0.1cm}\tablenotemark{11}}$&$12.^{+4.4}_{-2.0}$&  \citet{Johnson2011}\\
$\bf{HD 163607}$&66&2005 Jul 19&2015 Feb 4&3487&$\bf{2.3^{+0.37}_{-0.39}}$&$4.4^{+0.50}_{-0.43}$&  \citet{Giguere2012}\\
HD 164509&57&2005 Jul 19&2014 Sep 8&3338&$-3.3^{+0.56}_{-0.53}$&$6.2^{+0.76}_{-0.64}$&  \citet{Giguere2012}\\
HD 164922&166&1996 Jul 11&2015 Feb 8&6786&$-0.030^{+0.10}_{-0.10}$&$3.0^{+0.20}_{-0.18}$&  \citet{Vogt2006}\\
$\bf{HD 168443}$&139&1996 Jul 12&2014 Aug 11&6604&$\bf{-3.0^{+0.16}_{-0.16}}$&$3.6^{+0.27}_{-0.25}$&  \citet{Pilyavsky2011}\\
HD 168746&27&2000 Jul 30&2014 Jul 26&5109&$-0.24^{+0.30}_{-0.30}$&$2.6^{+0.77}_{-0.61}$ &  \citet{Vogt2006}\\
HD 169830&52&2000 Jul 30&2014 Sep 10&5155&$-0.30^{+0.30}_{-0.29}$&$4.5^{+0.64}_{-0.53}$&  \citet{Mayor2004}\\
HD 170469&42&2000 Jun 10&2014 Jun 22&5125&$0.93^{+0.50}_{-0.52}$&$4.5^{+0.69}_{-0.56}$&  \citet{Fischer2007}\\
HD 175541&81&1996 Jul 19&2014 Jul 25&6580&$0.66^{+0.40}_{-0.41}$&$6.4^{+0.61}_{-0.53}$&  \citet{Johnson2007}\\
HD 177830&121&1996 Jul 11&2014 Sep 6&6631&$0.097^{+0.27}_{-0.27}$&$4.7^{+0.36}_{-0.33}$&  \citet{Vogt2006}\\
HD 178911b&41&1999 Jun 12&2014 Aug 11&5539&$-0.070^{+0.47}_{-0.47}$&$5.5^{+0.82}_{-0.66}$&  \citet{Wittenmyer2009}\\
HD 179079&84&2004 Jul 11&2014 Sep 8&3711&$0.15^{+0.33}_{-0.33}$&$3.9^{+0.38}_{-0.33}$&  \citet{Valenti2009}\\
$\bf{HD 180902}$&26&2007 Aug 27&2014 Aug 11&2541&$\bf{470^{+5.7}_{-6.0}}$&$4.4^{+1.0}_{-0.77}$&  \citet{Johnson2010}\\
HD 181342&30&2007 Aug 27&2014 Aug 11&2541&$-0.43^{+1.5}_{-1.5}$&$12.^{+2.0}_{-1.6}$ &  \citet{Johnson2010}\\
HD 183263&73&2001 Jul 4&2014 Dec 11&4908&$-2.5^{+4.6}_{-2.6}$&$3.4^{+0.39}_{-0.34}$&  \citet{Wright2009}\\
HD 187123&113&1997 Dec 23&2014 Sep 6&6101&$-0.22^{+0.24}_{-0.22}$&$2.4^{+0.23}_{-0.21}$&  \citet{Wright2009}\\
HD 188015&63&2000 Jul 29&2014 Sep 6&5152&$-0.21^{+0.33}_{-0.33}$&$4.7^{+0.54}_{-0.47}$&  \citet{Vogt2006}\\
HD 189733&28&2003 Jul 12&2014 Aug 24&4061&$-0.63^{+1.0}_{-1.0}$&$14.^{+2.6}_{-2.0}$&  \citet{Bouchy2005}\\
HD 190228&31&2002 Aug 28&2013 Dec 11&4123&$-0.36^{+0.61}_{-0.61}$&$4.7^{+0.90}_{-0.70}$&  \citet{Wittenmyer2009}\\
HD 190360&150&1996 Oct 9&2014 Jun 22&6465&$-0.32^{+0.15}_{-0.15}$&$2.8^{+0.21}_{-0.20}$&  \citet{Wright2009}\\
HD 192263&39&1998 Jun 19&2014 Sep 7&5924&$-0.38^{+0.69}_{-0.66}$&$8.1^{+1.2}_{-0.97}$&  \citet{Vogt2006}\\
HD 192310&112&2004 Aug 20&2014 Sep 10&3673&$0.26^{+0.14}_{-0.14}$&$2.1^{+0.19}_{-0.17}$&  \citet{Pepe2011}\\
HD 195019&57&1998 Sep 12&2014 Sep 11&5843&$1.1^{+0.37}_{-0.36}$&$4.7^{+0.58}_{-0.50}$&  \citet{Vogt2006}\\
HD 200964&58&2007 Oct 26&2014 Sep 11&2512&$-0.38^{+0.55}_{-0.53}$&$5.1^{+0.65}_{-0.54}$&  \citet{Johnson2011}\\
$\bf{HD 206610}$&38&2007 Aug 1&2014 Aug 9&2565&$\bf{-8.8^{+0.68}_{-0.63}}$&$4.9^{+0.80}_{-0.65}$&  \citet{Johnson2010}\\
HD 207832&63&2004 Jul 4&2013 Oct 20&3395&$-2.3^{+1.3}_{-0.89}$&$7.7^{+0.99}_{-0.82}$&  \citet{Haghighipour2012}\\
HD 209458&81&1999 Jun 11&2014 Aug 19&5548&$0.10^{+0.32}_{-0.33}$&$5.9^{+0.56}_{-0.49}$ &  \citet{Torres2008}\\
HD 210277&139&1996 Jul 12&2014 Jul 22&6584&$0.36^{+0.16}_{-0.16}$&$3.6^{+0.26}_{-0.24}$&  \citet{Vogt2006}\\
HD 212771&30&2007 Aug 27&2014 Aug 14&2544&$2.1^{+1.1}_{-1.2}$&$8.3^{+1.5}_{-1.2}$&  \citet{Johnson2010}\\
HD 217107&123&1998 Sep 12&2014 Sep 8&5840&$-0.36^{+0.50}_{-0.47}$&$3.6^{+0.40}_{-0.30}$&  \citet{Wright2009}\\
HD 222582&51&1997 Dec 23&2014 Aug 4&6068&$-0.21^{+0.32}_{-0.32}$&$3.3^{+0.47}_{-0.40}$&  \citet{Vogt2006}\\
HD 224693&38&2004 Jul 4&2014 Aug 14&3693&$0.66^{+0.47}_{-0.47}$&$6.1^{+1.1}_{-0.85}$&  \citet{Johnson2006}\\
HD 231701&28&2004 Jul 4&2014 Aug 11&3690&$0.067^{+0.79}_{-0.80}$&$6.4^{+1.5}_{-1.1}$&  \citet{Fischer2007}
\enddata
\tablenotetext{10}{Because this system has a new outer planet whose period is just covered by the RV baseline, we fix the trend to zero.}
\tablenotetext{11}{Because the RV accelerations in systems HD 50499, HD 68988, HD 72659, HD 75898, HD 92788, and HD 158038 have some curvature, we fit them with a two planet solution.  Since the partially resolved orbit and linear trend are degenerate, we fix the slope to zero in these fits.  During these fits, we also fix the poorly constrained eccentricity of the outer planet to zero.  One caveat is that we assume that the residual RV signals are due to a single body, even though they could be the sum of multiple bodies.}
\tablecomments{Systems with $3\sigma$ trends and above are listed in bold.}
\end{deluxetable*}

\begin{deluxetable}{llllll}
\tabletypesize{\scriptsize}
\tablecaption{Summary of AO Observations}
\tablewidth{0pt}
\tablehead{
\colhead{Target} & \colhead{UT Obs. Date} & \colhead{Filter} & \colhead{Array} & \colhead{$T_{\mathrm{int}}$ [s]} & \colhead{$N_{\mathrm{exp}}$}
}
\startdata
HD 3651 & 2013 Aug 19 &  $K_{\rm cont}$ & 256 & 9.0 & 12 \\
HD 4208 & 2013 Nov 17 & $K_{\rm cont}$ & 1024 & 10.0 & 15 \\
HD 11506 & 2013 Nov 17 & $K_{\rm cont}$ & 1024 & 10.0 & 15 \\
HD 24040 & 2015 Jan 10	&  $K_{\rm cont}$	& 1024 & 13.6 & 12\\
HD 28678 & 2014 Oct 04 & $K_{\rm cont}$ & 1024 & 13.6 & 12 \\
HD 38801 & 2014 Dec 7 &  $K_{\rm cont}$& 1024 & 12.5 & 12 \\
HD 38801 & 2014 Jan 12 &  $K_p$ & 1024 & 9.0 & 9 \\
HD 50499 & 2014 Nov 07 & $K_{\rm cont}$& 1024 & 13.6 & 12 \\
HD 50554 & 2013 Dec 18 & $K_{\rm cont}$ & 1024 & 10.0 & 12 \\
HD 66428 & 2013 Dec 18 &  $K_{\rm cont}$& 1024 & 10.0 & 12 \\
HD 68988 & 2013 Dec 18 &  $K_{\rm cont}$& 1024 & 10.0 & 12 \\
HD 72659 & 2014 Jan 12 & $K_{\rm cont}$ & 1024 & 9.0 & 15 \\
HD 72659 & 2014 Nov 10 &  $K_{\rm cont}$ & 1024 & 13.6 & 12 \\
HD 75898 & 2014 May 21 & Kc & 1024 & 12.5 & 12 \\
HD 75898 & 2014 May 21 &  Jc & 1024 & 12.5 & 12 \\
HD 86081 & 2013 Dec 18 &  $K_{\rm cont}$ & 1024 & 10.0 & 12 \\
HD 86081 & 2014 Dec 5 & $K_{\rm cont}$ & 1024 & 12.0 & 12 \\
HD 92788 & 2014 Dec 5 & $K_{\rm cont}$& 1024 & 13.6  &  12\\
HD 109749 & 2014 Jun 09 &  $K_{\rm cont}$ & 1024 & 12.5 & 12 \\
HD 158038 & 2013 Jul 17 &  BrG & 1024 & 2.8 & 25 \\
HD 163607 & 2013 Aug 19 &  $K_{\rm cont}$ & 1024 & 9.0 & 12 \\
HD 168443 & 2013 Aug 19 &  $K_{\rm cont}$ & 512 & 10.0 & 12 \\
HD 180902 & 2014 Jul 12 & $K_{\rm cont}$ & 1024 & 13.6 & 12 \\
HD 206610 & 2013 Aug 19 &  $K_{\rm cont}$& 1024 & 9.0 & 12
\enddata
\tablecomments{The ``Array'' column denotes the horizontal width, in pixels, of the section of the detector used to capture the image. All PHARO images are taken in the full 1024$\times$1024 array. The NIRC2 array dimensions used in this survey were 1024$\times$1024 (the full array), 512$\times$512, or 256$\times$264. These dimensions are constrained by NIRC2's readout software. The $T_{\rm int}$ column indicates the total integration time of a single exposure, in seconds, and the $N_{\rm exp}$ column indicates the number of exposures used in the final stacked image. System HD158038 was imaged using PHARO; the rest were imaged using NIRC2.}
\end{deluxetable}

\section{Analysis}

\subsection{Radial Velocity Fitting}

The presence of a distant, massive companion manifests as a long-term acceleration for observations with baselines significantly shorter than the companion's orbital period (e.g. Crepp et al 2012).  To detect and quantify the significance of these ``trends", we performed a uniform analysis of these systems using a Markov Chain Monte Carlo technique.  

The initial set of parameter values for the MCMC run were determined using a $\chi^2$ minimization fitting procedure.  For a single-planet system, the MCMC algorithm simultaneously fit eight free parameters to the RV data - six orbital parameters (the velocity semi-amplitude, the period of the orbit, the eccentricity of the orbit, the argument of periastron, the true anomaly of the planet at a given time, and the arbitrary RV zero point), a linear velocity trend, and a stellar jitter term \citep{Isaacson2010}.  This additional error term is added to the internal uncertainty of each radial velocity measurement in quadrature.  All parameters had uniform priors.  While it is formally correct to use log priors for parameters such as the velocity semi-amplitude, jitter term, and linear trend, we find that our use of uniform priors has a negligible effect on our posterior PDFs.  We initialize our MCMC chains using the published parameters for the inner planets in these systems, which are typically quite close to our final best-fit parameters.  Furthermore, we note that the choice of prior should only affect the posterior probability distributions in the data-poor regime; in this case the data provide good constraints on the parameters in question, and as a result the posterior PDF is effectively independent of our choice of prior.  The likelihood function used in this analysis is given in Equation 1, where $\sigma_i$ is the instrumental error, $\sigma_{jit}$ is the stellar jitter, $v$ are the data, and $m$ is the model.  

\begin{equation}
\mathscr{L} = \frac{1}{\sqrt{2\pi}\sqrt{\sigma_i^2 + \sigma_{jit}^2}}\exp\bigg(-0.5\bigg(\frac{(v - m)^2}{\sigma_i^2 + \sigma_{jit}^2}\bigg)\bigg)
\end{equation}

\noindent The confidence intervals on each parameter were obtained from their posterior distribution functions.

On August 19 2004, the HIRES CCD was upgraded, leading to a different RV zero point for data taken before and after this date.  For systems with Keck HIRES RVs obtained prior to 2004, we include an offset parameter between the two datasets as an additional free parameter.  Although there is some evidence that the post-upgrade jitter is lower than the pre-upgrade jitter by approximately 1 m/s (e.g. Howard et al. 2014), we find that this change is much smaller than the average jitter level for the majority of our targets, and our decision to fit a single jitter term across both epochs is therefore unlikely to have a significant effect on our conclusions.  Approximately 30$\%$ of our targets have no pre-upgrade data at all, while an additional 50$\%$ have fewer than ten data points pre- or post-upgrade, making it difficult to obtain meaningful constraints on the change in jitter between these two epochs (e.g., Fulton et al. 2015).  We therefore conclude that a uniform approach to these fits is preferable to a more customized approach in which we include two separate jitter terms for the approximately 20$\%$ of systems where such an approach is feasible.

In addition to reproducing the published solutions of confirmed exoplanets, we detected eight new long-period planets with fully resolved orbits in systems GJ 317, HD 4203, HD 33142, HD 95089, HD 99706, HD 102329, HD 116029, and HD 156279.  Trends were previously mentioned for GJ 317 \citep{Anglada2012}, HD 4203 \citep{Vogt2006}, HD 95089 \citep{Johnson2010}, HD 99706 \citep{Johnson2011}, and HD 116029 \citep{Johnson2011}.  

We note that the two planets in HD 116029 are in 3:2 period commensurability.  To assess whether a dynamical model fit was needed, we used the Mercury integrator to numerically integrate the orbits of both planets in HD 116029 in order to determine the magnitude of the change in orbital parameters.  We found that over the observational window of $\sim 8$ years, the orbital elements of both planets varied by less than a fraction of a percent.  Thus we conclude that a Keplerian model fit is sufficient to characterize the planets in HD 116029.  Relevant characteristics of the new outer planets are listed in Table 5, and the corresponding RV solutions are plotted in Figures 3 through 10.  RV measurements for these eight systems are listed in Table 4.

\begin{figure}[h]
\includegraphics[width=3.5 in]{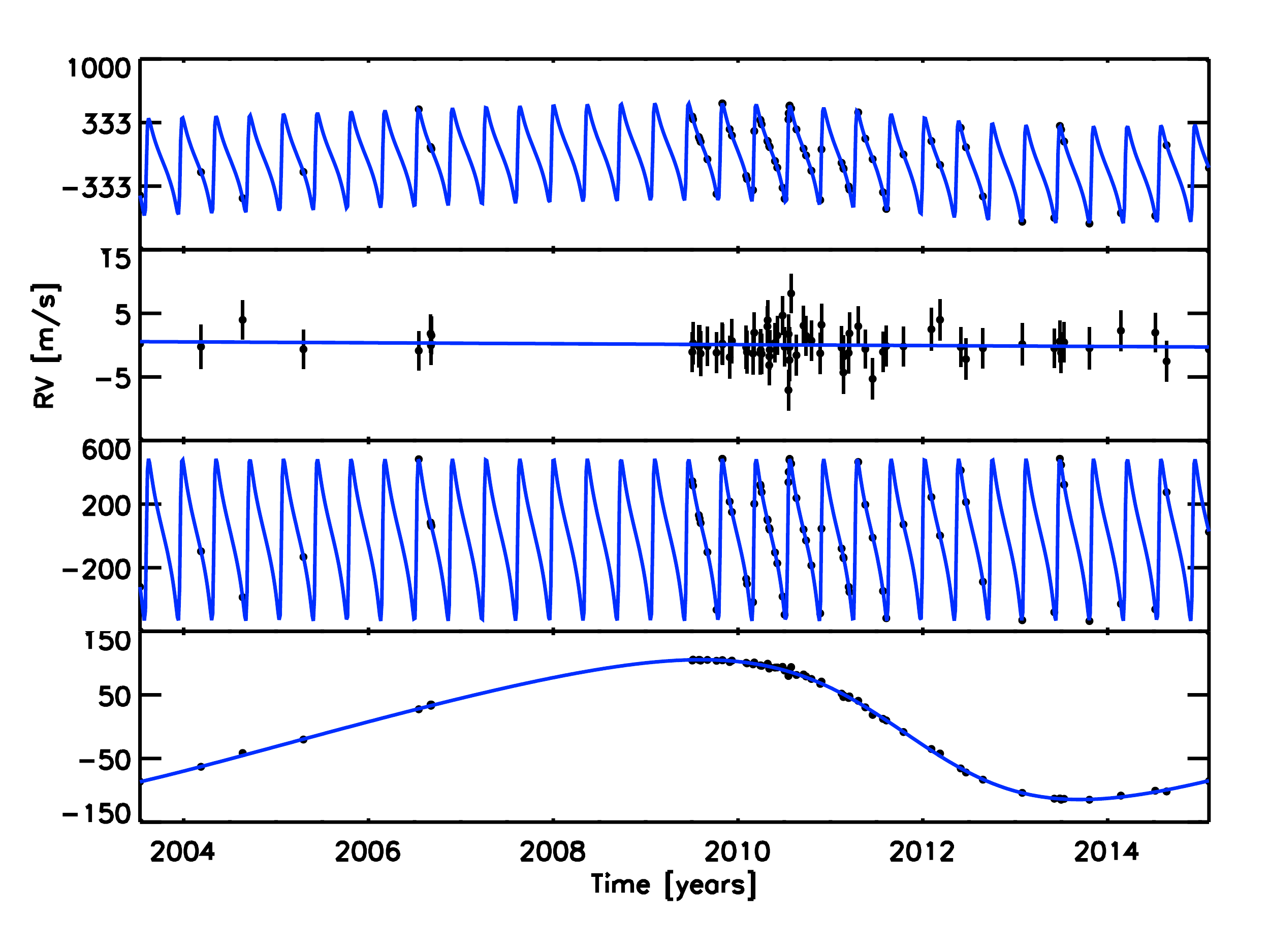}
\caption{RV measurements and best fit models for HD 156279.  The first and second panels show the combined two planet orbital solution and the residuals of that fit, respectively.  The third plot shows the orbital solution for the inner planet after the outer planet solution and trend were subtracted, while the fourth plot shows the outer planet orbital solution with the inner planet and trend subtracted.  }
\end{figure}

\begin{figure}[h]
\includegraphics[width=3.5 in]{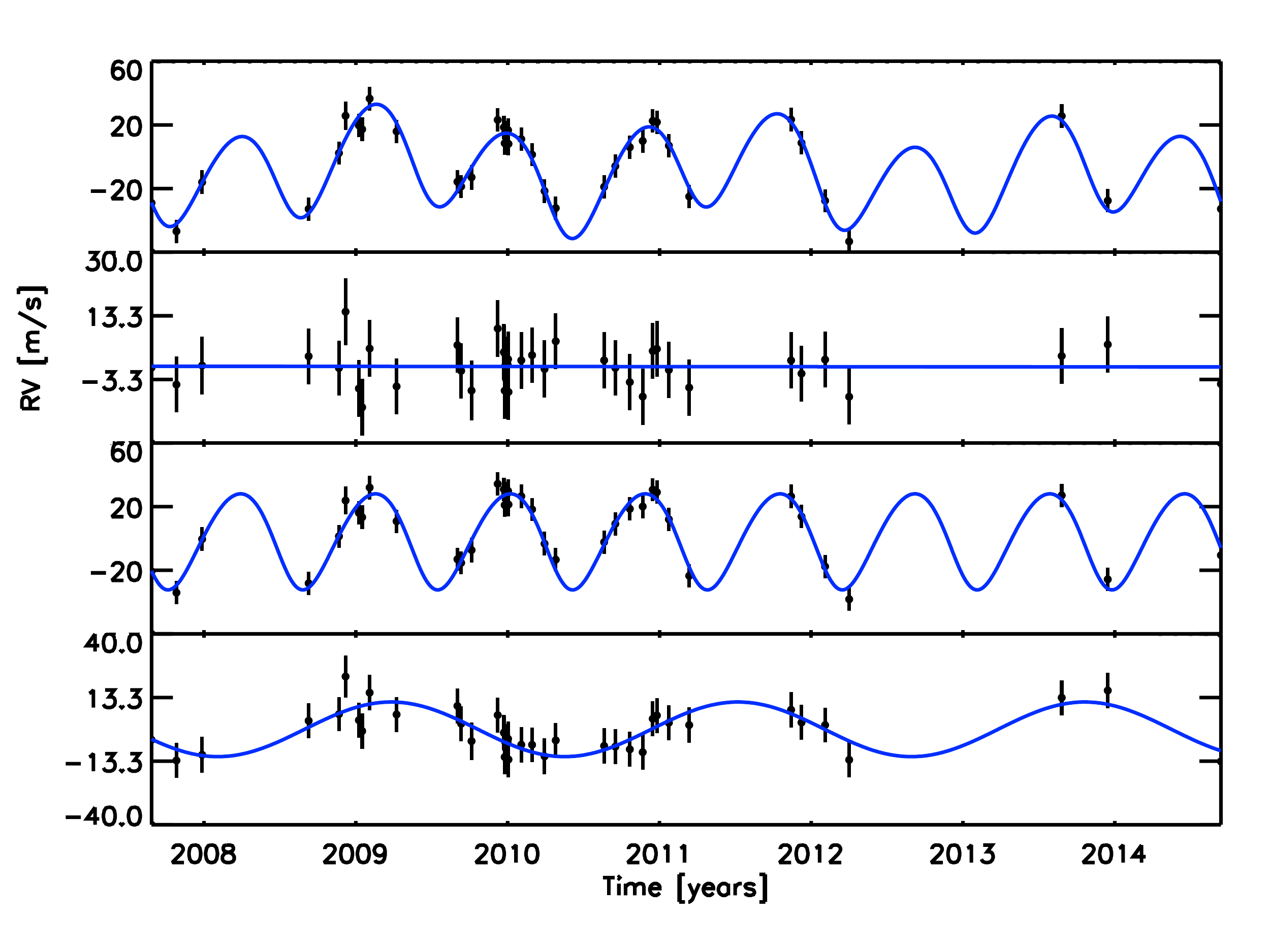}
\caption{RV measurements and best fit models for the systems HD 33142.  See caption to Figure 3 for more information.}
\end{figure}

\begin{figure}[h]
\includegraphics[width=3.5 in]{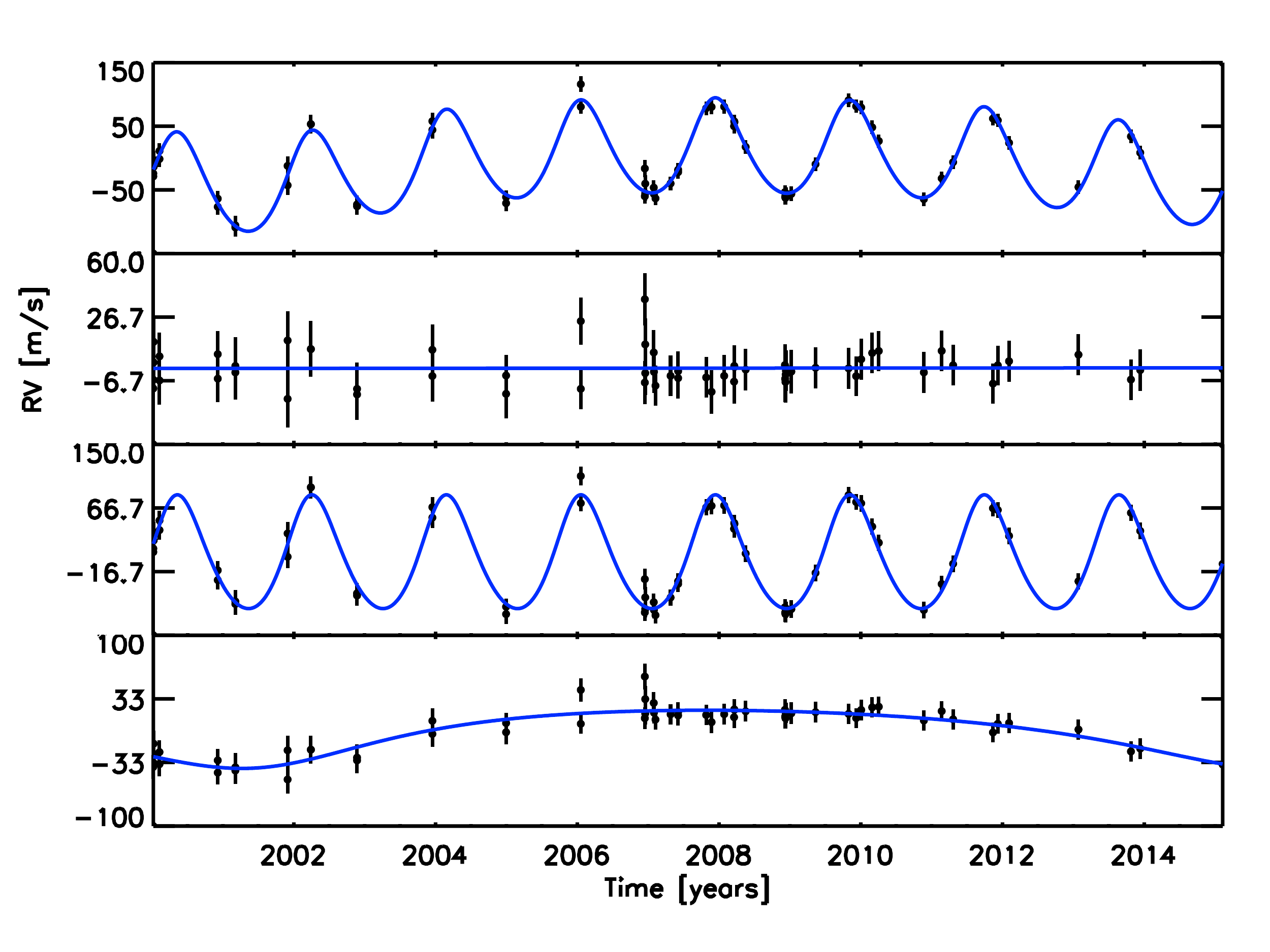}
\caption{RV measurements and best fit models for GJ 317.  See caption to Figure 3 for more information.}
\end{figure}

\begin{figure}[h]
\includegraphics[width=3.5 in]{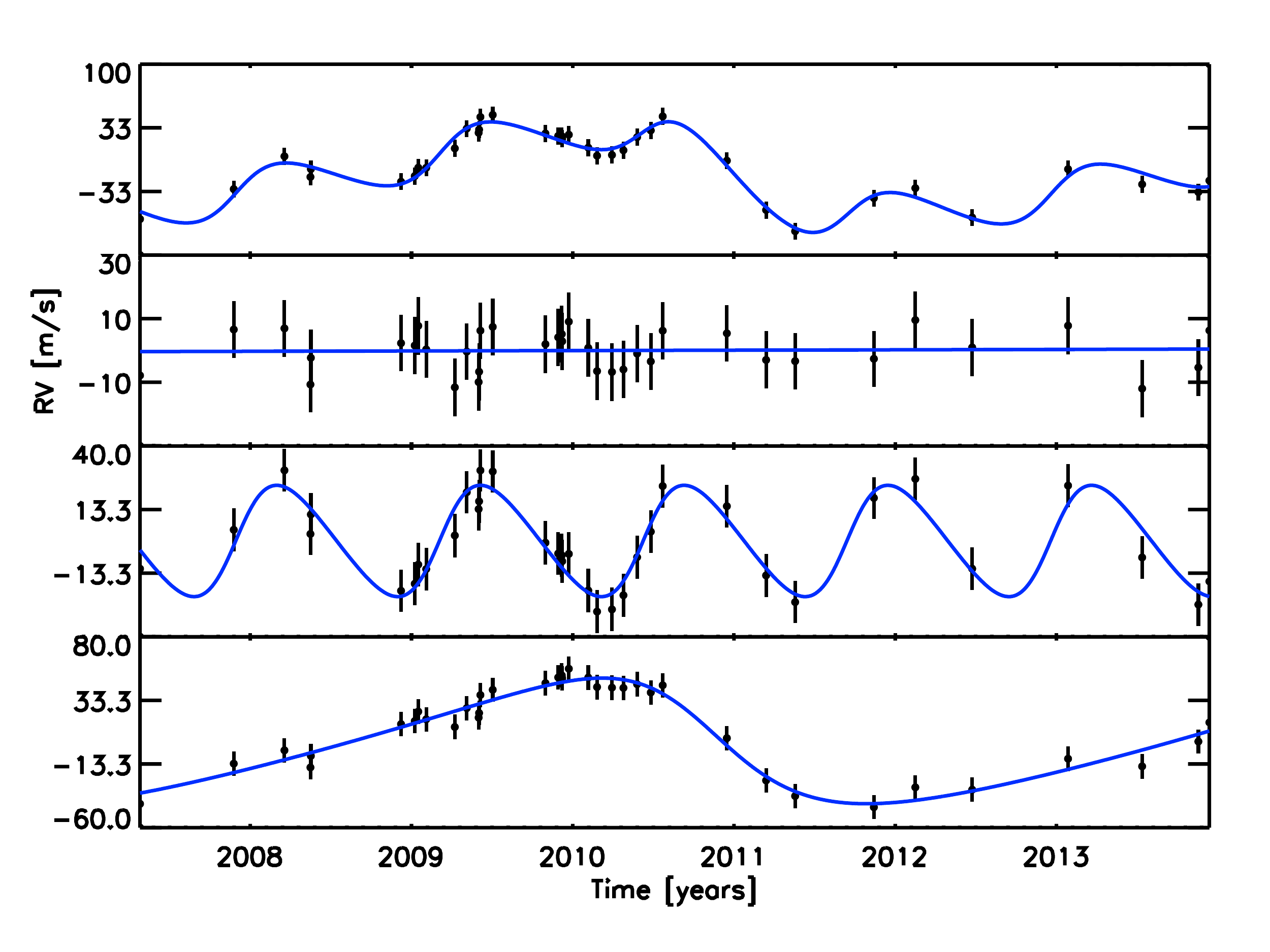}
\caption{RV measurements and best fit models for HD 95089.  See caption to Figure 3 for more information.}
\end{figure}

\begin{figure}[h]
\includegraphics[width=3.5 in]{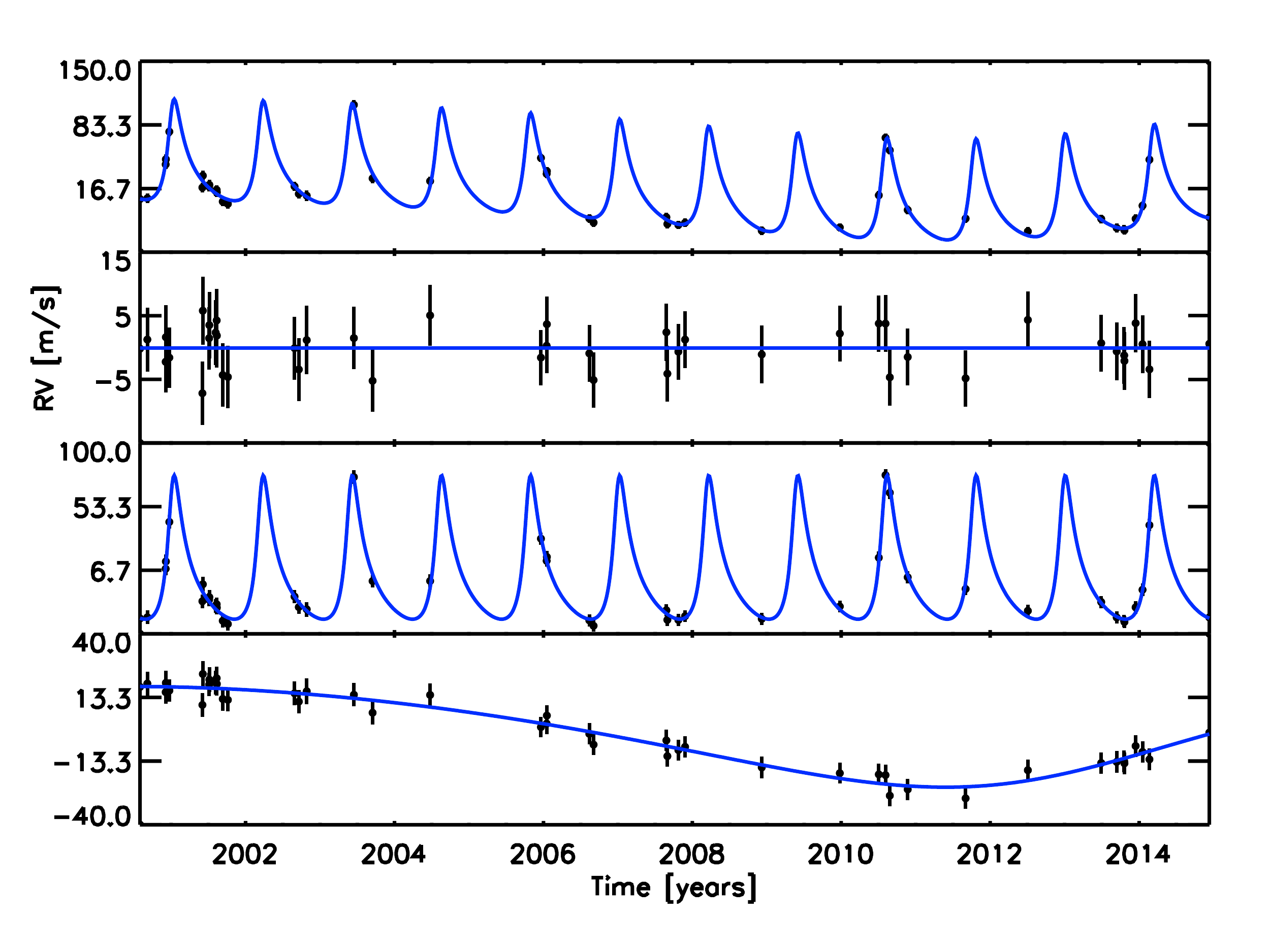}
\caption{RV measurements and best fit models for HD 4203.  See caption to Figure 3 for more information.}
\end{figure}

\begin{figure}[h]
\includegraphics[width=3.5 in]{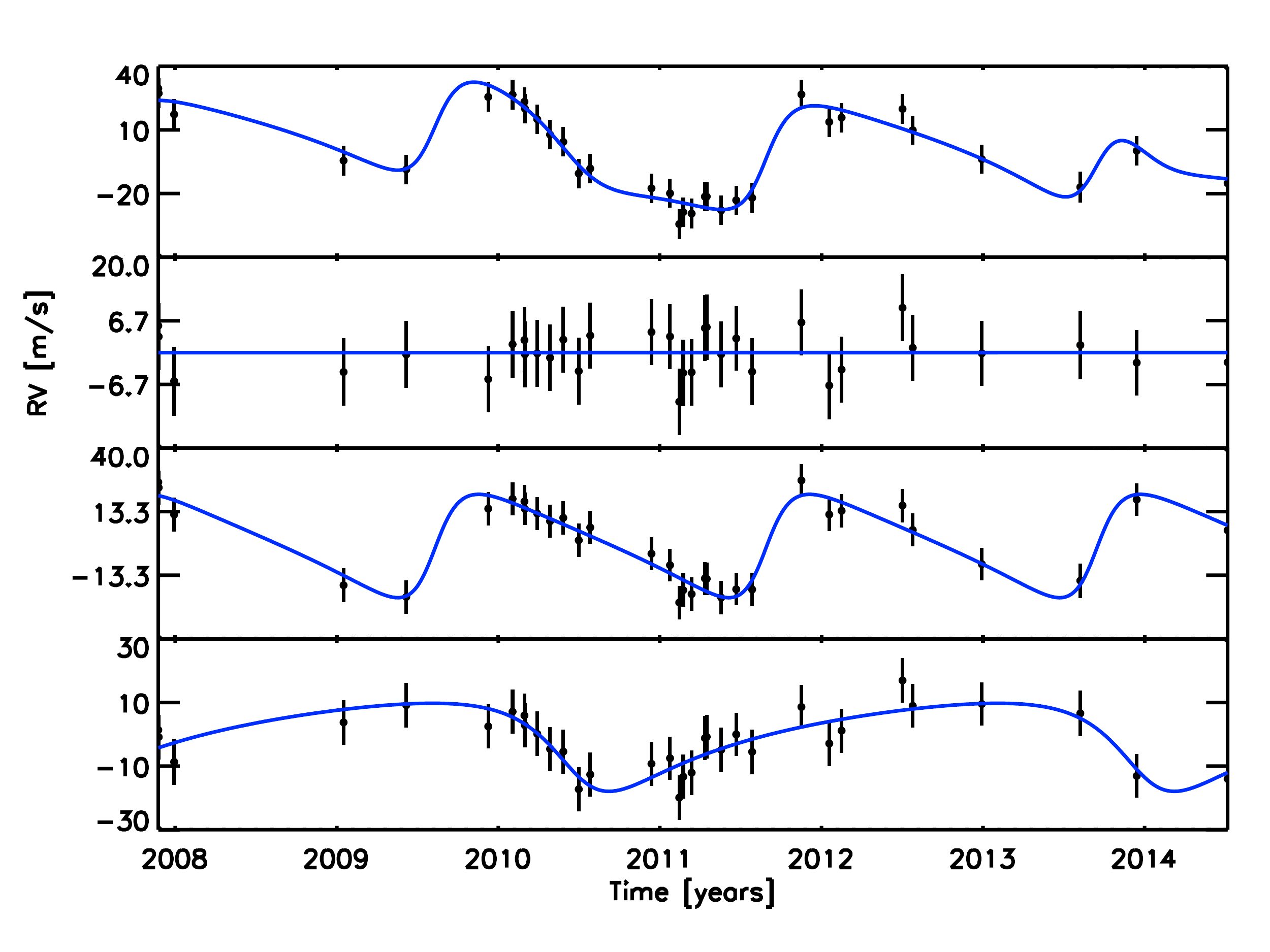}
\caption{RV measurements and best fit models for HD 99706.  See caption to Figure 3 for more information.}
\end{figure}

\begin{figure}[h]
\includegraphics[width=3.5 in]{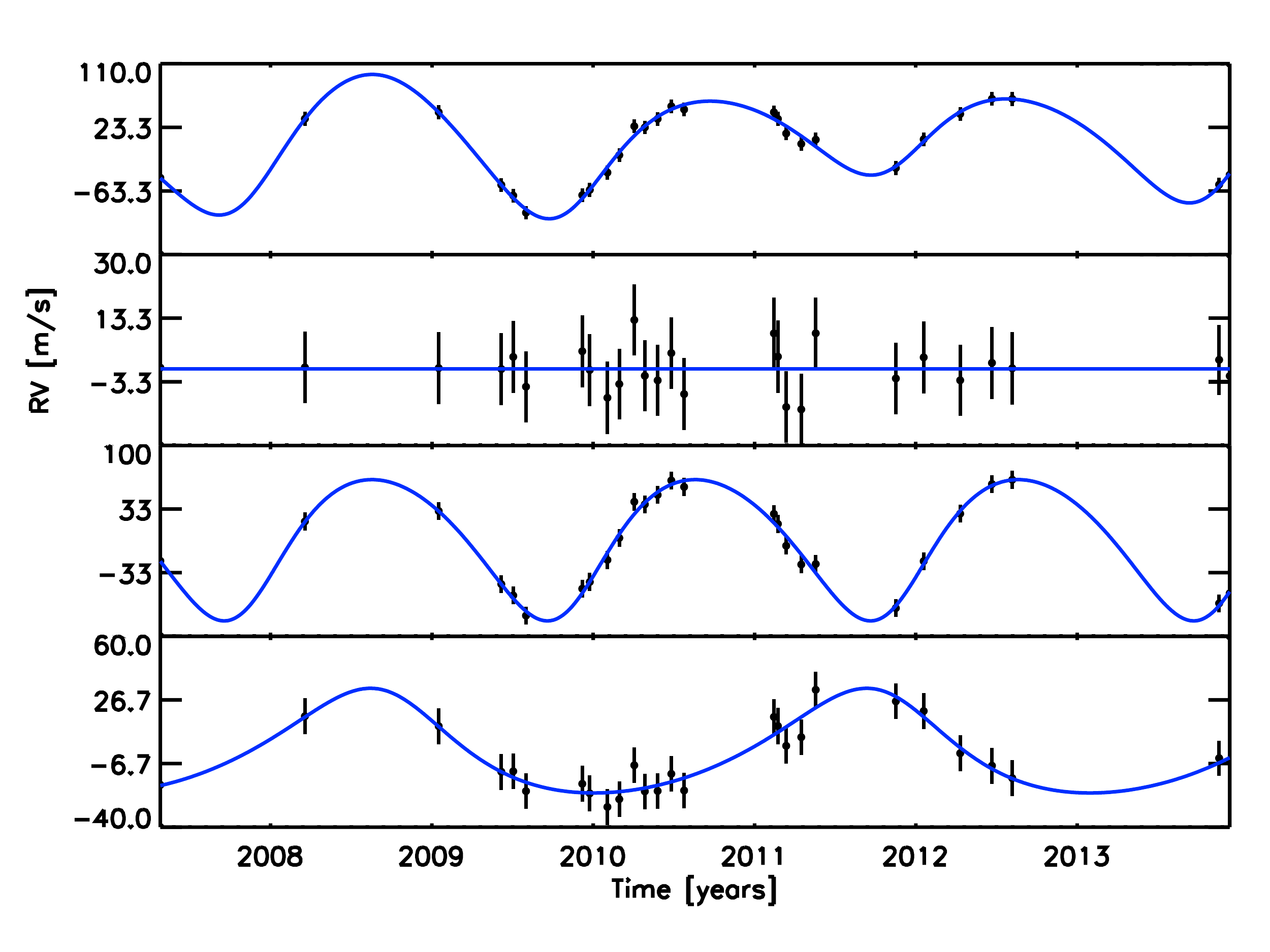}
\caption{RV measurements and best fit models for HD 102329. See caption to Figure 3 for more information.}
\end{figure}

\begin{figure}[h]
\includegraphics[width=3.5 in]{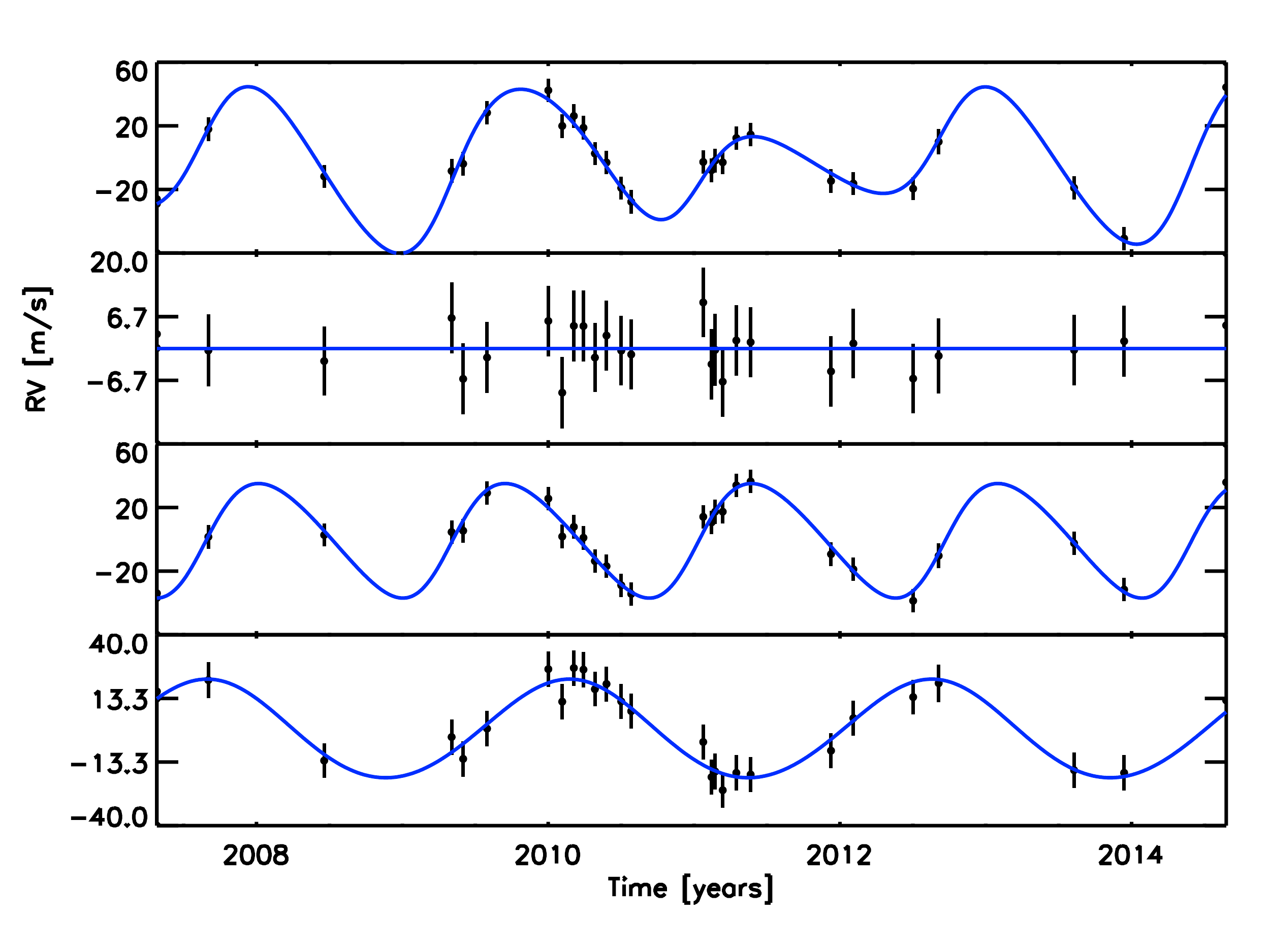}
\caption{RV measurements and best fit models for HD 116029.  See caption to Figure 3 for more information.}
\end{figure}


We considered a linear trend detection to be statistically significant if the best-fit slope differed from zero by more than 3$\sigma$, and report best-fit trend slopes and stellar jitter values for all systems in Table 2.  The nominal values quoted in this table are taken from the $\chi^2$ fits, and the errors come from the MCMC analysis.  We detected 20 statistically significant trends due to the presence of an outer companion.  We find that all but 16 of our orbital solutions for the known inner planets in these systems were consistent with the published orbits at the 2$\sigma$ level or better.  

Of the solutions that changed, the majority were systems with long-period planets for which our newly extended baseline provided a more tightly constrained orbital solution.  This longer baseline was particularly important for systems with both long-period planets and RV accelerations, such as HD 190360.  We present updated orbital solutions for all of the planets outside 3 AU in Table 5.  We defer the publication of updated orbits for planets inside 3 AU and individual radial velocities for all systems to future publications, as these systems are the subject of other research projects currently in progress.

\begin{deluxetable}{lccc}
\tabletypesize{\scriptsize}
\tablecaption{RVs for Systems With New Planets:  HD 156279, HD 33142, GJ 317, \\HD 90589, HD 4203, HD 99706, HD 102329, HD 116029}
\tablewidth{0pc}
\tablehead{
\colhead{System} &
\colhead{JD - 2,440,000}&
\colhead{RV [m s$^{-1}$]}&
\colhead{$\sigma_{RV}$ [m s$^{-1}$]}  
}
\startdata
HD 156279 &12832.9&-436.3&1.084\\
HD 156279 &13074.1&-191.6&1.302\\
HD 156279 &13238.8&-465.2&0.870\\
HD 156279 &13479.0&-184.7&0.874\\
HD 156279 &13934.9&471.5&0.896\\
HD 156279 &13981.8&77.0&0.814\\
HD 156279 &13982.9&67.7&0.843\\
HD 156279 &13983.8&61.7&0.845\\
HD 156279 &13984.9&55.8&0.785\\
HD 156279 &15016.0&395.2&0.951
\enddata
\tablecomments{The full set of RVs for each of these systems are available as electronic tables online.}

\end{deluxetable}

\begin{deluxetable*}{lccccccc}
\tabletypesize{\scriptsize}
\tablecaption{Updated Orbital Solutions for Planets Outside 3 AU and 8 New Planets}
\tablewidth{0pc}
\tablehead{
\colhead{Planet}         &
\colhead{Period [days]}   &
\colhead{T$_P$ -2,440,000 [days]} &
\colhead{Eccentricity}   &
\colhead{$\omega$ [deg]} &
\colhead{K $[\rm m s^{-1}]$}&
\colhead{Mass [$M_{\rm Jup}$]} &
\colhead{Stellar Mass [$M_{\odot}$]}
}
\startdata
HD 13931 b&$4460^{+77}_{-67}$&  $13359^{+1592}_{-826}$ & $0.033^{+0.030}_{-0.017}$&$18^{+123}_{-67}$&$23.92^{+0.90}_{-0.85}$&$1.92^{+0.08}_{-0.07}$  &  $1.022^{+0.020}_{-0.022}$\\
HD 24040 b&$3498^{+23}_{-23}$&  $12264^{+467}_{-348}$&$0.010^{+0.015}_{-0.009}$&$332^{+48}_{-36}$&$51.4^{+1.4}_{-1.4}$&$4.08^{+0.11}_{-0.11}$  &  $1.18^{+0.10}_{-0.10}$\\
HD 33636 b&$2112.6^{+1.6}_{-1.6}$&$13305.9^{+3.5}_{-3.4}$&$0.488^{+0.005}_{-0.005}$&$336.18^{+0.90}_{-0.88}$&$160.9^{+1.0}_{-1.1}$&$8.98^{+0.06}_{-0.06}$  &  $1.017^{+0.032}_{-0.032}$\\
HD 50499 b&$2453^{+27}_{-27}$&  $13612^{+65}_{-67}$&$0.334^{+0.059}_{-0.059}$&$241^{+15}_{-14}$&$18.4^{+1.6}_{-1.3}$&$1.36^{+0.12}_{-0.10}$  &  $1.280^{+0.034}_{-0.080}$\\
HD 66428 b&$2280.4^{+6.6}_{-6.6}$&  $12277^{+20}_{-20}$&$0.448^{+0.016}_{-0.015}$&$179.7^{+3.1}_{-3.1}$&$51.4^{+1.5}_{-1.4}$&$3.09^{+0.07}_{-0.07}$  &  $1.061^{+0.070}_{-0.056}$\\
HD 72659 b&$3506^{+40}_{-38}$&  $15301^{+54}_{-59}$ &$0.249^{+0.028}_{-0.027}$&$272.7^{+8.4}_{-6.9}$&$39.0^{+2.4}_{-2.1}$&$2.99^{+0.19}_{-0.17}$  &  $1.068^{+0.022}_{-0.022}$\\
HD 73534 b&$1707^{+37}_{-35}$&  $14981^{+808}_{-280}$ &$0.022^{+0.058}_{-0.037}$&$83^{+171}_{-60}$&$15.2^{+1.1}_{-1.0}$&$1.02^{+0.07}_{-0.07}$  &  $1.170^{+0.070}_{-0.070}$\\
HD 106270 b&$1872^{+20}_{-19}$&  $14774^{+32}_{-28}$&$0.197^{+0.035}_{-0.035}$&$7.5^{+6.1}_{-5.2}$&$137.3^{+4.4}_{-4.3}$&$9.78^{+0.28}_{-0.28}$  &  $1.330^{+0.050}_{-0.050}$\\
HD 117207 b&$2628^{+21}_{-20}$&  $13325^{+83}_{-83}$&$0.150^{+0.026}_{-0.027}$&$85^{+12}_{-12}$&$27.8^{+0.95}_{-0.94}$&$1.90^{+0.07}_{-0.06}$  &  $1.031^{+0.046}_{-0.040}$\\
HD 154345 b&$3267^{+33}_{-33}$&  $15278^{+197}_{-359}$&$0.038^{+0.027}_{-0.021}$&$341^{+22}_{-40}$&$17.05^{+0.48}_{-0.49}$&$1.15^{+0.03}_{-0.03}$  &  $0.893^{+0.038}_{-0.038}$\\
$\bf{GJ}$ $\bf{317}$ $\bf{c}$  &  $5312^{+758}_{-1248}$  &  $17424^{+1913}_{-3660}$&$0.308^{+0.065}_{-0.079}$  &  $194^{+27}_{-31}$  &  $30^{+36}_{-14}$  &  $1.54^{+1.26}_{-0.57}$  &  $0.240^{+0.040}_{-0.040}$\\
$\bf{HD}$ $\bf{4203}$ $\bf{c}$&$7053^{+1624}_{-2324}$&  $16179^{+1365}_{-1733}$&$0.182^{+0.124}_{-0.172}$&$232.2^{+30.7}_{-32.5}$&$12.5^{+11.0}_{-5.0}$&$1.51^{+0.98}_{-0.57}$  &  $1.130^{+0.028}_{-0.100}$\\
HD 11964A c&$1956^{+26}_{-25}$&  $14189^{+682}_{-341}$&$0.073^{+0.051}_{-0.037}$&$158^{+125}_{-64}$&$9.00^{+0.45}_{-0.45}$&$0.583^{+0.029}_{-0.029}$  &  $1.080^{+0.028}_{-0.012}$\\
$\bf{HD}$ $\bf{33142}$ $\bf{c}$ & $834^{+29}_{-24}$  &  $15664^{+326}_{-117}$&$0.05^{+0.172}_{-0.114}$  &  $322^{+139}_{-53}$  &  $ 11.4^{+2.0}_{-1.9}$  &  $5.97^{+1.04}_{-0.80}$  &  $1.620^{+0.090}_{-0.090}$\\
HD 37605 c&$2455^{+468}_{-148}$&  $14285^{+151}_{-213}$&$0.^{+0.055}_{-0.029}$&$136^{+18}_{-28}$&$426^{+9.1}_{-3.1}$&$3.37^{+0.83}_{-0.26}$  &  $1.00^{+0.50}_{-0.50}$\\
HD 38529 c&$2132.4^{+3.2}_{-3.2}$&  $14398.1^{+8.0}_{-8.0}$&$0.342^{+0.007}_{-0.007}$&$19.9^{+1.5}_{-1.5}$&$171.1^{+1.5}_{-1.5}$&$13.23^{+0.11}_{-0.12}$  &  $1.340^{+0.020}_{-0.020}$\\
HD 74156 c&$2460^{+14}_{-15}$&  $13440^{+16}_{-16}$&$0.370^{+0.016}_{-0.016}$&$267.1^{+3.3}_{-3.2}$&$109.4^{+2.4}_{-2.3}$&$7.77^{+0.16}_{-0.16}$&  $1.238^{+0.040}_{-0.044}$\\
$\bf{HD}$ $\bf{95089}$ $\bf{c}$ & $1860^{+370}_{-570}$  &  $15492^{+43}_{-50}$&$0.294^{+0.070}_{-0.067}$  &  $74.6^{+8.1}_{-9.8}$  &  $46.1^{+3.4}_{-4.7}$  &  $3.97^{+0.33}_{-0.59}$  &  $1.38^{+0.12}_{-0.12}$\\
$\bf{HD}$ $\bf{99706}$ $\bf{c}$ & $1278^{+151}_{-198}$  &  $15383^{+249}_{-140}$&$0.411^{+0.231}_{-0.178}$  &  $136^{+64}_{-64}$  &  $13.8^{+2.9}_{-2.5}$  &  $5.69^{+1.43}_{-0.96}$  &  $1.72^{+0.12}_{-0.12}$\\
$\bf{HD}$ $\bf{102329}$ $\bf{c}$ & $1123^{+79}_{-53}$  & $14736^{+569}_{-200}$&$0.209^{+0.231}_{-0.202}$  &  $21^{+165}_{-74}$  &  $27.4^{+6.8}_{-4.5}$  &  $1.52^{+0.30}_{-0.25}$ &  $1.30^{+0.15}_{-0.15}$ \\
HD 114783 c&$4319^{+151}_{-130}$&  $18112^{+422}_{-537}$&$0.^{+0.091}_{-0.085}$&$6.5^{+37.9}_{-44.4}$&$9.21^{+0.71}_{-0.68}$&$0.611^{+0.056}_{-0.053}$  &  $0.853^{+0.034}_{-0.038}$\\
$\bf{HD}$ $\bf{116029}$ $\bf{c}$ & $907^{+30.}_{-29.}$&  $15291^{+134}_{-86}$&$0.038^{+0.127}_{-0.075}$&$17.3^{+167.0}_{-49.7}$&$20.7^{+2.2}_{-2.2}$&$1.27^{+0.15}_{-0.15}$ &  $1.33^{+0.11}_{-0.11}$ \\
$\bf{HD}$ $\bf{156279}$ $\bf{c}$&$4191^{+270}_{-310}$&  $15912^{+17}_{-17}$&$0.231^{+0.018}_{-0.021}$&$101.0^{+2.3}_{-1.9}$&$110.2^{+4.8}_{-5.3}$&$8.60^{+0.50}_{-0.55}$  &  $0.930^{+0.040}_{-0.040}$\\
HD 169830 c&$1834.3^{+8.3}_{-8.2}$&  $15350^{+40}_{-39}$&$0^{+0.018}_{-0.019}$&$95.7^{+8.2}_{-7.9}$&$39.7^{+1.3}_{-1.3}$&$3.54^{+0.10}_{-0.10}$  &  $1.410^{+0.028}_{-0.112}$\\
HD 183263 c&$5048^{+433}_{-701}$&  $14952^{+77}_{-74}$&$0.073^{+0.025}_{-0.034}$&$284.9^{+6.1}_{-5.4}$&$85.2^{+9.1}_{-14.5}$&$9.0^{+1.1}_{-1.7}$  &  $1.121^{+0.064}_{-0.040}$\\
HD 187123 c&$3380^{+41}_{-40}$&  $13649^{+42}_{-44}$&$0.295^{+0.026}_{-0.025}$&$260.4^{+3.7}_{-3.7}$&$24.97^{+0.76}_{-0.70}$&$1.80^{+0.06}_{-0.06}$  &  $1.037^{+0.026}_{-0.024}$\\
HD 190360 c&$2889^{+14}_{-14}$&  $13548^{+32}_{-25}$&$0.301^{+0.020}_{-0.020}$&$17.9^{+4.7}_{-3.8}$&$21.95^{+0.50}_{-0.49}$&$1.45^{+0.03}_{-0.03}$  &  $0.983^{+0.026}_{-0.048}$\\
HD 217107 c&$5178^{+74}_{-67}$&  $15951^{+49}_{-59}$&$0.376^{+0.014}_{-0.014}$&$206.2^{+2.7}_{-2.7}$&$53.2^{+1.9}_{-1.7}$&$4.48^{+0.20}_{-0.18}$  &  $1.108^{+0.034}_{-0.052}$
\enddata
\tablecomments{New planet names are in bold.}

\end{deluxetable*}

\subsection{Non-Planetary Sources of RV Trends}

There were two scenarios in which systems with statistically significant trend detections were excluded from further analysis.  In two systems, we found that the observed accelerations were correlated with stellar activity.  We compared the RV trends in each system to the measured emission in the Ca II H$\&$K lines, quantified by the $\sval$ index \citep{Wright2004, Isaacson2010}, to determine if the RV trends were caused by stellar activity instead of an outer companion \citep{Santos2010}.  Both HD 97658 and HD 1461 showed a clear correlation between the observed RV trend and the measured $\sval$ values, and we therefore excluded them from subsequent analysis.  

We also excluded systems with a linear acceleration that could have been caused by a nearby directly imaged stellar companion.  We first examined our K band AO images for all stars with statistically significant radial velocity trends in order to determine which systems contained a directly imaged stellar companion.  HD 164509 has a companion 0.75$\arcsec$ away, and HD 195109 has a companion 3.4$\arcsec$ away.  To determine whether these companions could have caused the RV trends in these systems, we compared the minimum mass estimate from the RV trend to the companion mass estimate from the AO image.  We calculated the minimum companion mass using the equation from Torres (1999):

\begin{align}
M_{\rm comp} = 5.34 \times 10^{-6} M_{\odot} \bigg(\frac{d}{\rm pc}\frac{\rho}{\rm arcsec}\bigg)^2 \nonumber \\
\times \bigg|\frac{\dot v}{\rm m s^{-1}yr^{-1}}\bigg|F(i,e,\omega,\phi).
\end{align}

\noindent In this equation, $d$ is the distance to the star, $\rho$ is the projected separation of the companion and the star on the sky, $\dot v$ is the radial velocity trend, and $F(i,e,\omega,\phi)$ is a variable that depends on the orbital parameters of the companion that are currently unconstrained.  We use a value of $\sqrt{27}/2$ for $F$, which is the minimum value of this function calculated in Liu et al (2002).  

HD 164509 is 52 pc away and has a companion located at a separation of 0.75$\arcsec$.  With a radial velocity trend of 3.4 m s$^{-1}$ yr$^{-1}$, this trend corresponds to a minimum companion mass of $0.072$ $M_{\odot}$.  To estimate the mass of the companion from the AO image, the brightness of the companion in K band relative to the primary is used, as described in section 3.4.  With a relative K band magnitude of 3.59, we find that the estimated mass from the AO data is 0.33  $M_{\odot}$.  Since the companion mass calculated from the AO data is greater than the minimum mass needed to explain the RV trend, we therefore conclude that this companion may indeed be responsible for the observed trend and exclude this system from subsequent analysis.

HD 195109 is 38.5 pc away and has a companion located at a separation of 2.4$\arcsec$.  With a radial velocity acceleration of 1.9 m s$^{-1}$ yr$^{-1}$, a stellar companion at the observed AO separation must have a mass of at least 0.44 $M_{\odot}$ in order to cause the observed trend.  With a relative K-band magnitude of 2.66, we find that the estimated mass from the AO data is 0.58 $M_{\odot}$.  We conclude that the imaged companion could have caused the RV acceleration, and thus removed this system from future analyses.  We note that this companion was previously reported in \citet{Mugrauer2007}.

Howard et al (2010) imaged a faint M-dwarf companion located $489.0\pm1.9$ mas from the primary star HD 126614.  With an absolute K-band magnitude of 6.72, the authors estimated the mass of this companion to be $0.324 \pm 0.004$ $M_{\odot}$.  From Equation 1, the estimated minimum mass of the companion inducing the RV trend, given a distance of 72.6 pc and a trend of 14.6 ms$^{-1}$yr$^{-1}$, is 0.26 $M_{\odot}$.  Since the minimum estimated RV mass is lower than the estimated AO mass, we conclude that the imaged AO companion could cause the RV trend, and thus remove this system from subsequent analyses.  Note that none of these AO companions have second epoch data, and thus have not been confirmed as bound to their respective primaries.  However, at these projected separations and contrast ratios the probability that the companion is a background star is relatively low, and we therefore proceed under the assumption that they are bound.

We also carried out a literature search to determine whether any of the remaining trend systems had additional stellar or substellar companions.  We found that HD 109749 has a known binary companion described in the published literature.  HD 109749 has a companion with K-band magnitude of 8.123 separated by 8.35$\arcsec$ \citep{Desidera2007}.  This visual binary lies outside the field of view for our AO observations.  After calculating the minimum companion mass from the measured RV trend and comparing this value to the estimated mass from the AO data found in the literature, we found that this companions cannot explain the accelerations observed in these systems.  

After removing stellar sources of RV trends, we find 20 systems with accelerations that have slopes at least $3\sigma$ away from zero.  The RV data and best-fit accelerations for each of these systems are plotted in Figures 11 and 12.  Six of these trends were previously reported in the published literature:  HD 24040 \citep{Boisse2012}, HD 168443 \citep{Pilyavsky2011}, HD 180902 \citep{Johnson2010}, HD 68988 \citep{Vogt2006}, HD 158038 \citep{Johnson2011}, and HD 50499 \citep{Vogt2005}.

\begin{figure*}
\begin{tabular}{ c  c  c}
\includegraphics[width=0.34\textwidth]{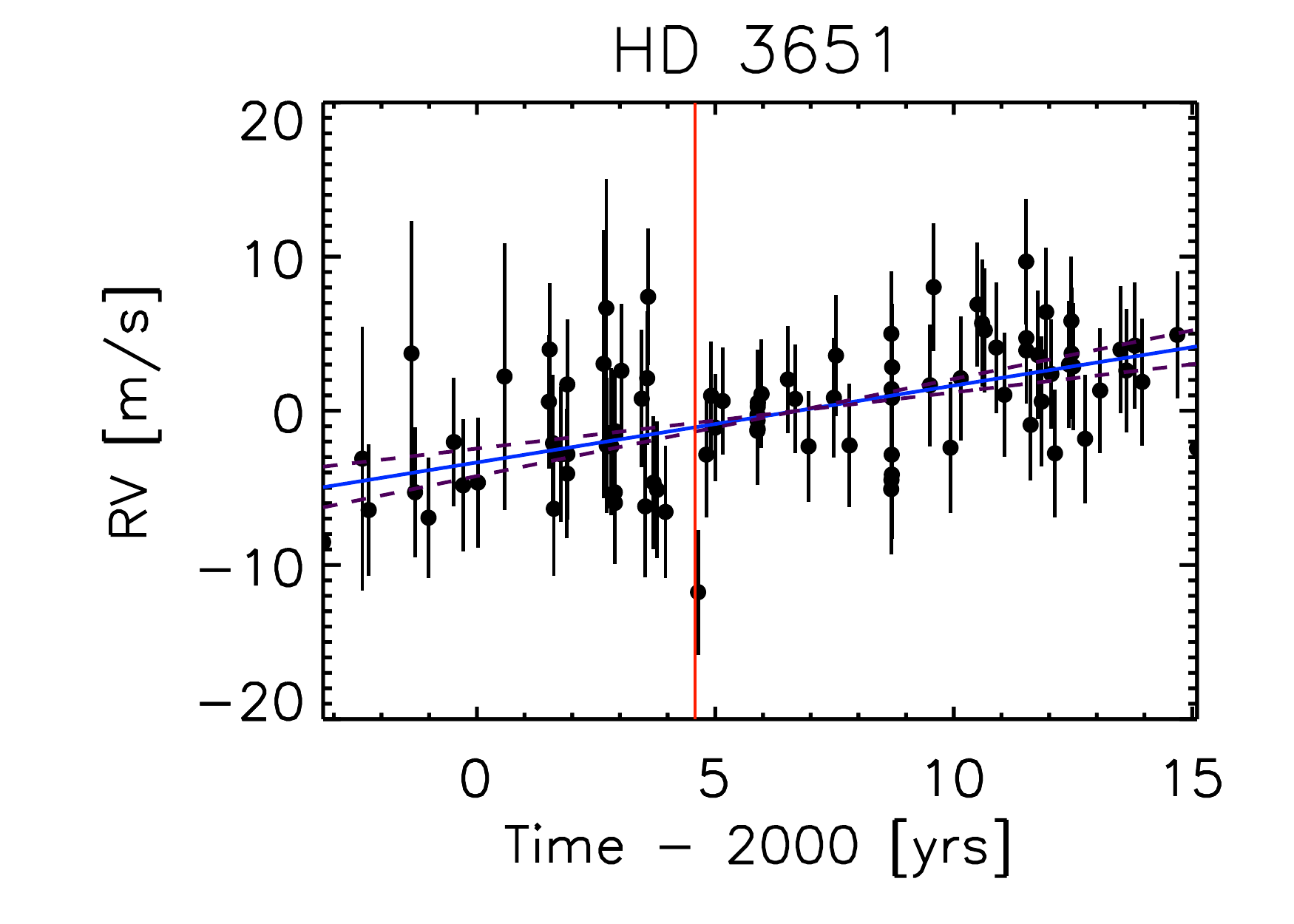} &
\includegraphics[width = 0.34\textwidth]{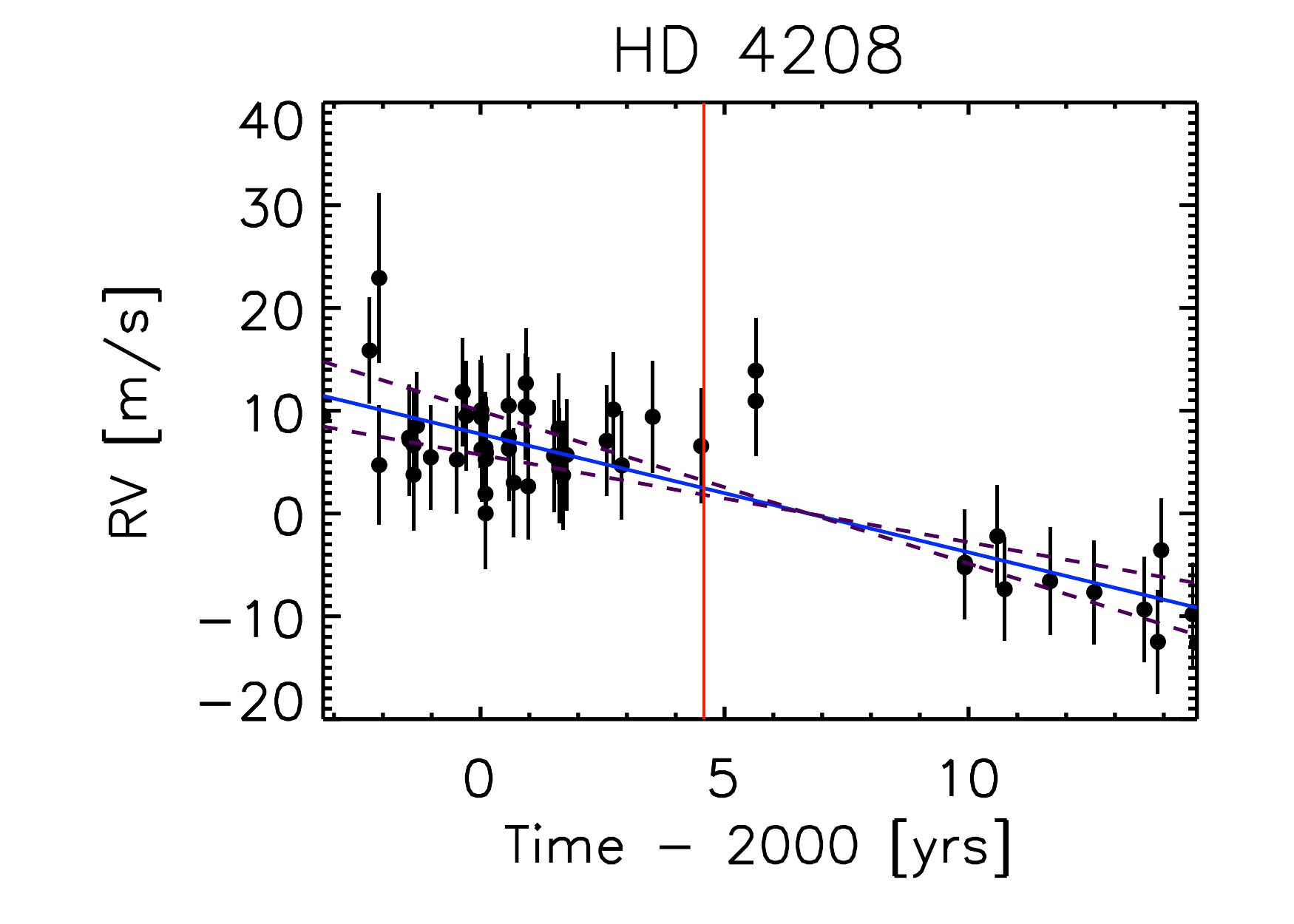}&
\includegraphics[width = 0.34\textwidth]{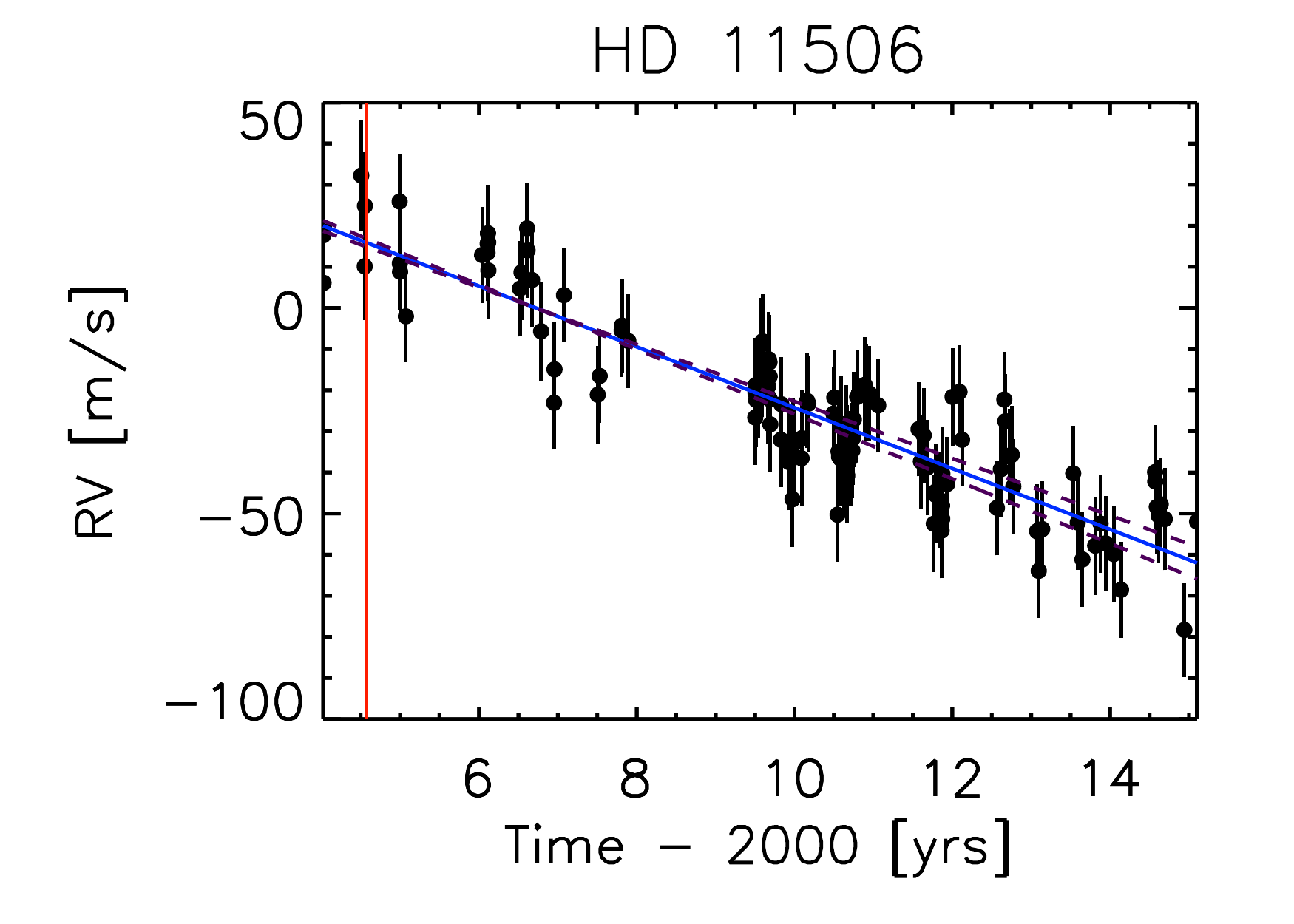} \\
\includegraphics[width=0.34\textwidth]{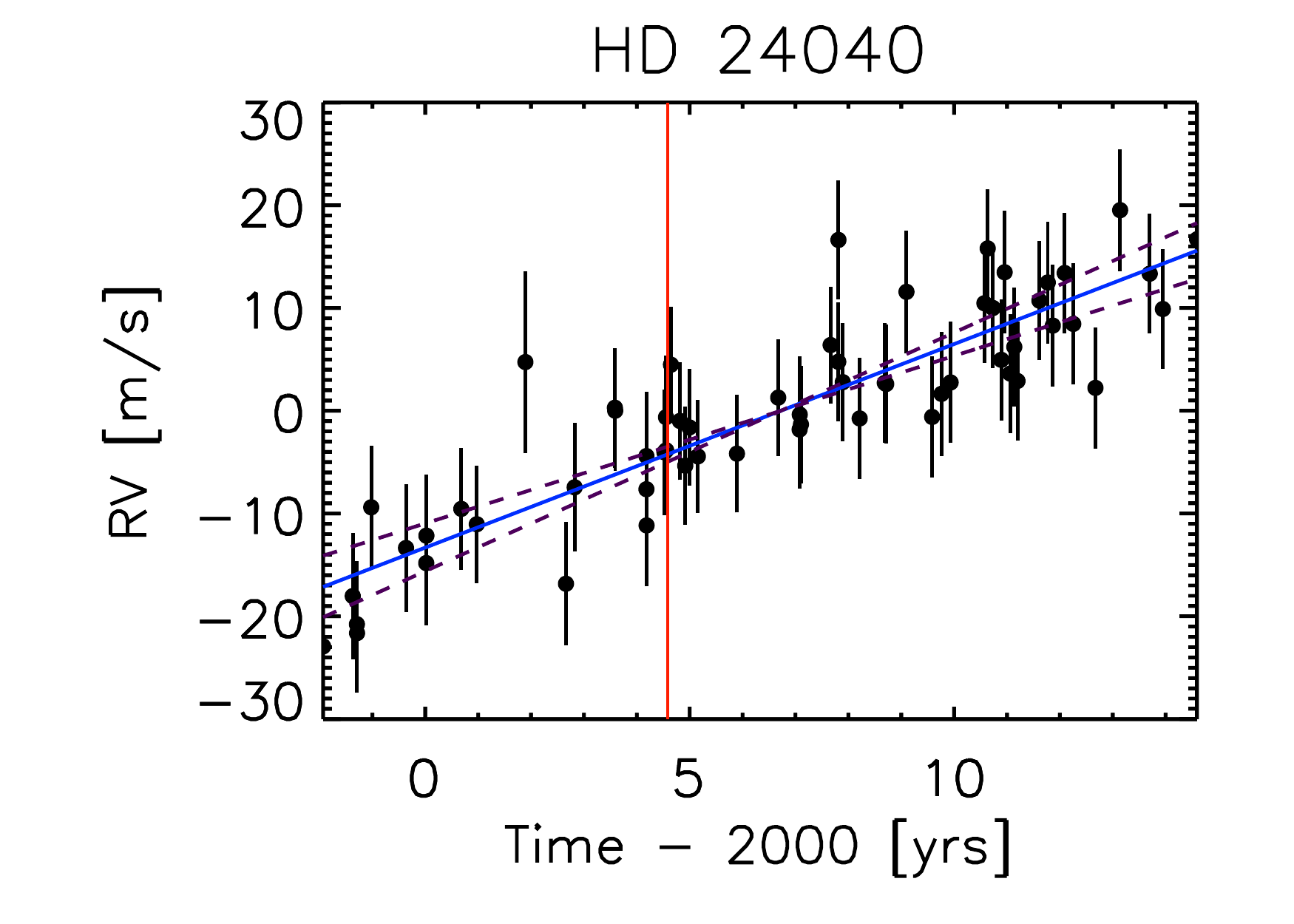} &
\includegraphics[width=0.34\textwidth]{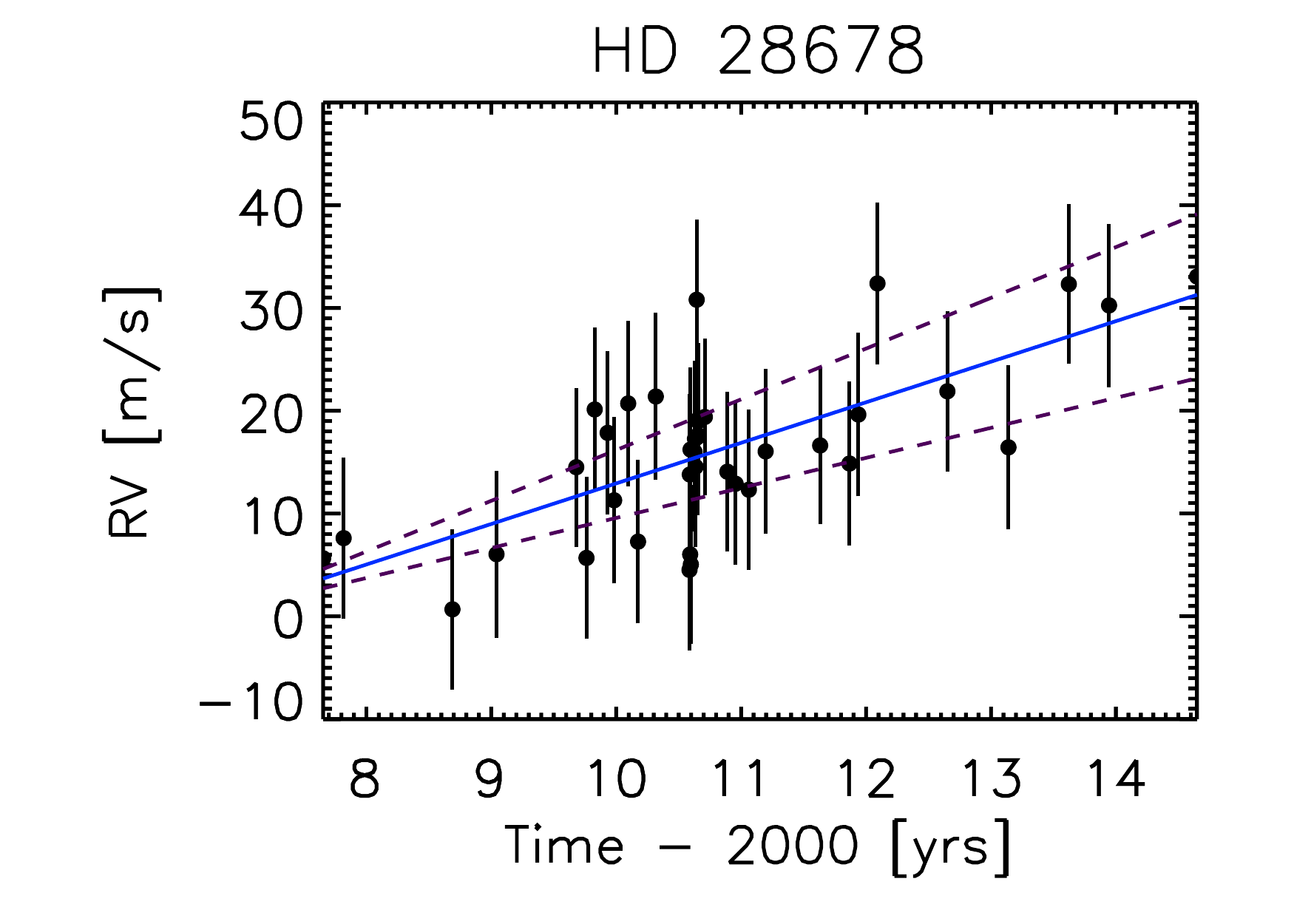} &
\includegraphics[width=0.34\textwidth]{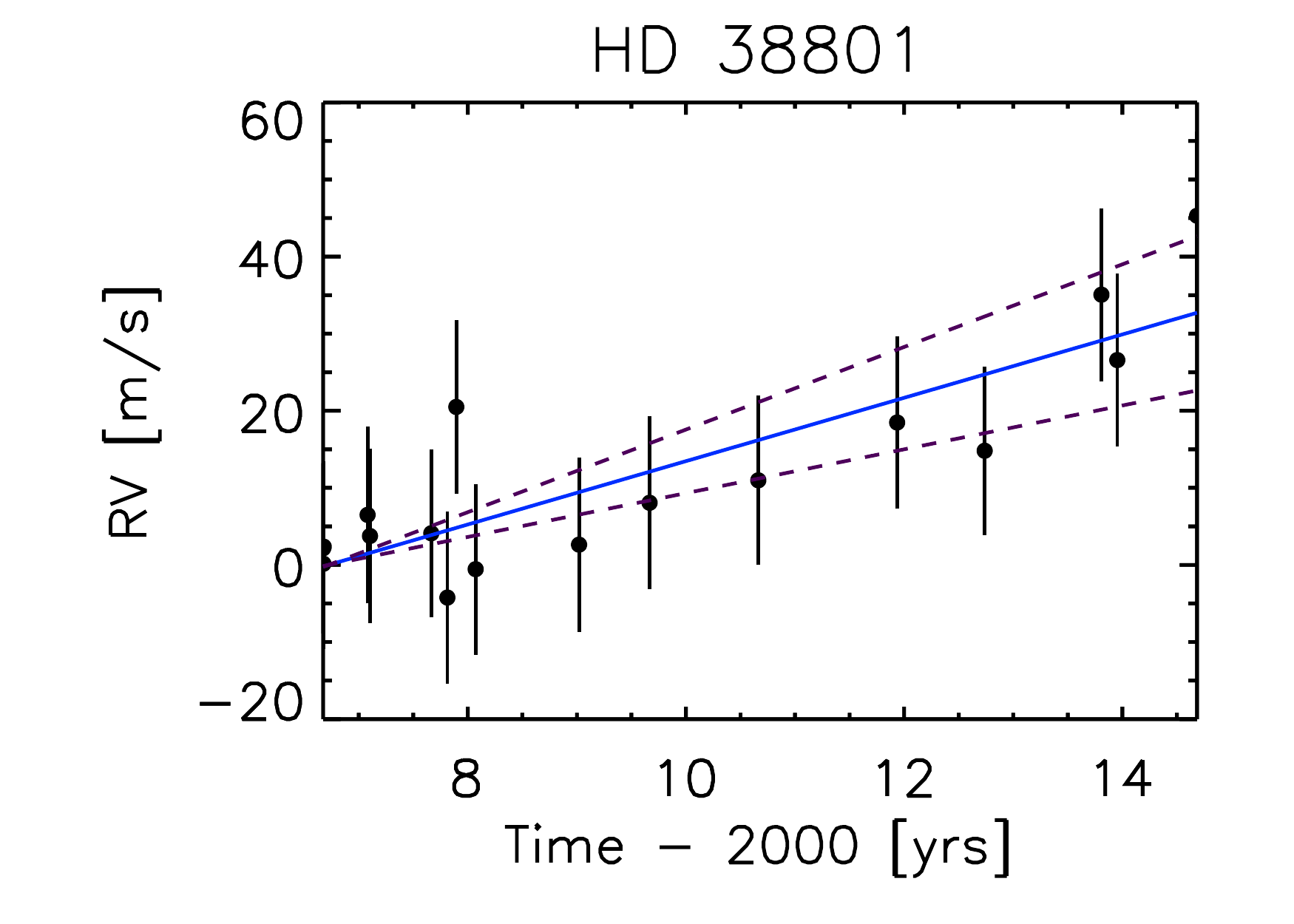} \\
\includegraphics[width=0.34\textwidth]{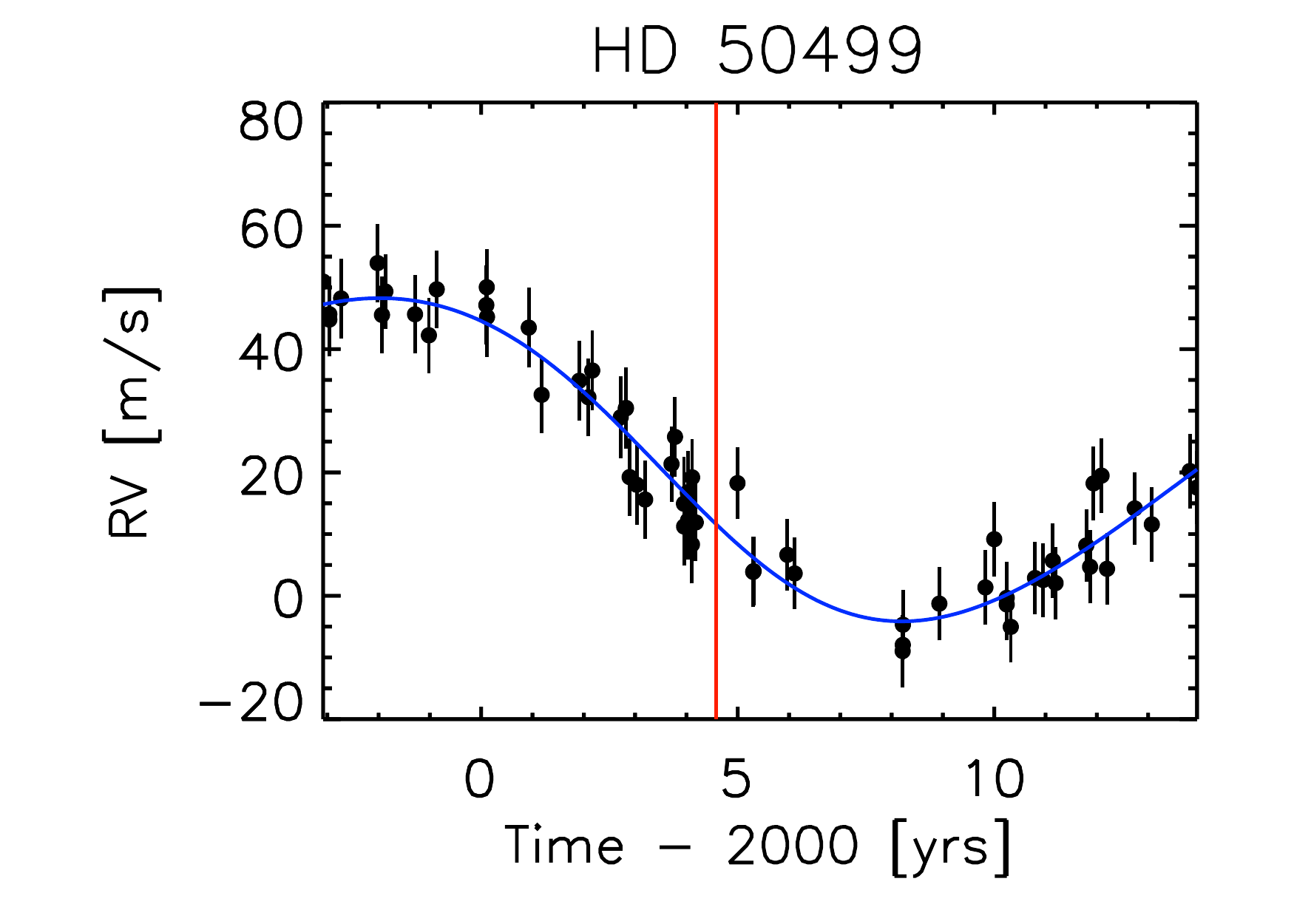} &
\includegraphics[width=0.34\textwidth]{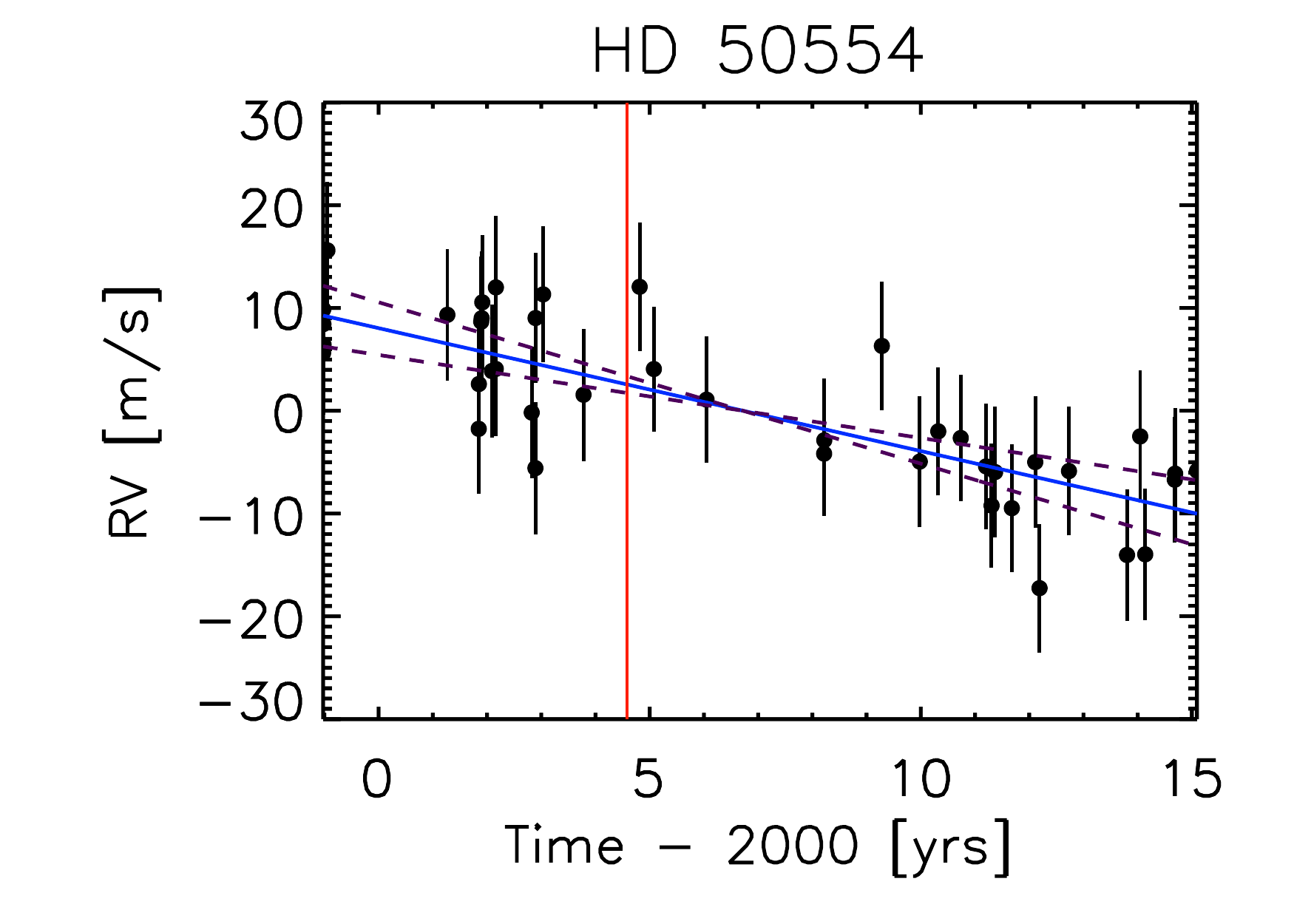} &
\includegraphics[width=0.34\textwidth]{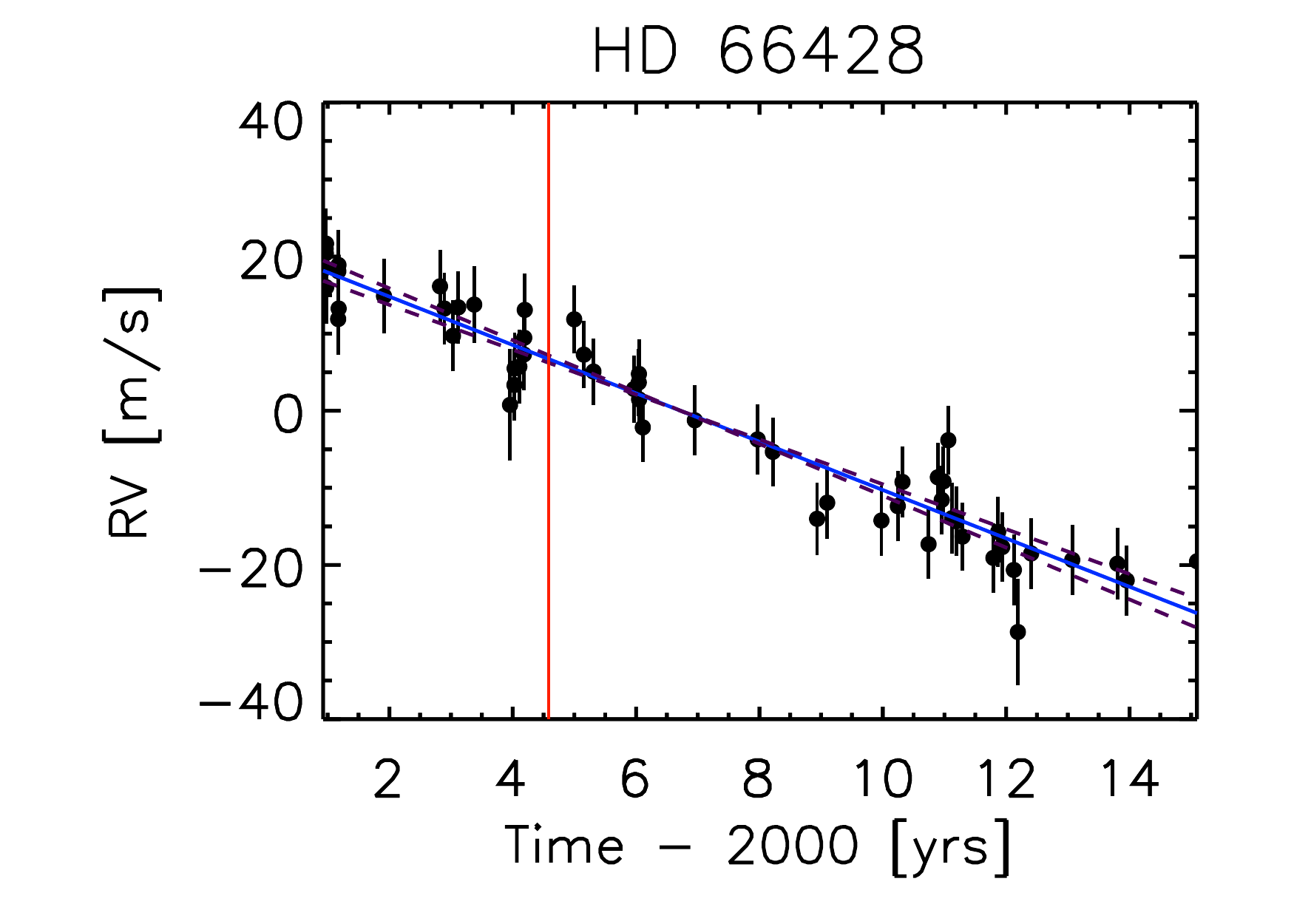}\\
\includegraphics[width=0.34\textwidth]{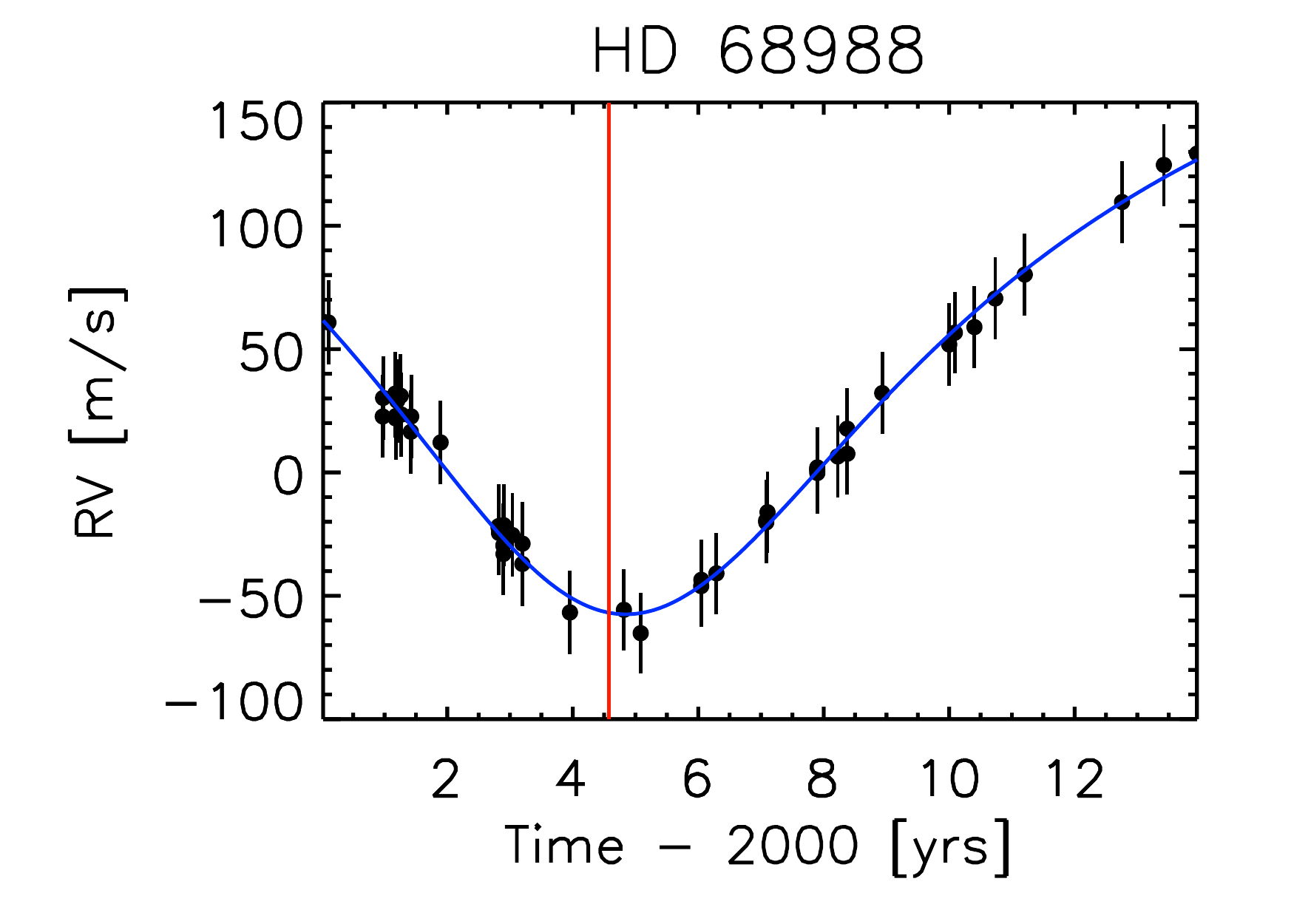} &
\includegraphics[width=0.34\textwidth]{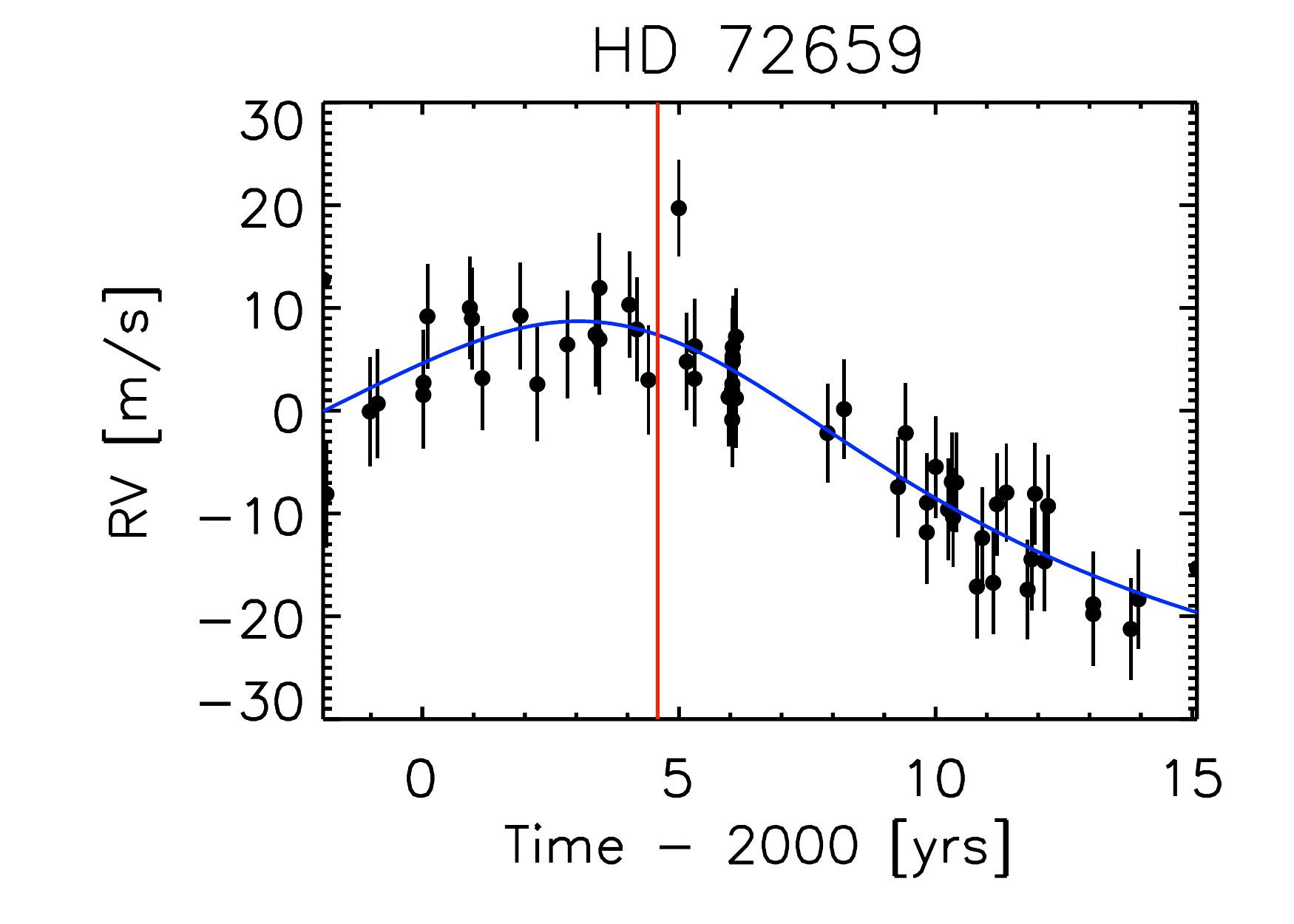} &
\includegraphics[width=0.34\textwidth]{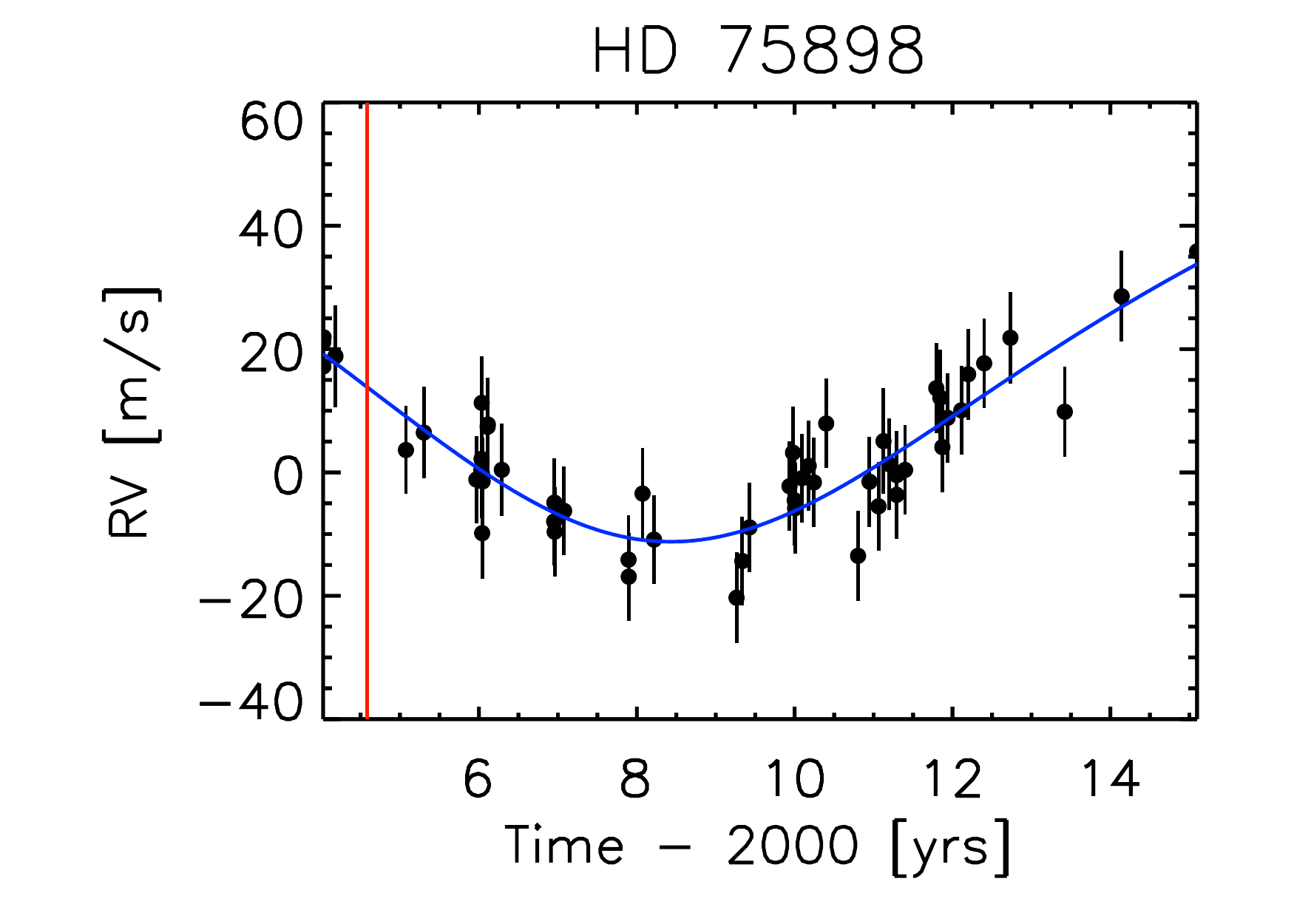}

\end{tabular}
\caption{Best fit accelerations to the radial velocity data with a $3\sigma$ trend.  The best fit trend is shown as a solid blue line, the errors on the slope are presented as dashed purple lines.  The solid red line marks the date when the HIRES detector was replaced, which caused an offset in the measured RVs for the stars in our sample.}  The confirmed planet orbital solutions have been subtracted from both the RV data and from the best fit orbital solution to yield the trends.  Systems with curved trends include HD 50499, HD 68988, HD 72659, HD 75898, HD 92788, and HD 158038.  The plots with the curved trends show the best fit one planet orbital solution to the data after the inner planet solution was subtracted.

\end{figure*}

\begin{figure*}
\begin{tabular}{c c c}
\includegraphics[width=0.34\textwidth]{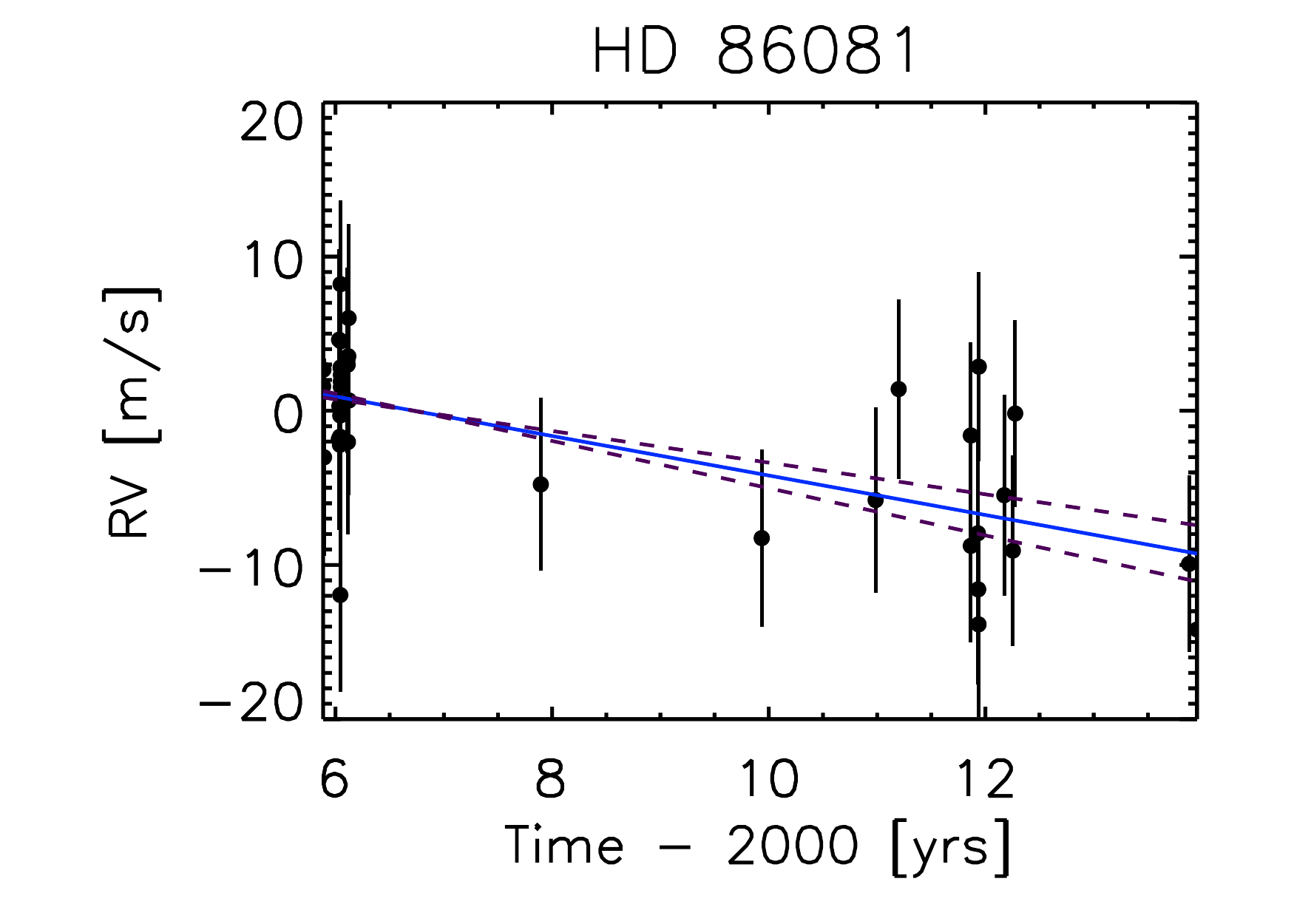}&
\includegraphics[width=0.34\textwidth]{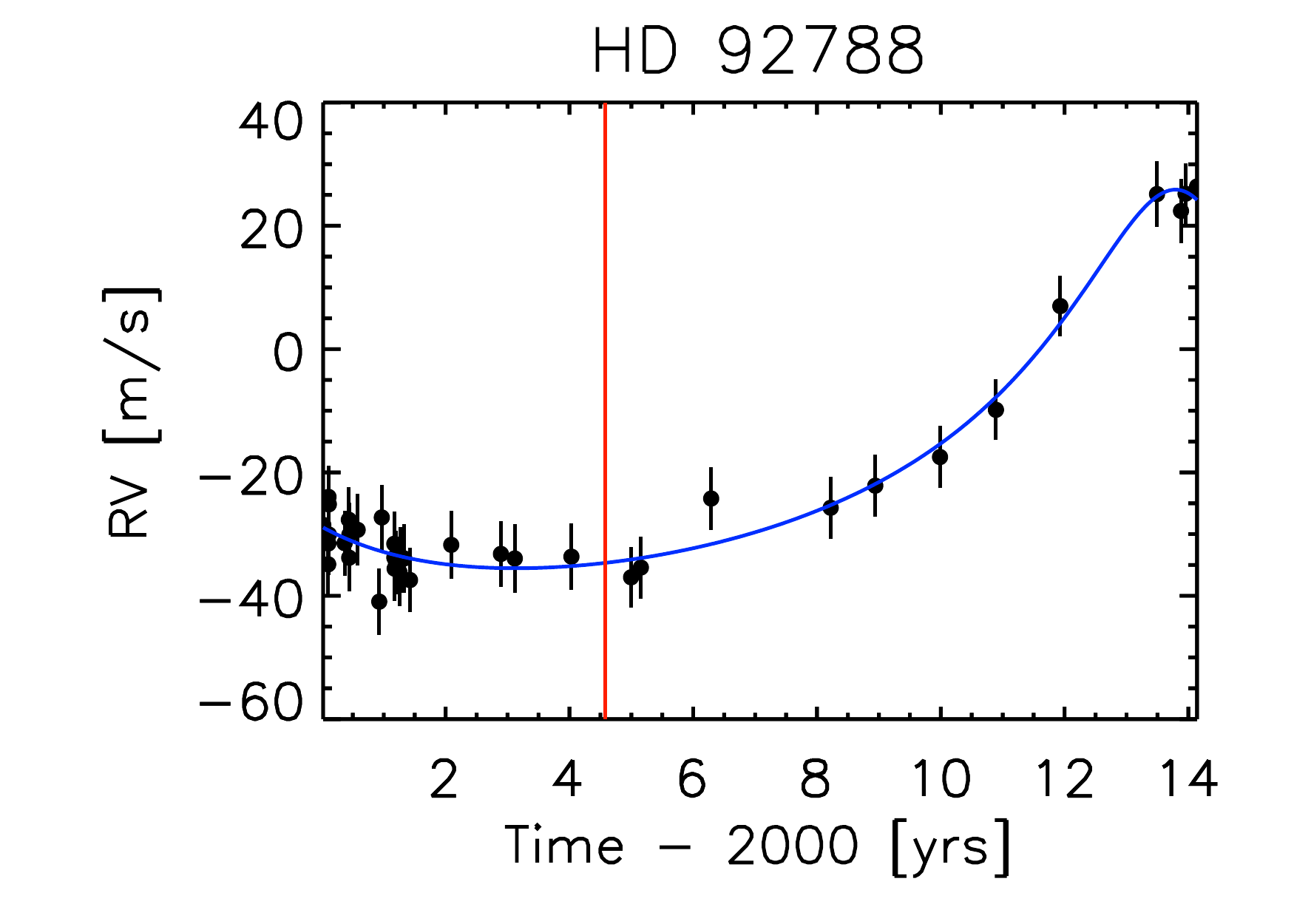}&
\includegraphics[width=0.34\textwidth]{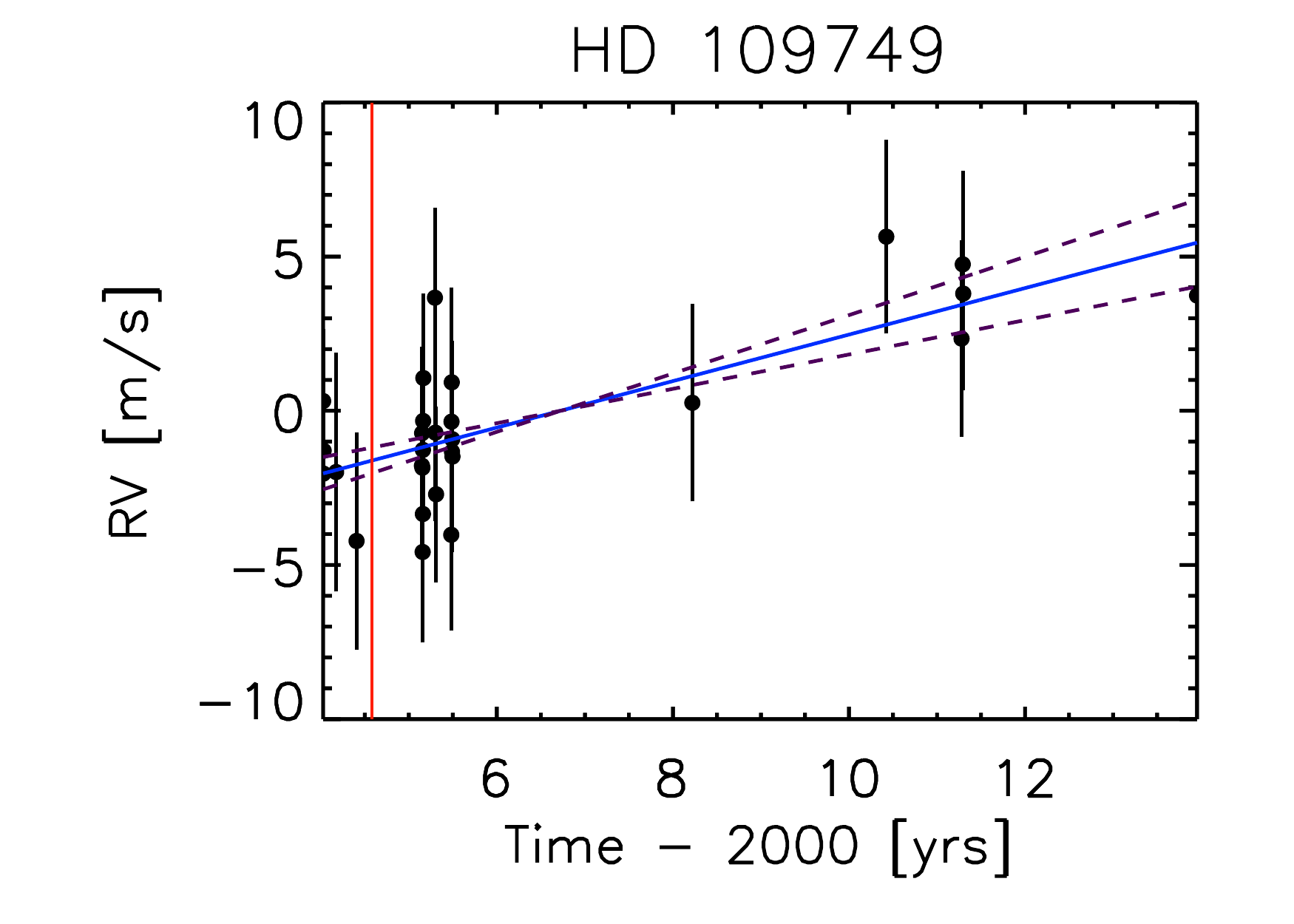} \\
\includegraphics[width=0.34\textwidth]{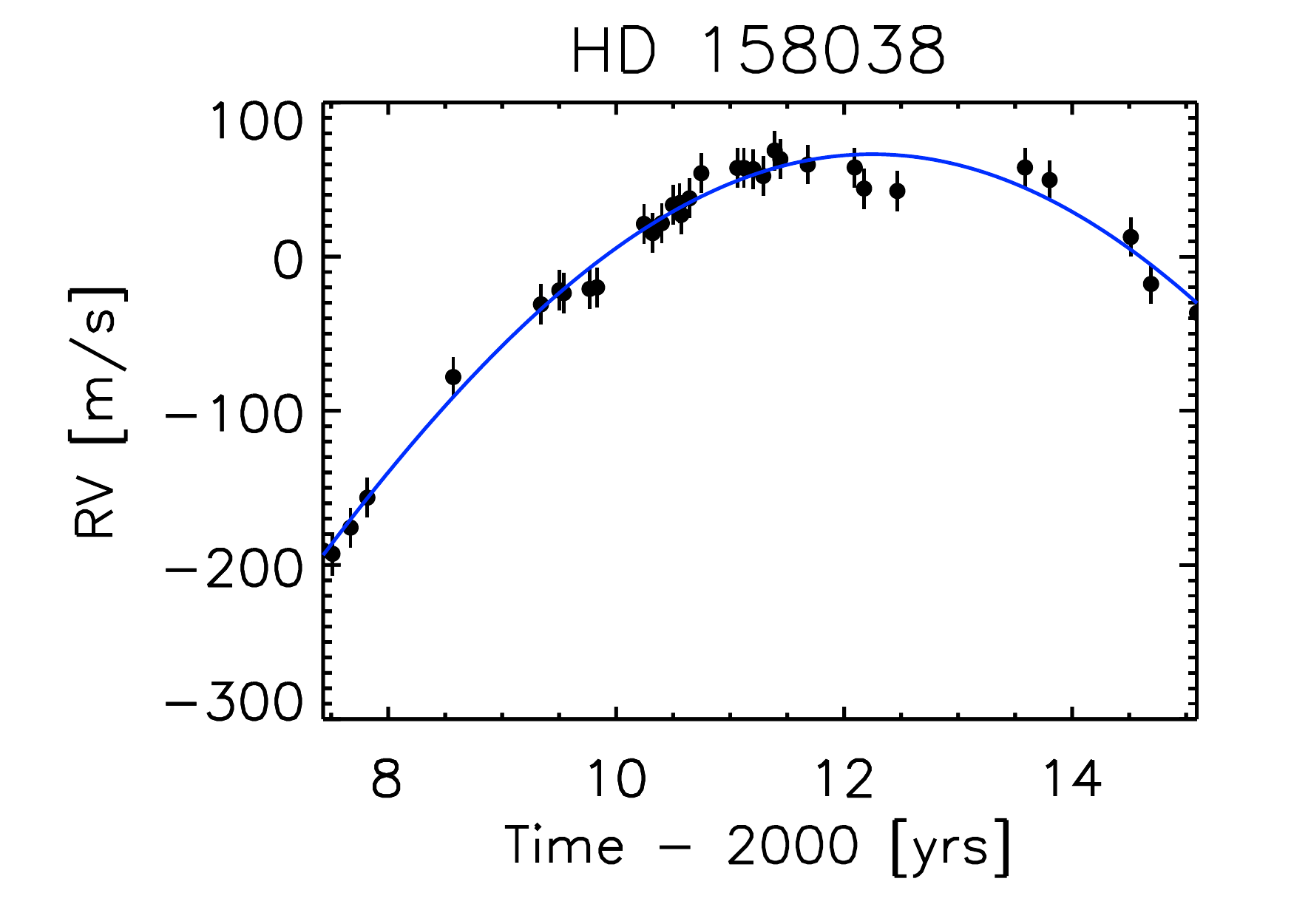}&
\includegraphics[width=0.34\textwidth]{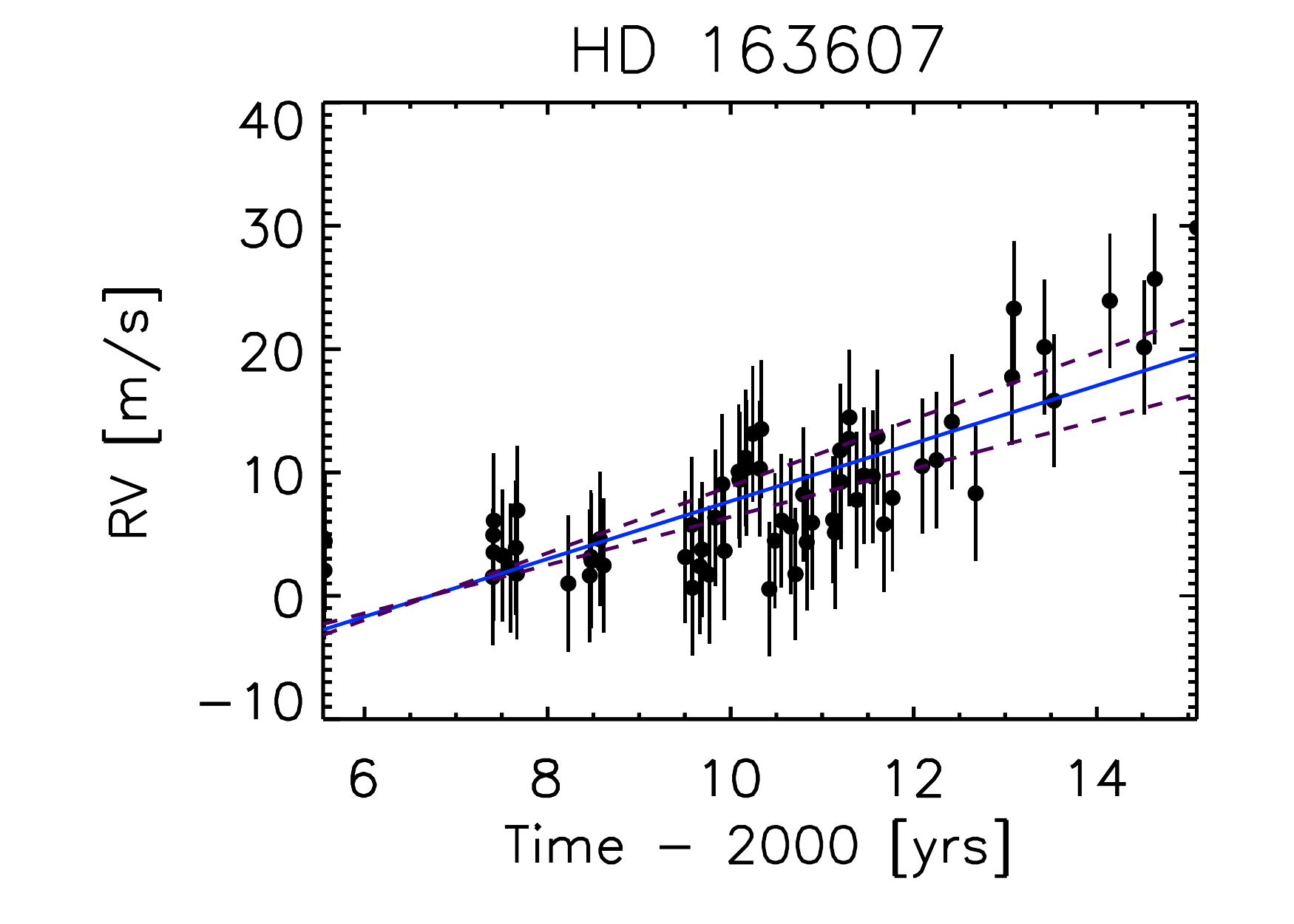}&
\includegraphics[width=0.34\textwidth]{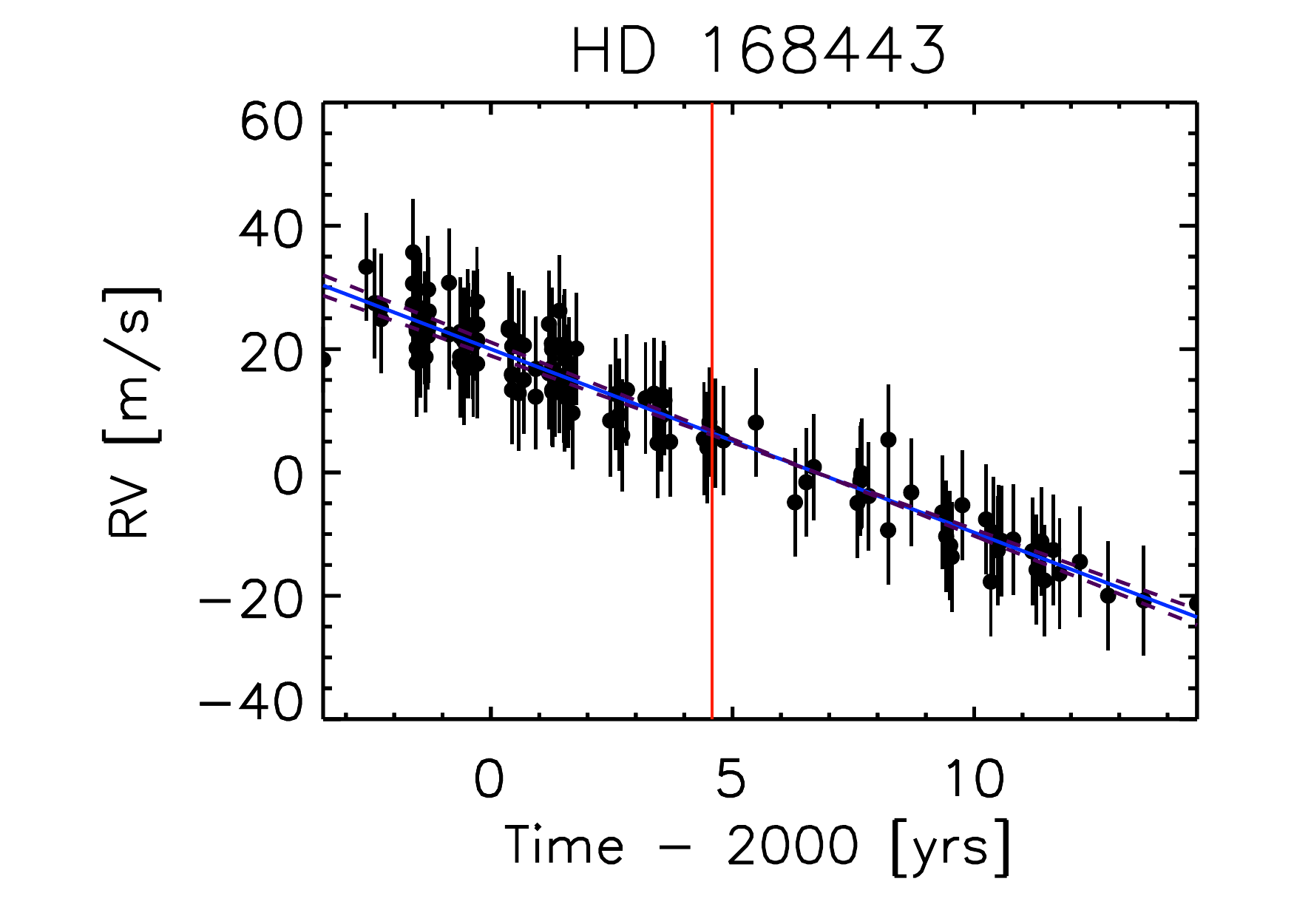} \\

\includegraphics[width=0.34\textwidth]{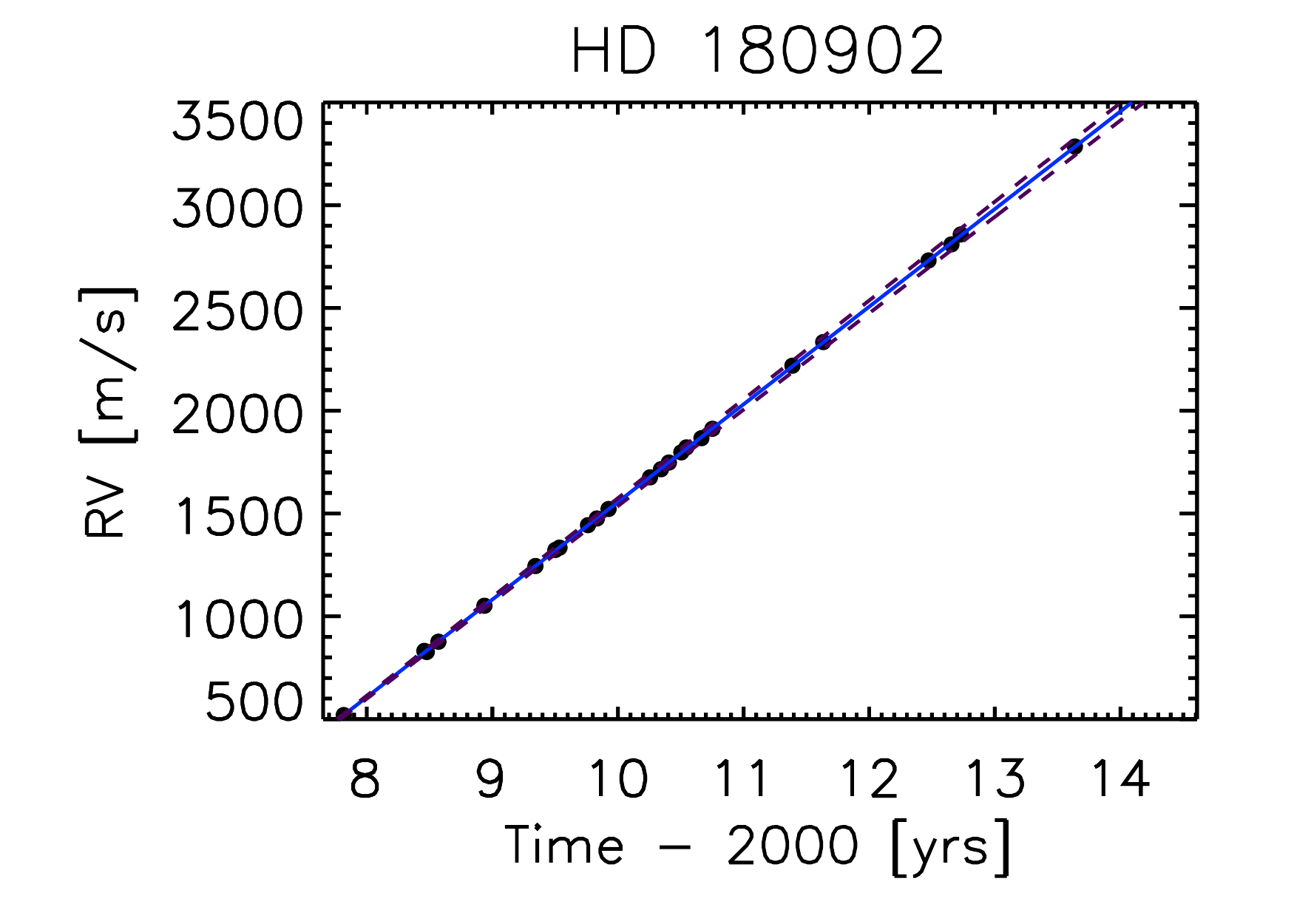} &
\includegraphics[width=0.34\textwidth]{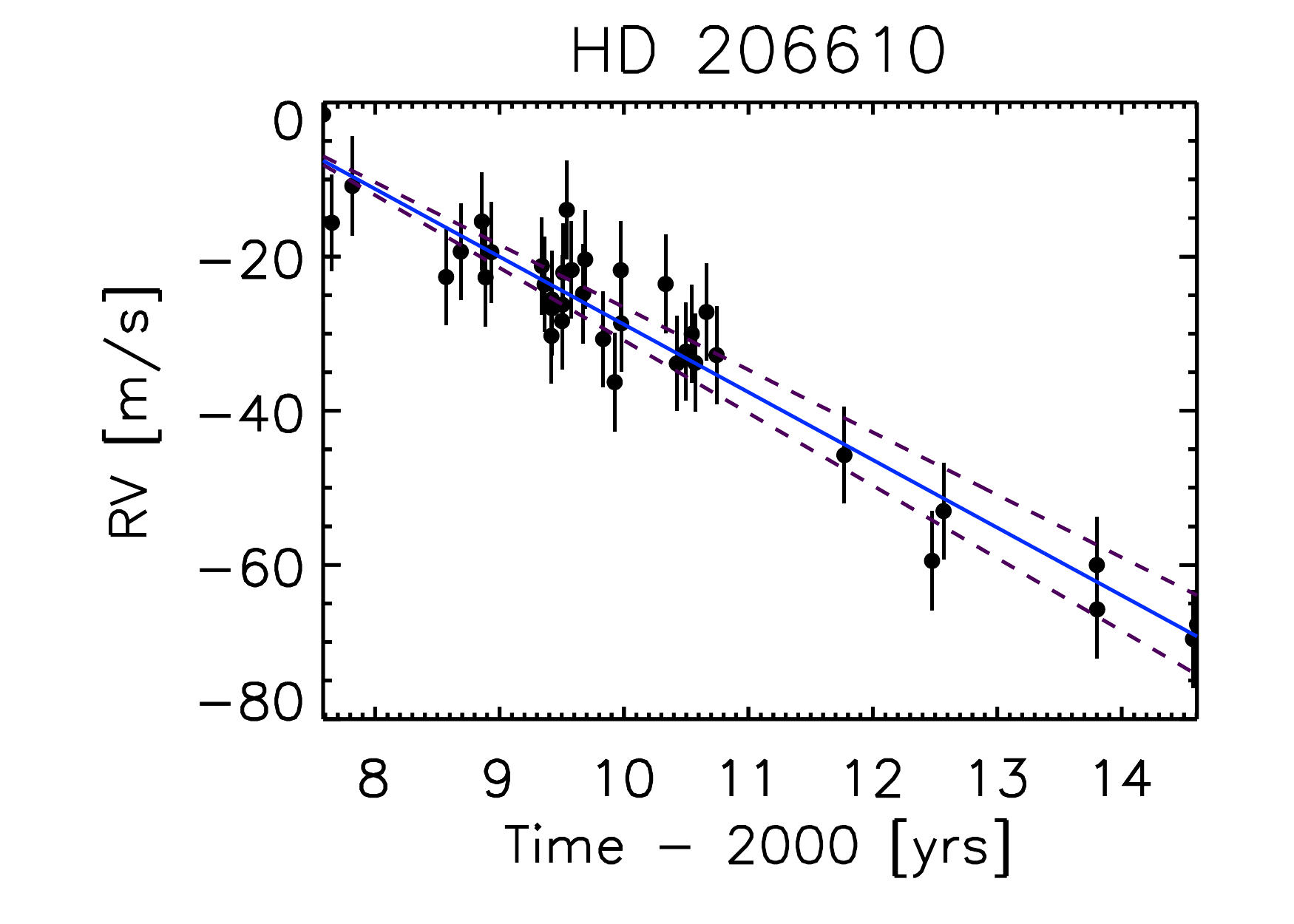}

\end{tabular}
\caption{Remaining best fit accelerations to the radial velocity data with a $3\sigma$ trend.}
\end{figure*}

\subsection{Contrast Curves}

We used contrast curves from our AO observations to put limits on the masses and separations that a companion in each system could have.  We calculate contrast curves for our target stars as follows.  First, we measure the full width at half max (FWHM) of the central star's point spread function in the stacked and combined image, taking the average of the FWHM in the $x$ and $y$ directions as our reference value.  We then create a box with dimensions equal to the FWHM and step it across the array, calculating the total flux from the pixels within the box at a given position.  The $1\sigma$ contrast limit is then defined as the standard deviation of the total flux values for boxes located within an annulus with a width equal to twice the FWHM centered at the desired radial separation.  We convert absolute flux limits to differential magnitude units by taking the total flux in a box of the same size centered on the peak of the stellar point spread function and calculating the corresponding differential magnitude at each radial distance.  We show the resulting $5\sigma$ average contrast curve for these observations in Figure 13; although our field of view extends farther in some directions than the maximum separations shown here, we have limited our calculations to radial separations with data available at all position angles.

We next use our contrast curves to place limits on the allowed masses of stellar companions as a function of projected separation.   We interpolate the PHOENIX stellar atmosphere models \citep{husser13} in the available grid of solar metallicity models to produce a model that matches the effective temperatures and surface gravities of the primary star.  For the proposed low-mass main sequence companions, we create PHOENIX models with radii and effective temperatures drawn from \citet{baraffe98}.  We then calculate the corresponding contrast ratio between the primary and secondary by integrating over the appropriate bandpass (either $K_p$ or $K_s$), adjusting the mass of the secondary downward until we match the $5\sigma$ limit from our contrast curve.  We discuss the merits of this approach as compared to other methods commonly utilized in AO imaging searches in Knutson et al. (2014).

\begin{figure}
\includegraphics[width=0.5\textwidth]{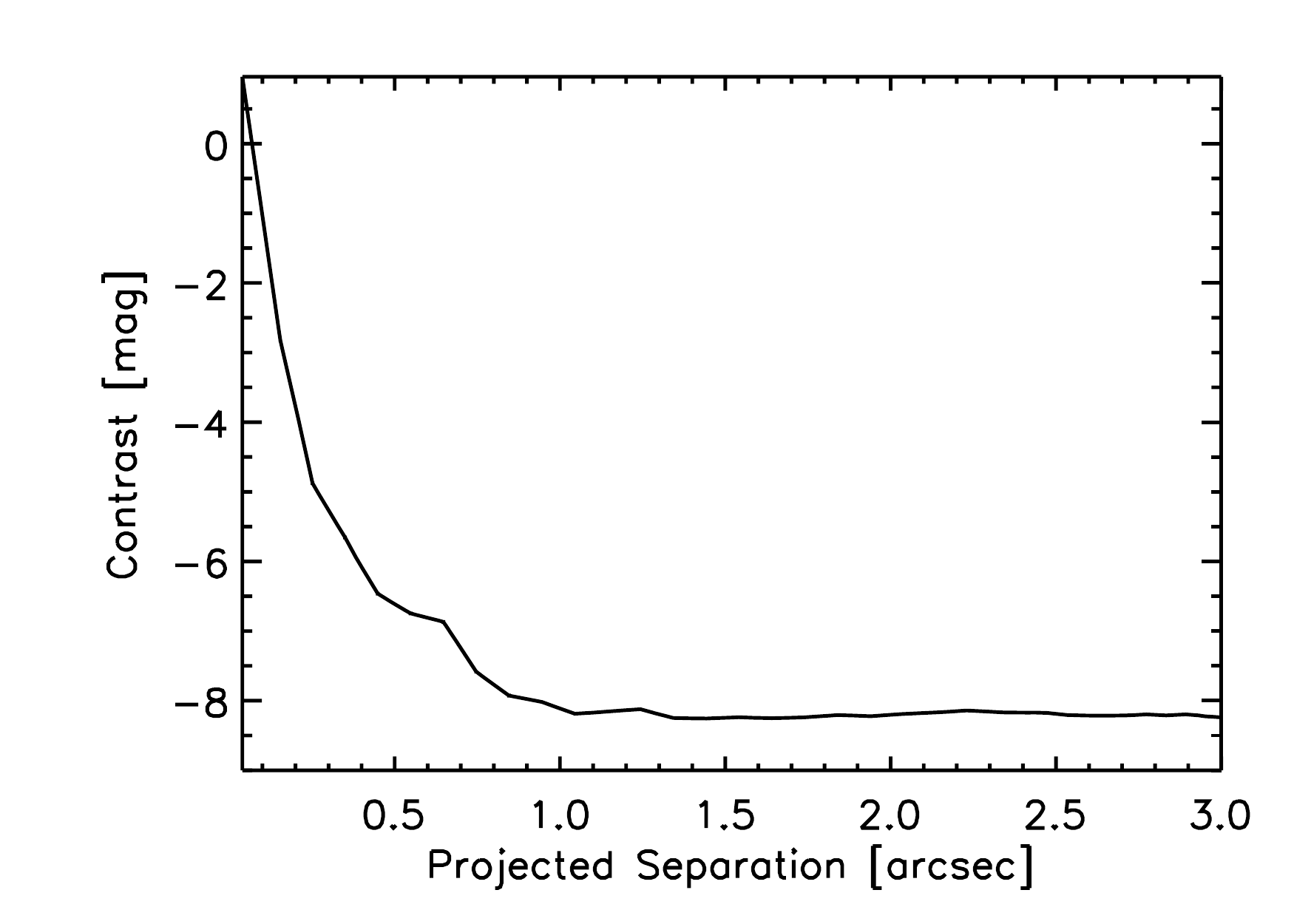}
\caption{Mean contrast curve from the K-band AO observations described in this study.}
\end{figure}

\subsection{Companion Probability Distributions}

We combine our AO and RV observations in order to constrain the allowed range of masses and semi-major axes for the observed companions.  The duration and shape of the RV trend places a lower limit on the mass and semi-major axis of the companions.  Similarly, a non-detection in AO gives a complementary upper limit on these quantities.  We create a two dimensional probability distribution for each companion, by defining an equally spaced 50$\times$50 grid of logarithmic companion mass (true mass) and semi-major axis ranging from 1-500 AU and $0.05-1000$ $M_{\rm Jup}$.   We then subtract off the orbital solutions of the confirmed inner planets, leaving only the trends due to the companions.  At each grid point in mass and semi-major axis, we inject 500 simulated companions.  While the semi-major axis and mass of the companion remain fixed at each point, we drew a new inclination of the orbit each time from a uniform distribution in $cos(i)$, and a new eccentricity each time from the beta distribution \citep{Kipping2013}.  This distribution is defined in Equation 3, where $P_{\beta}$ is the probability of a given eccentricity, $\Gamma$ is the gamma function, and $a' = 1.12$ and $b' = 3.09$ are constants calculated from the known population of long period giant planets.

\begin{equation}
P_{\beta}(e;a',b') = \frac{\Gamma (a'+b')}{\Gamma (a') \Gamma(b')}e^{a'-1}(1 - e)^{b'-1}.
\end{equation}
Given this fixed mass, semi-major axis, and eccentricity for each simulated companion, we fit the remaining orbital parameters to the RVs using a least squares algorithm, and we calculate a corresponding $\chi^2$ value.  We note that the probability distribution calculations are not particularly sensitive to the assumed eccentricity distribution.  We recalculate the probability distributions for 30 random systems within our sample assuming a uniform eccentricity distribution, and found that the $1\sigma$ semi-major axis and mass ranges, as presented in Table 6 for the $3\sigma$ trend systems, are generally consistent with each other to a couple of grid points.

We incorporate the constraints on potential companions from our AO observations using a method identical to the one described above.  Within each mass and semi-major axis box we first generate a set of 500 companions with randomly selected masses, semi-major axes, and an eccentricity drawn from Eq. 3.  We then fit for the remaining orbital parameters using the RV data, and use this best-fit orbit to calculate a set of 1000 projected separations for the companion sampled uniformly across the orbit.  We then use our AO contrast curve to determine whether or not a companion of that mass and projected separation could have been detected in our AO image for each of the 1000 time steps considered.  If the companion lies above our contrast curve we assume that it would have been detected, and if it lies below the curve we count it as a non-detection.  For companions with large enough projected separations our images do not span all position angles, and we therefore assume that companions that lie above our contrast curve would be detected with a probability equal to the fractional position angle coverage of our image at that separation.  We can then calculate the probability that a given companion would have been detected by determining the fraction of our 1000 time steps in which the companion lies above the contrast curve for that star.

The lower and upper limits on the mass/semi-major axis parameter space occupied by each companion  can be combined to form a two dimensional probability distribution.  After multiplying the $\chi^2$ cube in mass, semi-major axis, and eccentricity from the RV trends by the detection probability cube from the AO contrast curves, we marginalize this new cube over eccentricity to yield a two dimensional probability distribution.   Figure 14 shows the posterior distributions for the companions in each of the 20 systems with statistically significant RV trends.  Table 6 lists the $1\sigma$ mass and semi-major axis ranges derived for each companion from this analysis.  As expected, systems with strong curvature in the observed radial velocity accelerations have tighter constraints on the allowed mass and semi-major axis of the companion than those with linear trends.
\begin{figure*}
\begin{tabular}{cccc}
\includegraphics[width=0.25\textwidth]{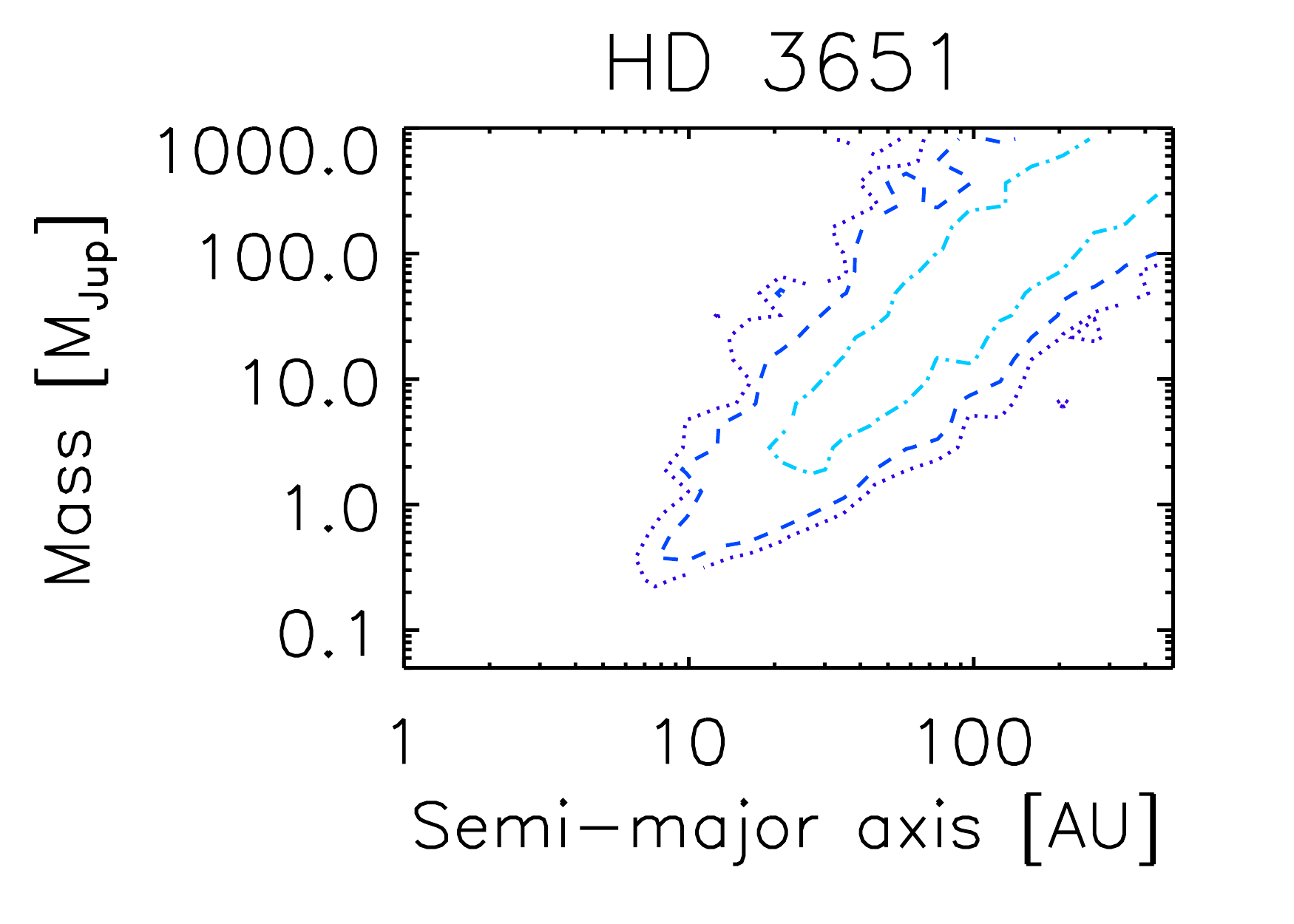} &
\includegraphics[width=0.25\textwidth]{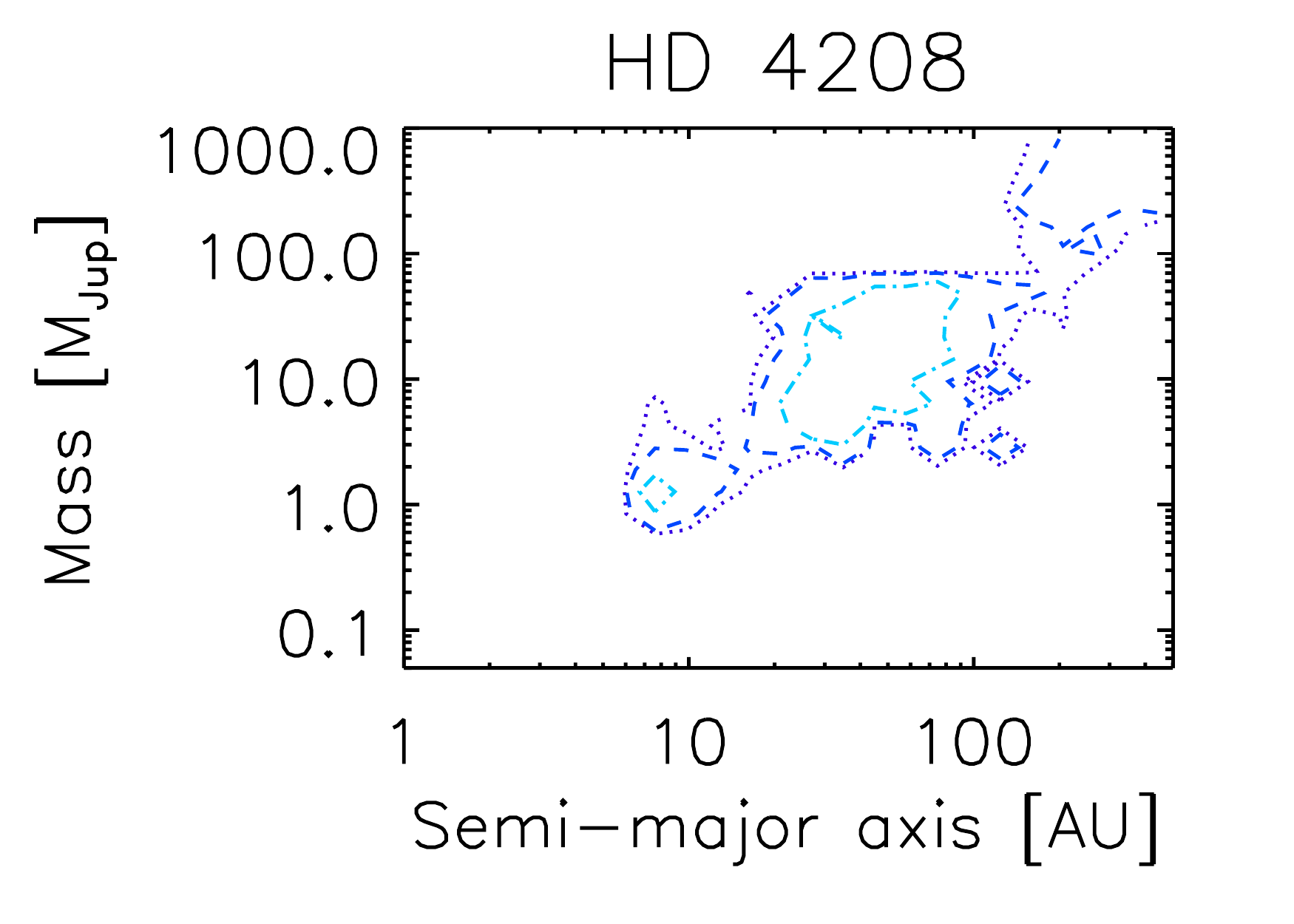} &
\includegraphics[width=0.25\textwidth]{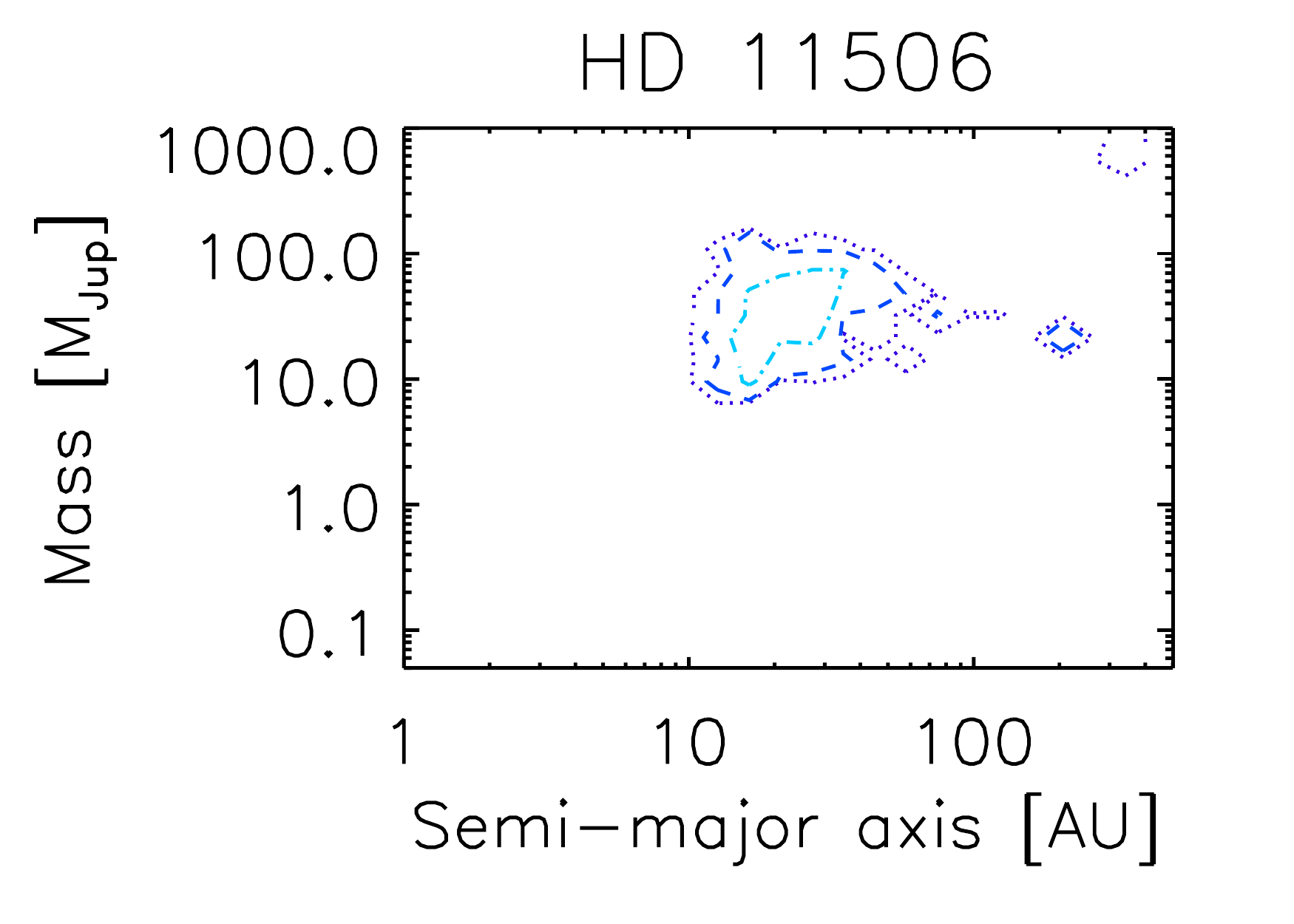} &
\includegraphics[width=0.25\textwidth]{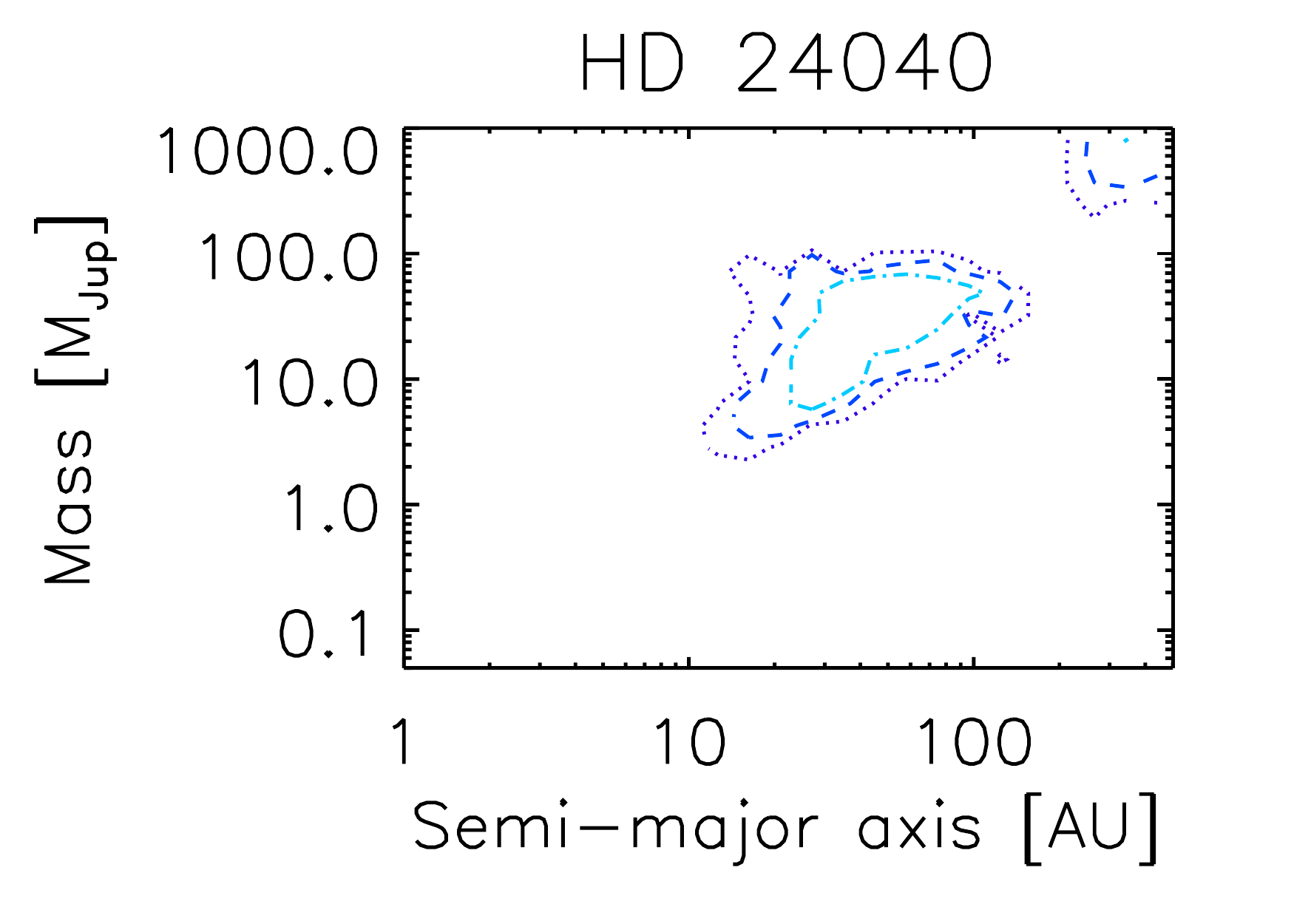} \\
\includegraphics[width=0.25\textwidth]{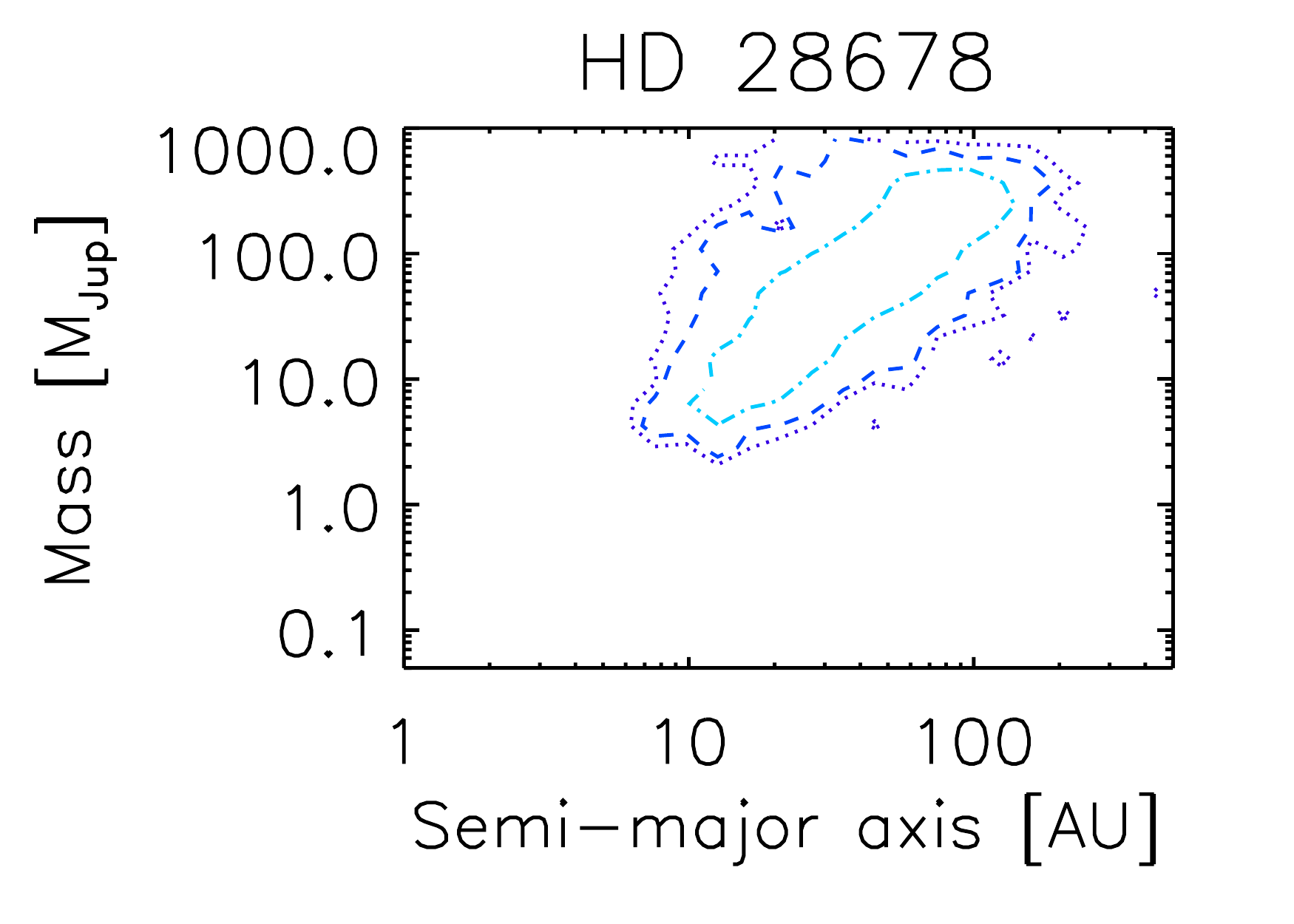} &
\includegraphics[width=0.25\textwidth]{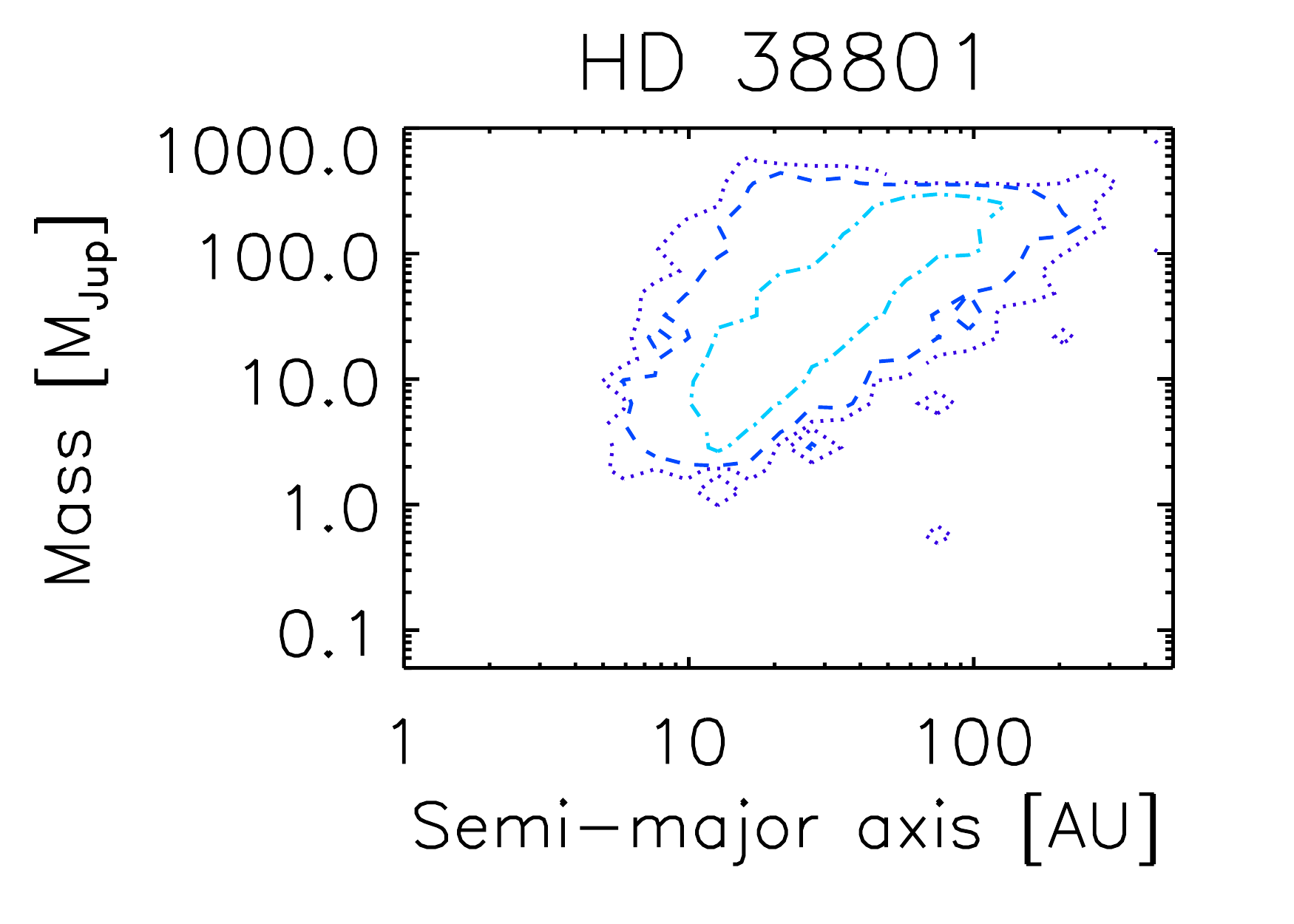} &
\includegraphics[width=0.25\textwidth]{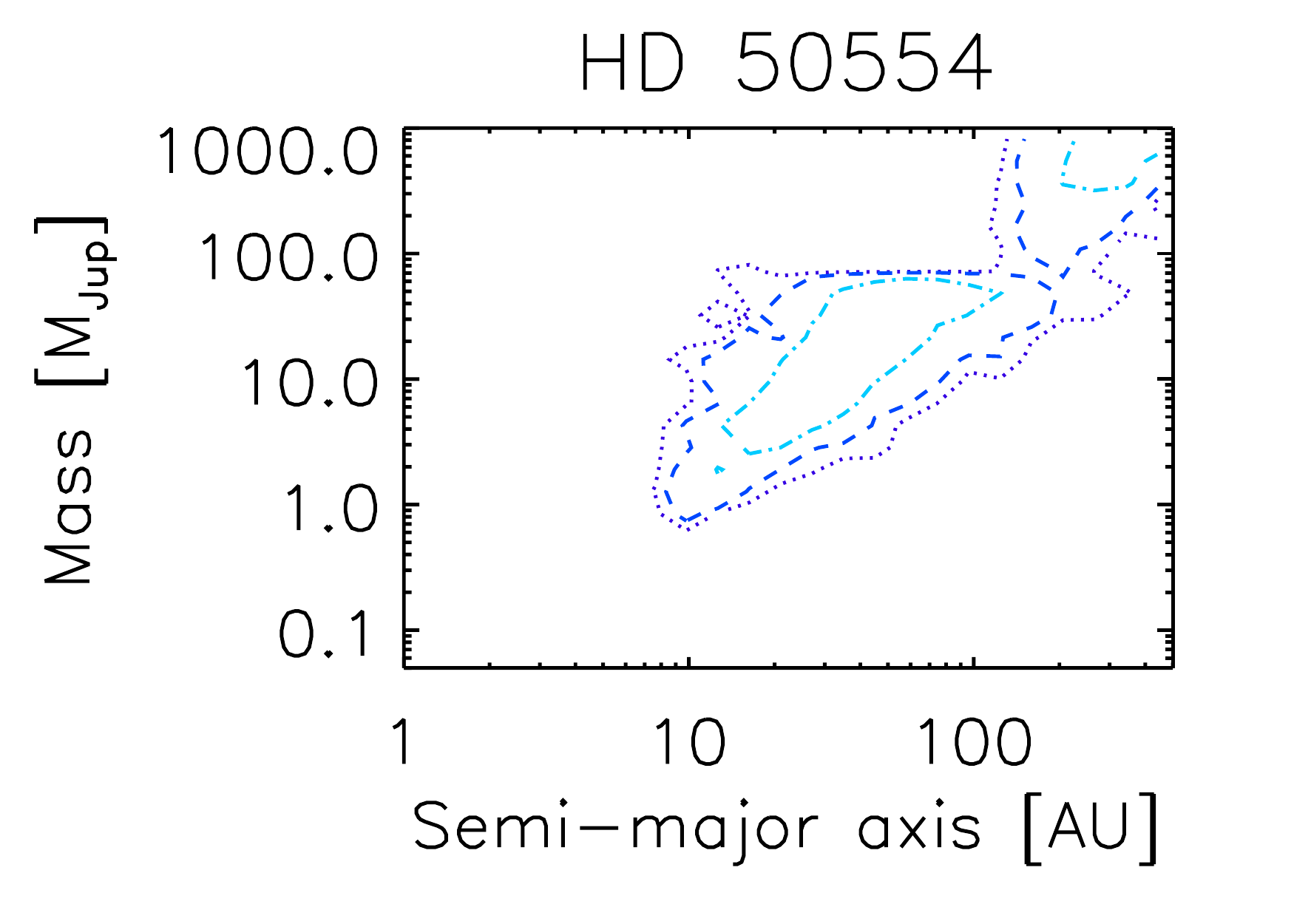} &
\includegraphics[width=0.25\textwidth]{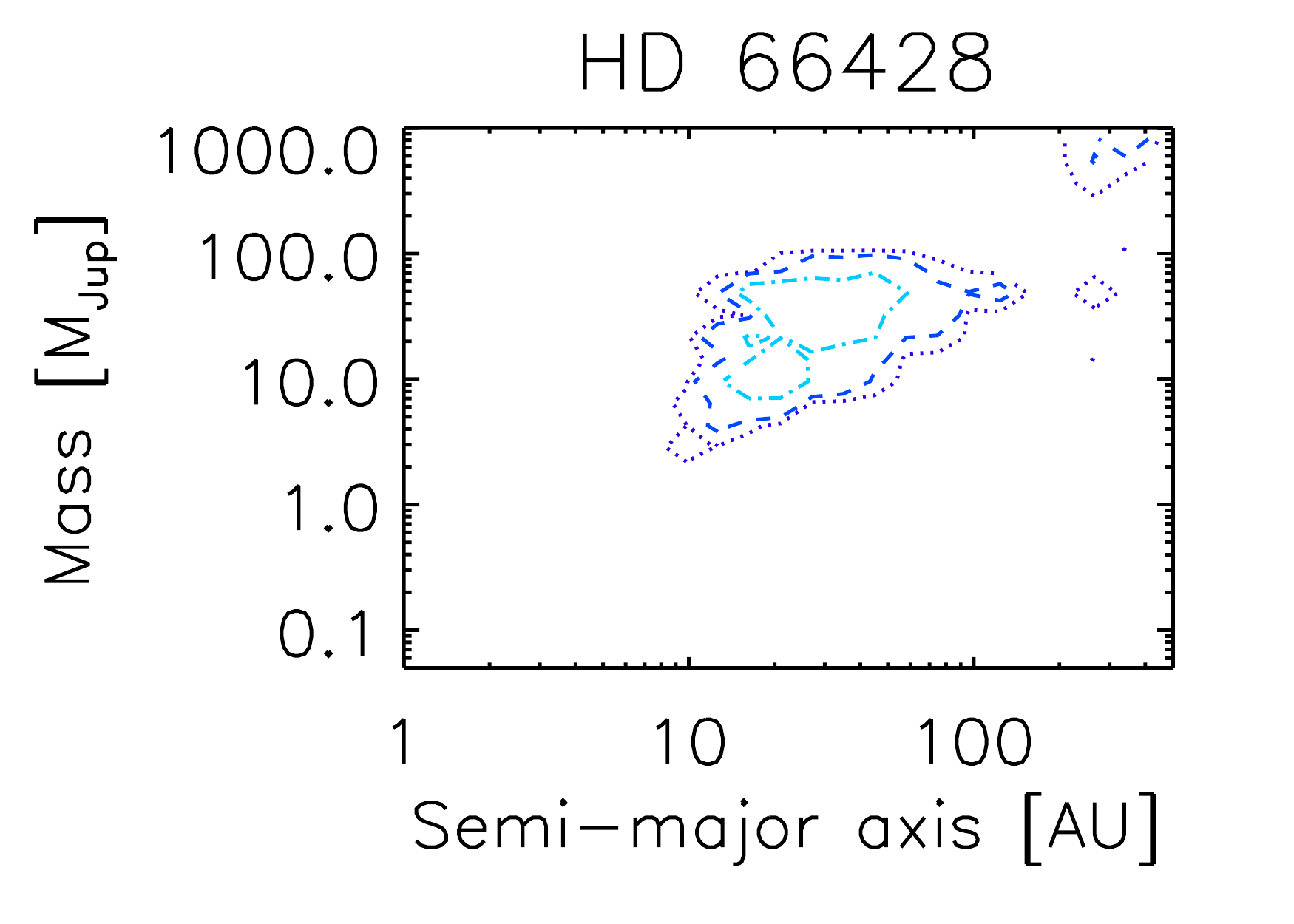}\\
\includegraphics[width=0.25\textwidth]{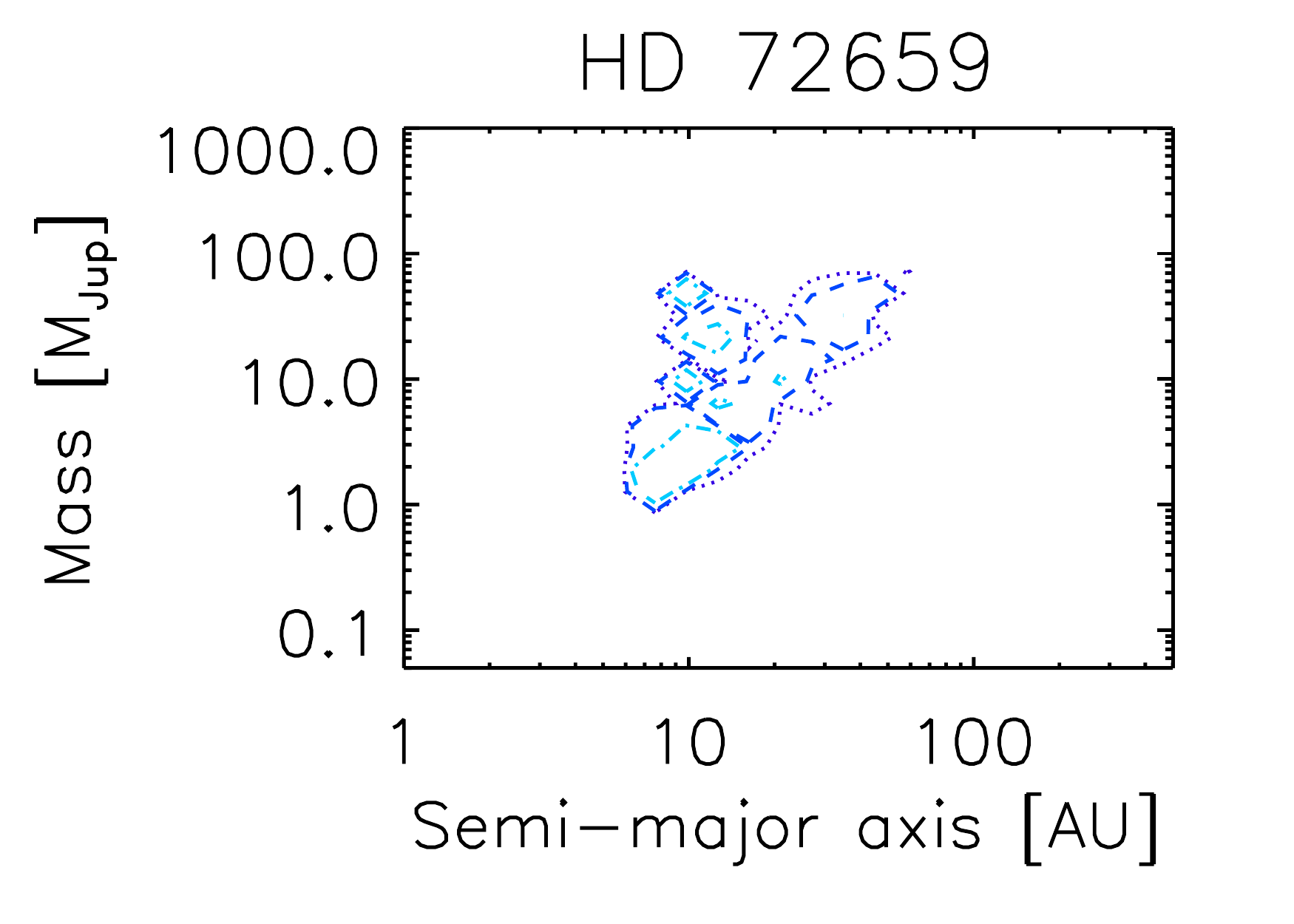} &
\includegraphics[width=0.25\textwidth]{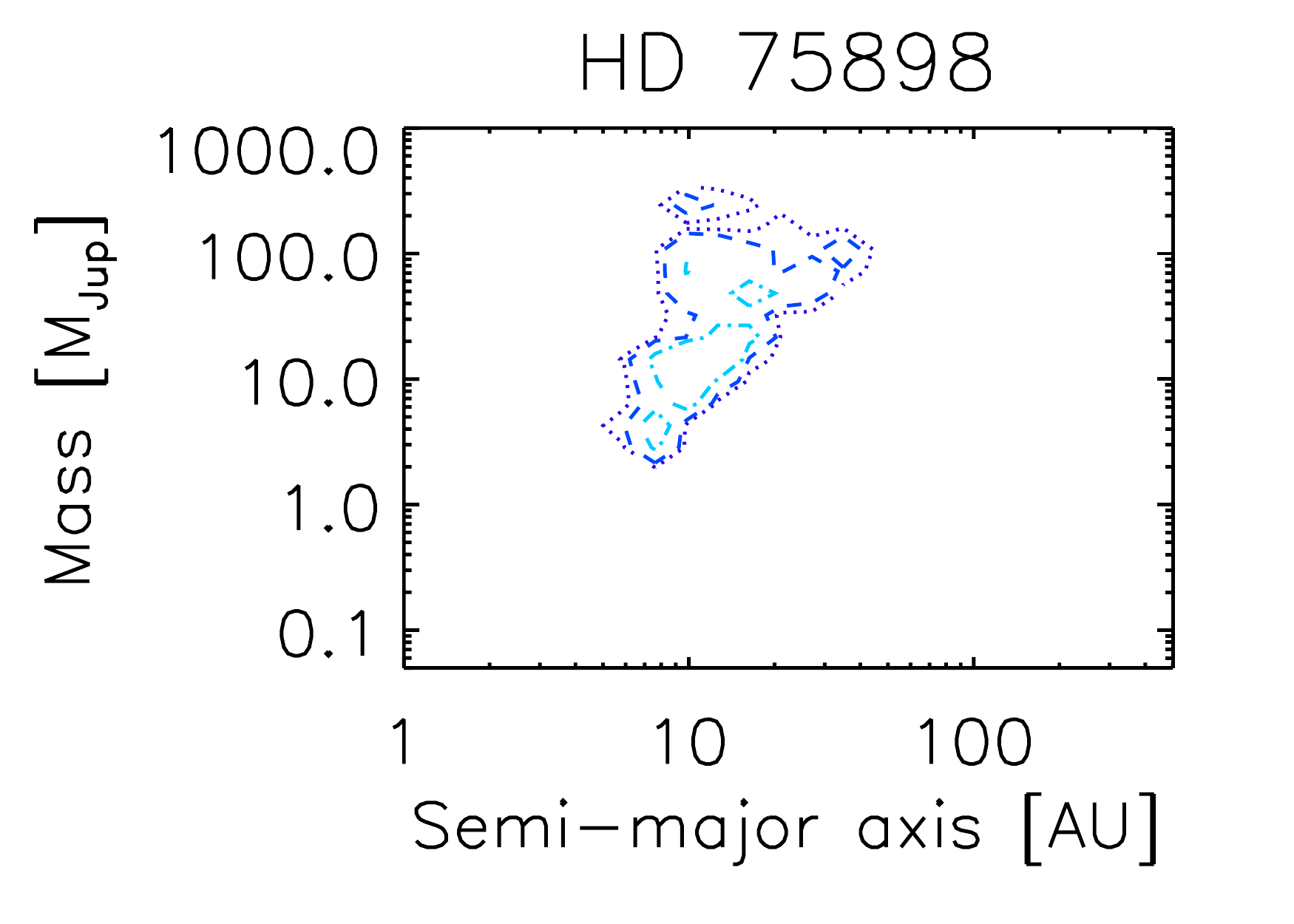}&
\includegraphics[width = 0.25\textwidth]{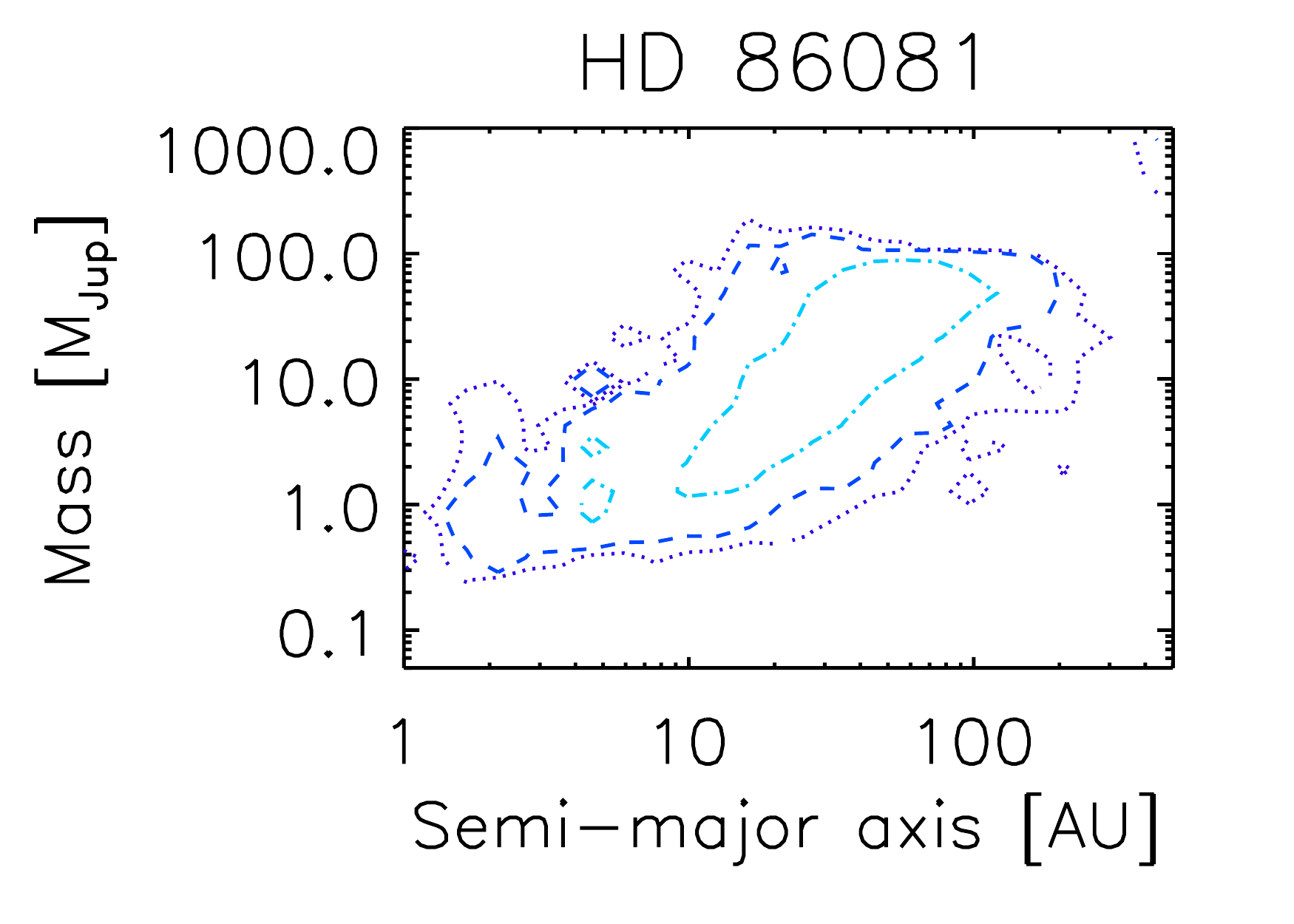}&
\includegraphics[width=0.25\textwidth]{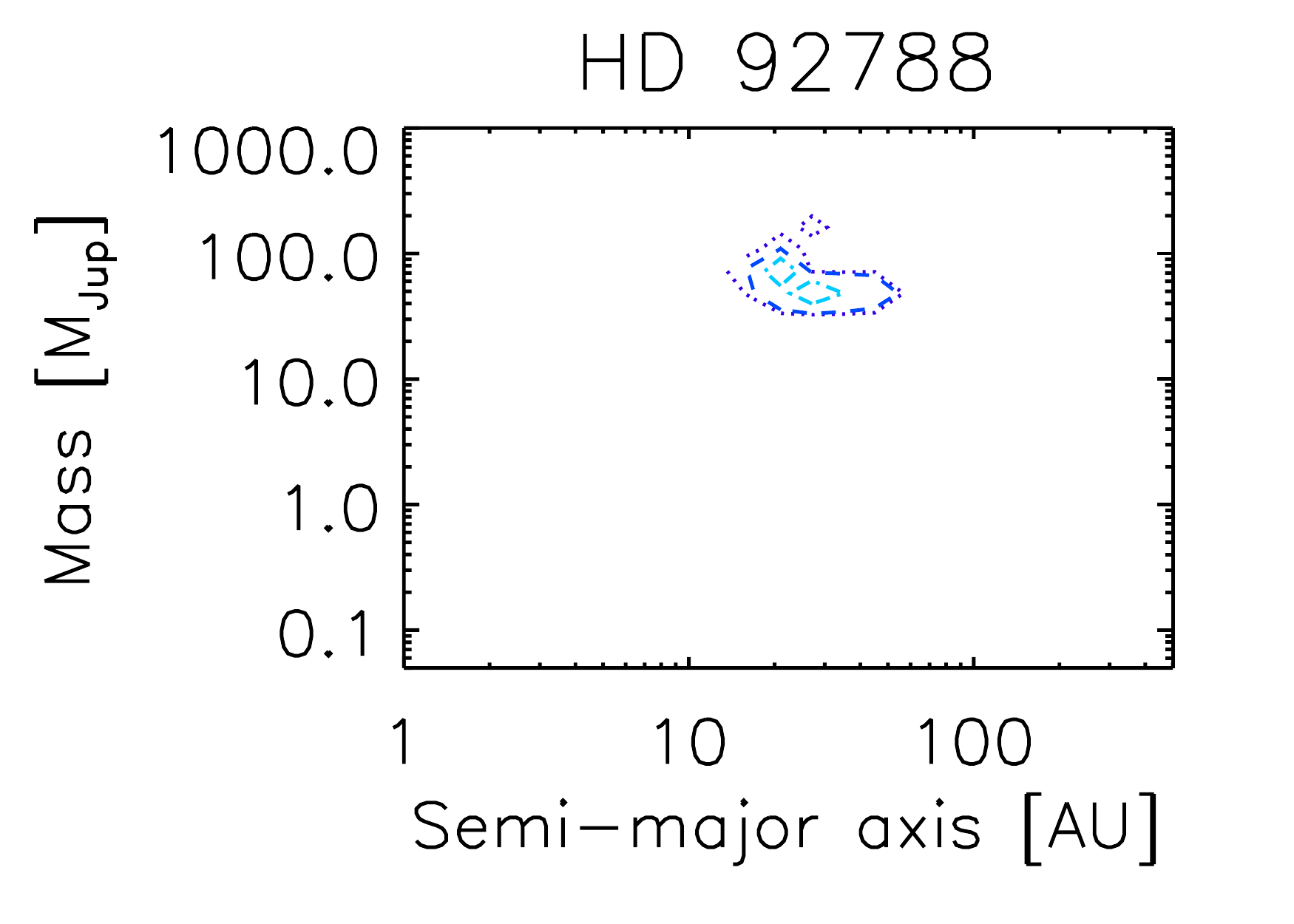} \\
\includegraphics[width=0.25\textwidth]{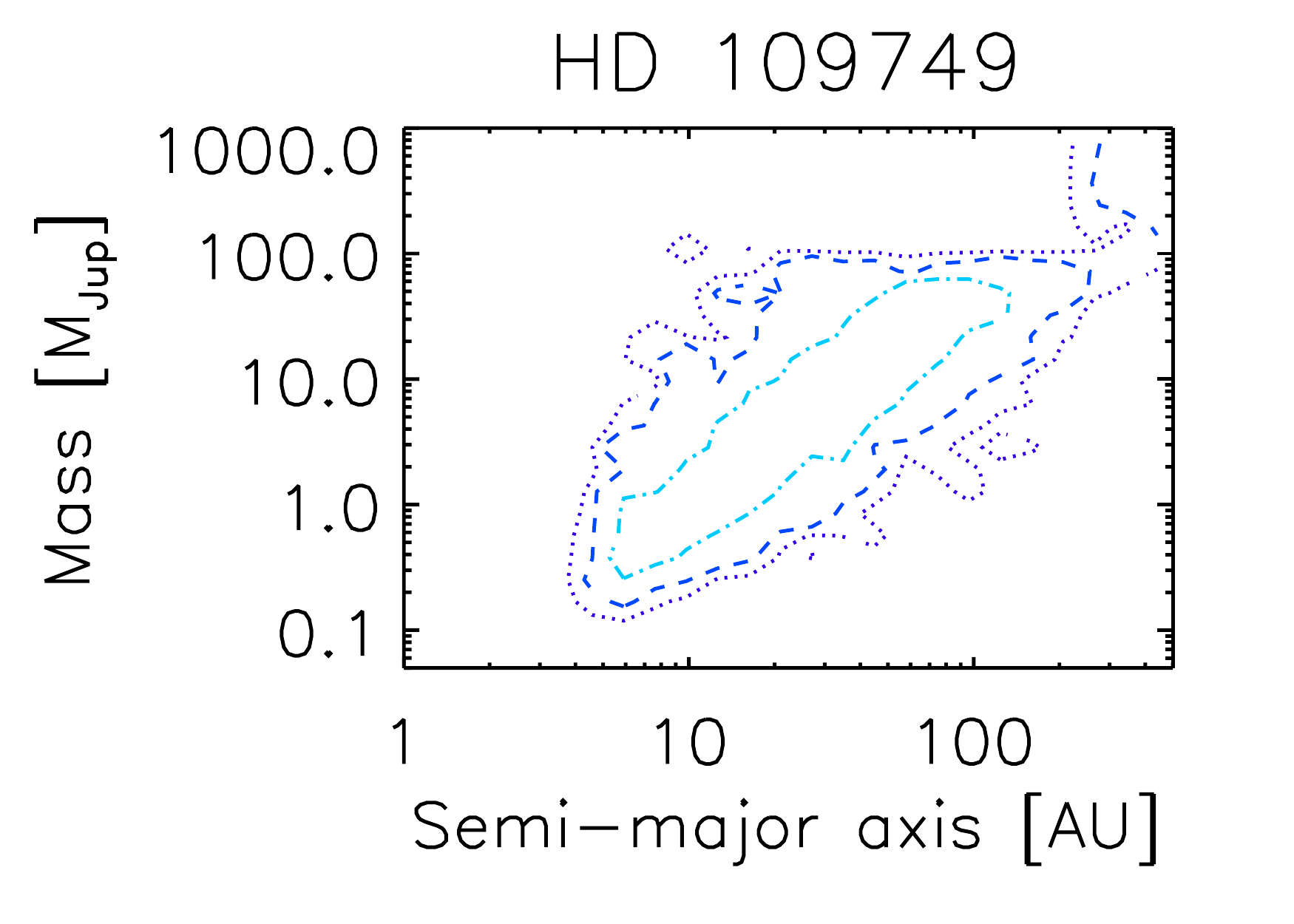}&
\includegraphics[width=0.25\textwidth]{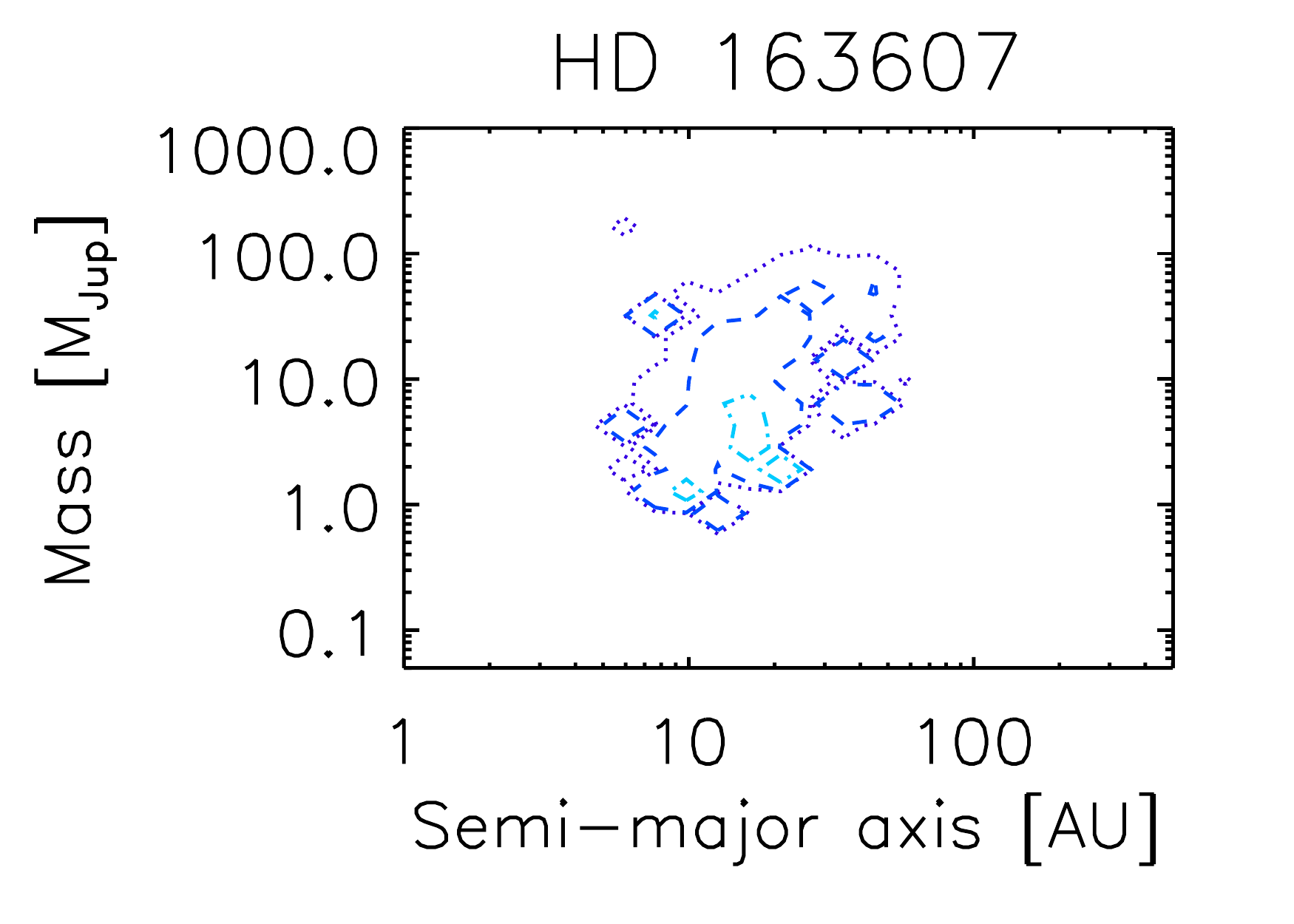} &
\includegraphics[width=0.25\textwidth]{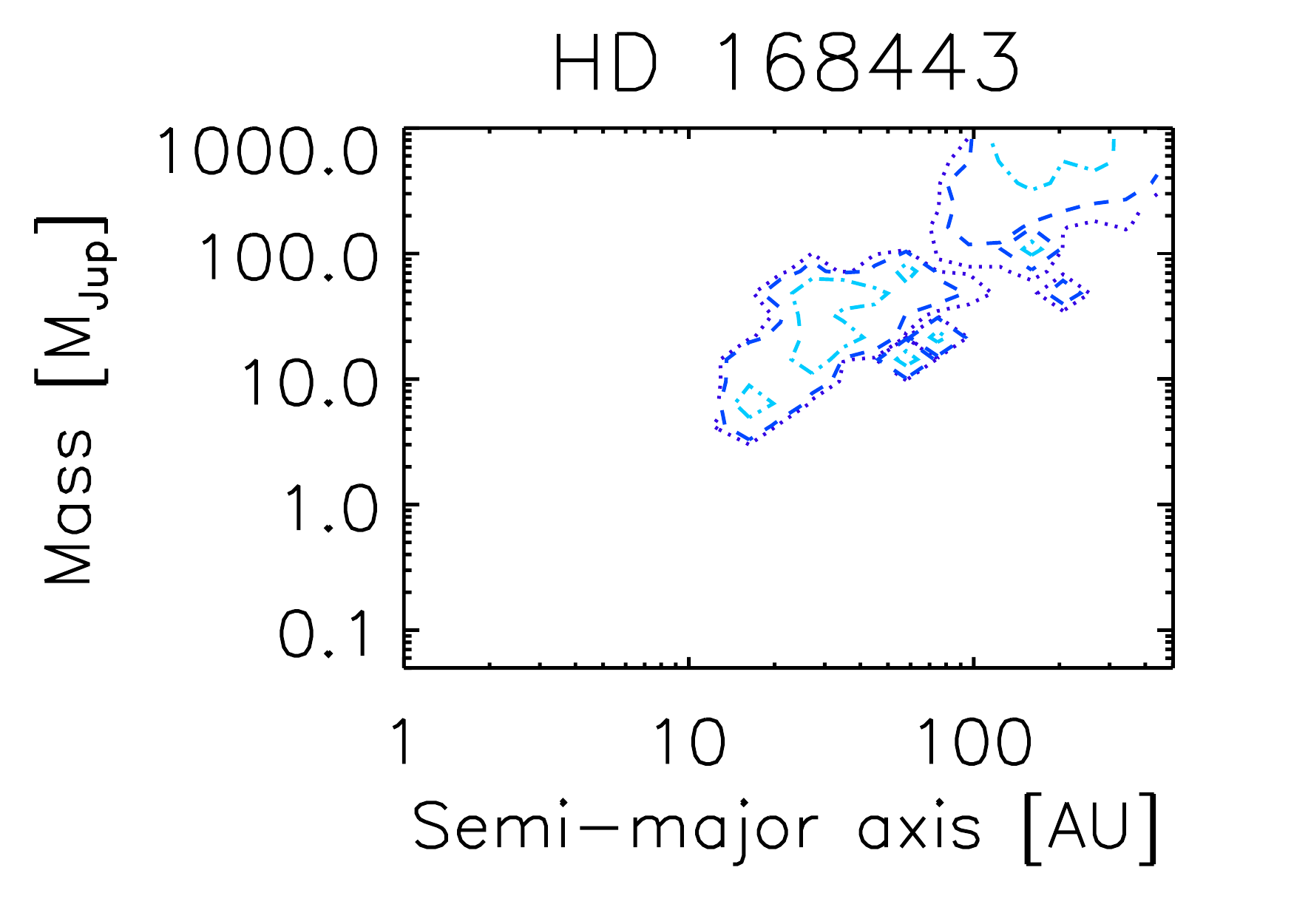} &
\includegraphics[width=0.25\textwidth]{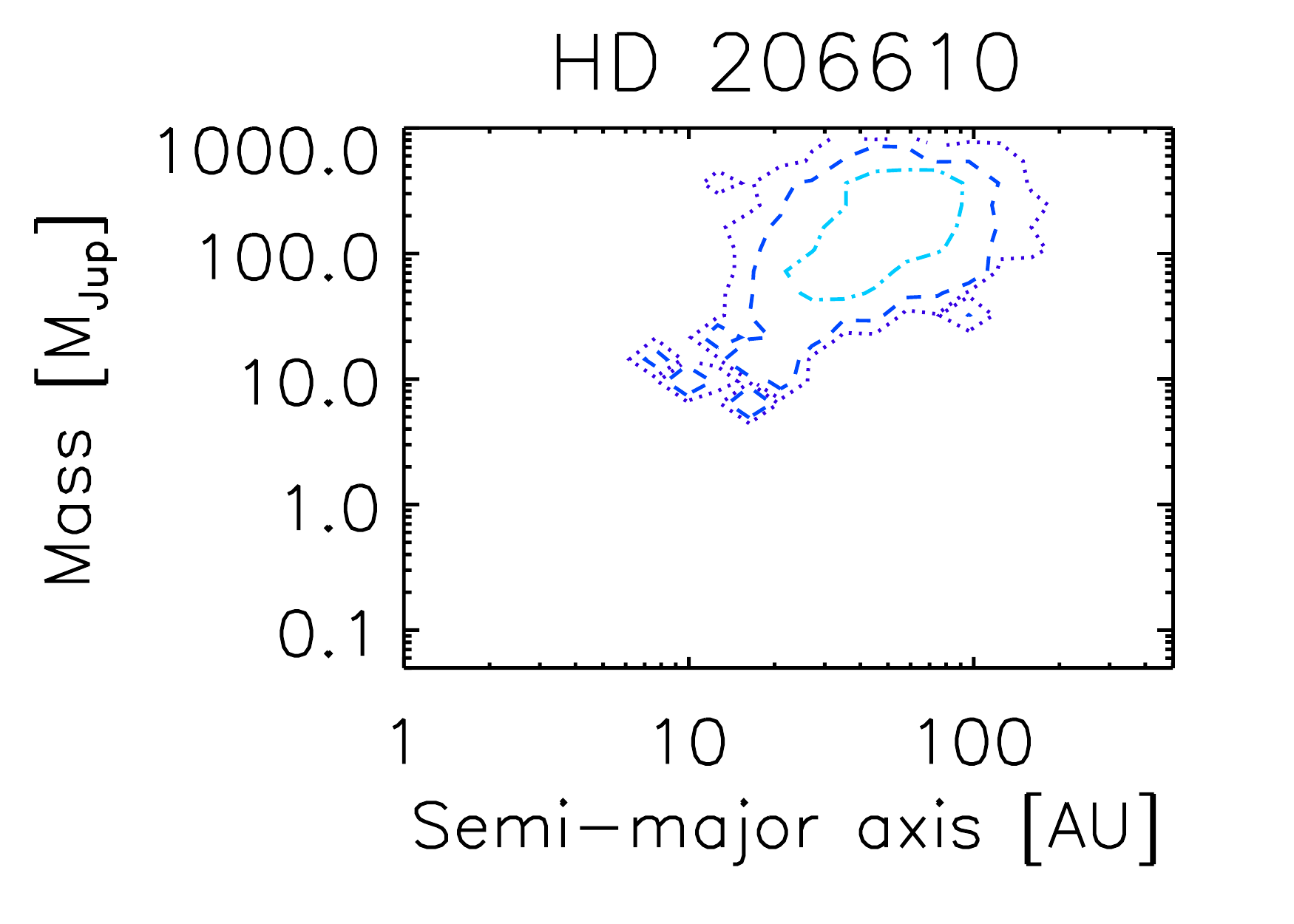} 

\end{tabular}

\caption{Companion probability distributions.  The three contours define the $1\sigma$, $2\sigma$, and $3\sigma$ levels moving outward.  While the radial velocity trends constrain these distributions on the low mass, low semi-major axis end, AO imaging constrains the high mass, high semi-major axis parameter space.  Note that the masses in these plots are true masses, not $M\sin i$.  Also note that the probability contours for HD 50499, HD 68988, HD 158038, and HD 180902 are not shown here.  This is due to the fact that the grid is too course to resolve the contours of these well-constrained systems (the probability density is concentrated in only a couple of grid points).  Finally, in some of these plots there is an apparent splitting of the contours at high mass and separation (e.g. HD 4208, HD 168443).  This is due to the fact that the constraints from the AO images were modified by the percentage of position angles covered at wide separations.}
\end{figure*}

\begin{deluxetable}{lcc}
\tabletypesize{\scriptsize}
\tablecaption{Constraints on Companion Properties}
\tablewidth{0pc}
\tablehead{
\colhead{Companion}         &
\colhead{Mass $[M_{Jup}]$}   &
\colhead{Semi-major axis [AU]}  
}
\startdata

HD 3651&$0.84-817$&$14-440$\\
HD 4208&$0.84-668$&$7.6-342$\\
HD 11506&$9.6-72$&$14-40$\\
HD 24040&$6.4-817$&$24-342$\\
HD 28678&$5.2-446$&$11-124$\\
HD 38801&$2.8-297$&$8.6-124$\\
HD 50554&$1.9-817$&$13-440$\\
HD 50499&$2.8-12$&$7.6-8.6$\\
HD 66428&$4.3-72$&$11-66$\\
HD 68988&$9.6-59$&$6.7-7.6$\\
HD 72659&$1.3-133$&$7.6-35$\\
HD 75898&$2.8-199$&$6.7-21$\\
HD 86081&$0.69-72$&$4.6-124$\\
HD 92788&$48-88$&$14-40$\\
HD 109749&$0.25-59$&$5.9-160$\\
HD 163607&$1.3-39$&$7.6-24$\\
HD 168443&$4.3-817$&$14-388$\\
HD 180902&$162-446$&$8.6-18$\\
HD 206610&$7.8-446$&$13-85$
\enddata 
 \tablecomments{The masses in this table are true masses, not $M\sin i$.}
 \end{deluxetable}

Based on the probability contours in Figure 14 and corresponding table of allowed companion masses, we conclude that the majority of companions are most likely gas giant planets, as field surveys indicate that the occurrence rate of brown dwarfs (13 -  80 $M_{\rm Jup}$) around sun-like stars is $3.2^{+3.1}_{-2.7}\%$ \citep{Metchev2009}.  We note that while the Metchev and Hillenbrand result is for brown dwarf companions to sun-like stars between 28-1590 AU, the brown dwarf parts of parameter space for our companions are typically outside of 28 AU.  Therefore, the comparison to the Metchev and Hillenbrand occurrence rate is appropriate.  For comparison, \citet{Cumming2008} states that $17\% - 20\%$ of solar type stars host a giant planet (0.3 - 10 $M_{\mathrm{Jup}}$) within 20 AU.

\subsection{Completeness Maps}

We quantified the sensitivity of this survey to companions over a range of masses and semi-major axes by determining the completeness of each system given the system's radial velocity baseline.  Once again, we defined a 50$\times$50 grid in log mass/semi-major axis space from 1-500 AU and 0.05-1000 $M_{\rm Jup}$.  In each defined grid box, we injected 500 simulated planets, each with a random mass and semi-major axis uniformly drawn from the grid box.  We draw the inclination of the orbit from a uniform distribution in $\cos i$, the eccentricity from the beta distribution, and the remaining orbital elements from a uniform distribution.  At each epoch that the star was observed, we calculated the expected RV signal caused by the injected companion.  We generated errors for these simulated data by drawing randomly from a normal distribution of width $\sqrt{\sigma_i^2 + \sigma_{jitter}^2}$, where $\sigma_i$ are the randomly shuffled measurement errors from the original radial velocities and $\sigma_{jitter}$ is the best-fit jitter value.  

To determine if a simulated companion would be detectable, we fit either a one planet orbital solution, a linear trend, or a flat line to the simulated RV observations over the observed baseline.  To determine which was the best fit, we used the Bayesian information criterion (BIC).   This is defined as:  $\mathrm{BIC} = -2L + k\ln{n}$, where $L$ is the likelihood of the model, k is the number of free parameters in the model, and $n$ is the number of data points in the observed data set.  While the likelihood can be increased by simply fitting models with more free parameters, BIC selects against these with a penalty term.  The lower the BIC value the better the model fit.  Comparing two models, if $\Delta BIC > 10$, this is very strong evidence for the model with the lower BIC \citep{Kass1995}.  Thus if the BIC values for the trend or the one-planet models were less than ten compared to the BIC value for the flat line, the simulated companion was ``detected", whereas if the flat line was the best fit, that companion was ``not detected".  This process was repeated for 500 simulated companions injected into each grid box, producing a completeness map of detection probability as a function of mass and semi-major axis.  Figure 15 shows the average completeness map of all of the systems.  

\begin{figure}
\includegraphics[width=0.5\textwidth]{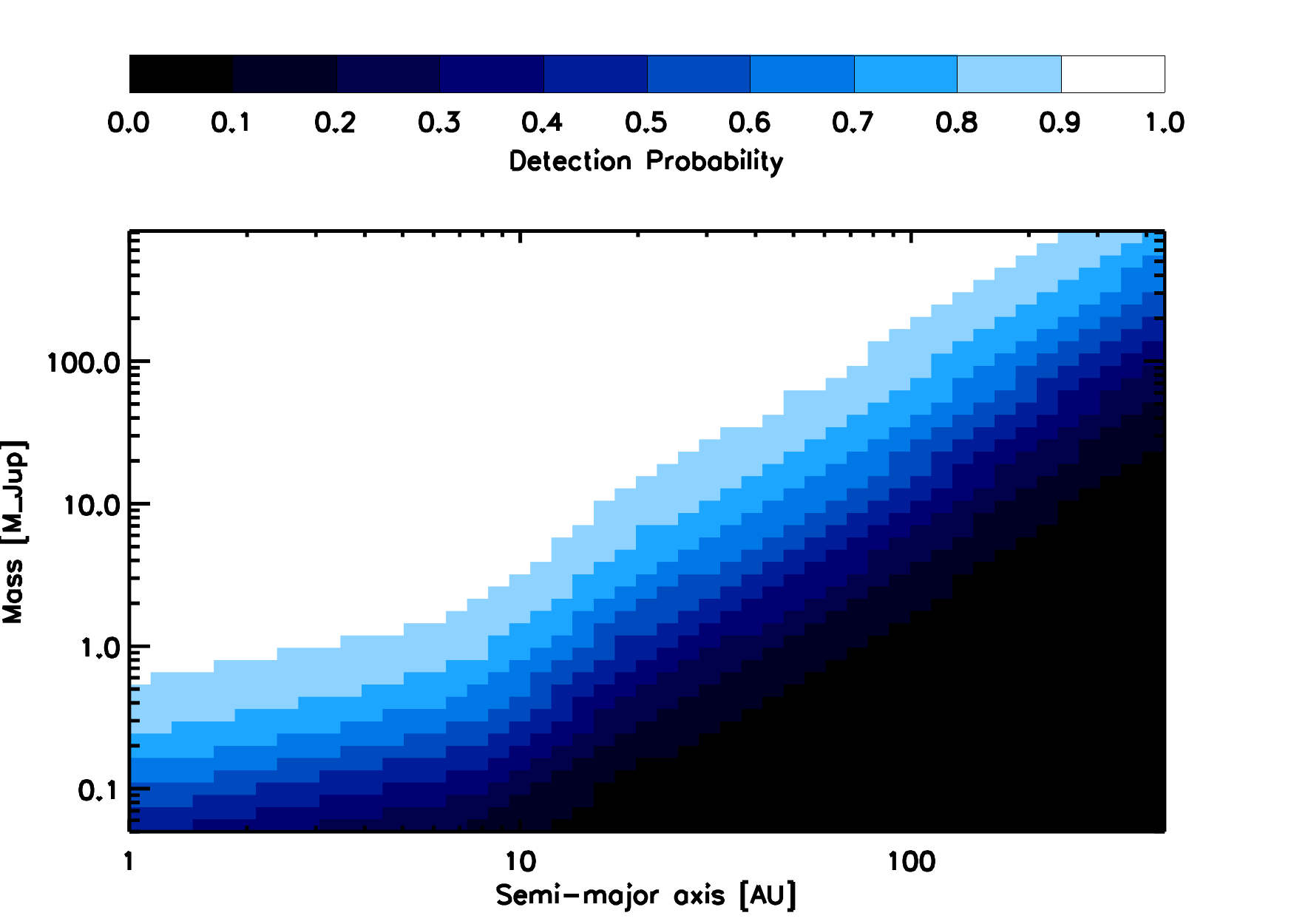}
\caption{Average completeness map for all systems.  Each color corresponds to a detection probability.  For example, companions occupying parameter space in the white areas of the map had a $90\%$ to a $100\%$ chance of being detected by this survey.}  
\end{figure}

Figure 16 shows the $50\%$ contour for the average of all the systems, for the least sensitive system, and for the most sensitive system.  The sensitivity of each system to planets with varying masses and semi-major axes depends on the length of the RV baseline, the magnitude of the measurement errors, and the number of data points for the system.  The longer the baseline, the smaller the errors, and the greater the number of data points, the more sensitive the system.  The least sensitive system is HD 5891, while the most sensitive system is HD 156668.  

\begin{figure}

\includegraphics[width=0.5\textwidth]{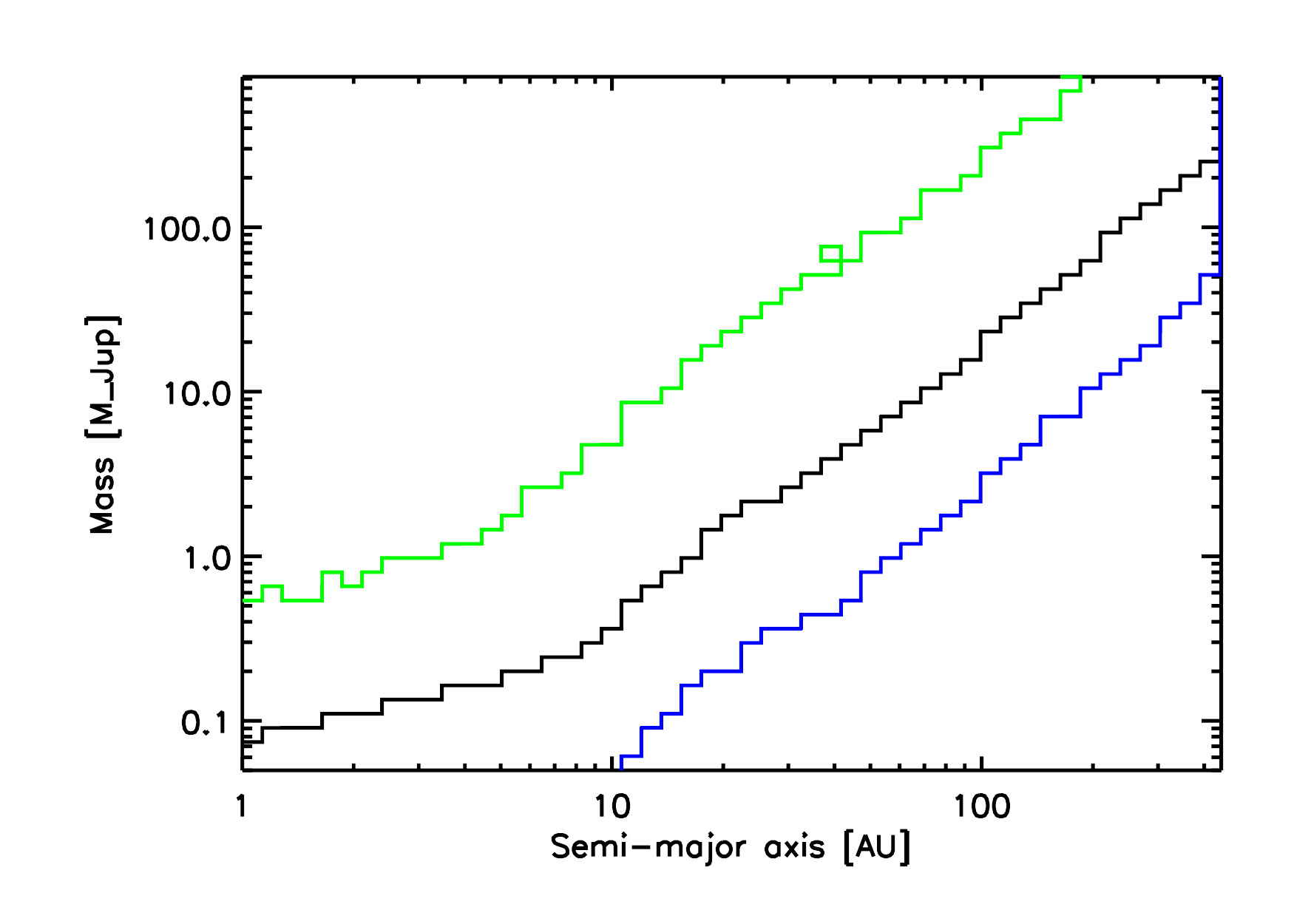}
\caption{Completeness contours corresponding to $50\%$ probability of detection.  The black contour corresponds to the average sensitivity for all the systems, the blue contour corresponds to HD 156668, the system with the greatest sensitivity, and the green contour corresponds to HD 5891, the system with the least sensitivity.}  
\end{figure}


\section{Discussion}

\subsection{The distribution of wide companions}

Now that we have determined the parameter space where each detected companion is most likely to reside, we can determine the most likely underlying distribution for these massive, long-period companions in confirmed exoplanet systems.  We assume that the companions are distributed in mass and semi-major axis space according to a double power law (e.g. Tabachnik $\&$ Tremaine 2002, Cumming et al 2008):

\begin{equation}
f(m,a) = Cm^{\alpha}a^{\beta}
\end{equation}

\noindent The total likelihood for a set of N exoplanet systems is given by:

\begin{equation}
\mathscr{L} =  \Pi^{N}_{i = 1}p(d_i | C, \alpha, \beta)
\end{equation}

\noindent  where the expression on the right is the probability of obtaining the set of data $d$ for a system $i$ given values for $C$, $\alpha$, and $\beta$.  We assume that each system can have at most one companion, and that the probability of obtaining the measured RV dataset for an individual star is therefore the sum of the probability that the system does contain a planet and the probability that the system does not contain a planet for each set of C, alpha, and beta values considered.  The probability of a system having zero planets is given by:

\begin{equation}
p(d_i, 0 | C, \alpha, \beta) = p(d_i | 0)[1 - Z]
\end{equation}

\noindent The quantity $p(d_i | 0)$ is the probability of obtaining the measured RV dataset given that there are no planets in the system.  $Z$ is the probability that the system contains a planet within the specified range in mass and semi-major axis space.  Here, $p(d_i | 0)$ and $Z$ are given by the following equations.

\begin{equation}
p(d_i | 0) = \Pi_j\frac{1}{\sqrt{2\pi}\sigma_j}exp\bigg[\frac{-1}{2}\bigg(\frac{d_j - m_j}{\sigma_j}\bigg)^2\bigg]
\end{equation}
\begin{equation}
Z = \int_{m_1}^{m_2} d\ln m \int_{a_1}^{a_2} d\ln a \hspace{0.1cm} C m^{\alpha} a^{\beta}
\end{equation}

\noindent In equation 7, $d_j$ is the jth datapoint in the dataset d for system i, $m_j$ is the corresponding model point, and $\sigma_j$ is the error on the jth datapoint.  

The probability of a system having one planet given values $C$, $\alpha$, and $\beta$ is:
\begin{equation}
p(d_i, 1 | C, \alpha, \beta) = \int_{a_1}^{a_2} d\ln a \int_{m_1}^{m_2} d\ln m \hspace{0.1cm} p(d_i | a, m) C m^{\alpha} a^{\beta}
\end{equation}

\noindent where $p(d_i | a, m)$ is the probability of a companion at a given mass and semi-major axis, which we know from the previously calculated two dimensional probability distributions.  We then combine these expressions in order to calculate the likelihood of a given set of C, $\alpha$, and $\beta$ values given the measured RV data for all the stars in our sample:

\begin{equation}
\mathscr{L} =  \Pi^{N}_{i = 1} \bigg[p_i(d_i, 0 | C, \alpha, \beta) + p_i(d_i, 1 | C, \alpha, \beta)\bigg]
\end{equation}

\noindent  Note that for this calculation we use the probability distributions for all systems, not just those with 3$\sigma$ trends.  To maximize $\mathscr{L}$, we varied the values of $C, \alpha,$ and $\beta$ using a grid search.  The $16\%$ - $84\%$ confidence intervals on these parameters were then obtained using the MCMC technique.

\subsection{Occurrence Rates}

The overall occurrence rate for the population of companions can be estimated by integrating $f(m,a)$ over a range of masses and semi-major axes.   In addition to the population of exoplanet systems described previously, we also included the 51 hot Jupiter systems published in \citet{Knutson2014}.  While we adopted the published RV model fits for each of the hot Jupiter systems, we recalculated probability distributions with the same grid spacing used for the 123 new systems described in this study for consistency.  

In \citet{Knutson2014}, we utilized a conservative approach in which we defined a given planet as a non-detection with $100\%$ probability whenever the measured trend slope was less than 3$\sigma$ away from zero.  Instead of using a binary picture of planet occurrence, our revised likelihood function is more statistically correct, as it considers the probability of hosting a planet in all of our systems.  We note that integrated companion occurrence rates calculated using this approach are particularly sensitive to the estimated jitter levels in our fits, where an underestimate of the true stellar jitter levels could result in an over-estimate of the corresponding companion occurrence rates.  As a test of this new method we re-calculate the companion occurrence rate for the sample of 51 transiting hot Jupiters presented in \citet{Knutson2014} and find a value of $70 \pm 8\%$ for companions between 1 - 13 $M_{\rm Jup}$ and 1 - 20 AU.  This is approximately 2$\sigma$ higher than the value of $51 \pm 10\%$ obtained for this sample of stars using our older, more conservative likelihood function.

We calculate the overall frequency of companions beyond 5 AU in our new expanded system of 174 planetary systems by integrating over our best-fit probability distributions.  We evaluate the companion frequency using a variety of different mass and period ranges in order to determine how sensitively this result is to the specific limits of integration selected.  The resulting total occurrence rates are presented in Table 7, and the corresponding values of $C$, $\alpha$, and $\beta$ are shown in Table 8.

\begin{deluxetable}{l|ccc}
\tabletypesize{\scriptsize}
\tablecaption{Total Occurrence Rates for Companions Beyond 5 AU}
\tablewidth{0pc}
\tablehead{
\colhead{}&
\colhead{5 - 20 AU}&
\colhead{5 - 50 AU}&
\colhead{5 - 100 AU}
}
\startdata
0.5 - 20 $M_{\rm Jup}$  & $59.2^{+5.1}_{-5.2}$ &  $66.5^{+5.6}_{-5.8}$  &    $62.1^{+5.4}_{-5.7}$ \\
0.5 - 13 $M_{\rm Jup}$  & $56.9^{+5.2}_{-5.3}$  &  $62.3^{+5.7}_{-5.8}$  &   $61.0^{+5.5}_{-5.8}$\\
1 - 20 $M_{\rm Jup}$  & $52.4^{+4.5}_{-4.7}$ &$59.6^{+5.4}_{-5.5}$ & $60.9^{+5.2}_{-5.6}$

\enddata
\end{deluxetable}

\begin{deluxetable*}{l|ccc}
\tabletypesize{\scriptsize}
\tablecaption{Power Law Coefficients for Companions Beyond 5 AU}
\tablewidth{0pc}
\tablehead{
\colhead{}&
\colhead{5 - 20 AU}&
\colhead{5 - 50 AU}&
\colhead{5 - 100 AU}
}
\startdata
0.5 - 20 $M_{\rm Jup}$  &  $C = 0.0036^{+0.0047}_{-0.0018}$ &$C = 0.0174^{+0.0174}_{-0.0085}$ & $C = 0.023^{+0.026}_{-0.012}$ \\
  & $\alpha = -0.04^{+0.13}_{-0.12}$ & $\alpha = 0.29^{+0.18}_{-0.16}$&  $\alpha = 0.53^{+0.25}_{-0.22}$ \\
  &  $\beta = 1.46^{+0.47}_{-0.37}$&  $\beta = 0.38^{+0.22}_{-0.22}$ & $\beta = 0.05^{+0.18}_{-0.19}$\\
0.5 - 13 $M_{\rm Jup}$ &  $C = 0.0063^{+0.0076}_{-0.0029}$ & $C = 0.015^{+0.031}_{-0.014}$  & $C = 0.019^{+0.039}_{-0.016}$ \\
  & $\alpha = 0.08^{+0.15}_{-0.14}$ &  $\alpha = 0.56^{+0.22}_{-0.19}$&  $\alpha = 0.86^{+0.28}_{-0.26}$\\
  &  $\beta = 1.22^{+0.33}_{-0.35}$&  $\beta = 0.38^{+0.21}_{-0.22}$ &  $\beta = 0.02^{+0.17}_{-0.20}$\\
 1 - 20 $M_{\rm Jup}$ & $C = 0.0020^{+0.0062}_{-0.0029} $&   $C = 0.0083^{+0.0084}_{-0.0038}$ &  $C = 0.0063^{+0.0072}_{-0.0029}$\\
  & $\alpha = -0.22^{+0.15}_{-0.15}$ &    $\alpha = 0.44^{+0.22}_{-0.23}$&  $\alpha = 0.86^{+0.26}_{-0.23}$ \\
  & $\beta = 1.82^{+0.25}_{-0.27}$ &   $\beta = 0.56^{+0.22}_{-0.22}$  &   $\beta = 0.26^{+0.14}_{-0.15}$
\enddata
\tablecomments{We note that the $\alpha$ and $\beta$ values presented here are strongly influenced by the slope of the probability distributions for companions with partially resolved orbits, and therefore should not be taken as reliable estimates of the actual companion distribution.  Please see the discussion below for further explanation.}
\end{deluxetable*}

We find that our values of $\alpha$ and $\beta$ vary significantly depending on the integration range chosen, and are therefore not accurate estimates of the power law coefficients for this population of long-period companions.  This dependence on integration range is due to the fact that many of the companions detected in our study have poorly constrained masses and orbits.  When we vary the range of masses and semi-major axes used in our fits we truncate the probability distributions for these companions at different points, therefore biasing our corresponding estimates of $\alpha$ and $\beta$. 
 
 Although it is difficult to obtain reliable estimates for the values of $\alpha$ and $\beta$ for long-period companions, we can nonetheless investigate whether or not this population increases in frequency as a function of increasing mass and semi-major axis by calculating the occurrence rate of this sample of systems using equal steps in log space to increase the semi-major axis and mass integration ranges.  When stepping in semi-major axis, we keep the mass range constant, 1 - 20 $M_{\rm Jup}$, and when stepping in mass, we keep the semi-major axis range constant, 5 - 20 AU.  We then compare the observed changes in companion frequency per step in log mass or log semi-major axis in order to determine empirically how the overall distribution of companions compares to predictions from various power law models.  For example, if the increase in frequency per log semi-major axis declines at larger separations this would imply a negative value for $\beta$, whereas the opposite would be true for a positive $\beta$.  We calculate the uncertainties on the changes in occurrence rates by adding the individual uncertainties on the occurrence rates in quadrature.

 We calculate the change in the integrated occurrence rate as a function of increasing semi-major axis (Figure 17) using a lower integration limit of 1 AU and including all planets in these systems, not just the outer companions.  We find that for small separations these rates increase relatively quickly as compared to the predictions of a power law model with $\beta = 0$ (i.e. a uniform distribution in semi-major axis), whereas for large separations these rates increase relatively slowly.  This suggests a positive $\beta$ value for giant planets at smaller separations and a negative $\beta$ value for outer companions at larger separations, with a broad peak in the distribution between 3 - 10 AU.  When we examine the corresponding change in occurrence rate for companions beyond 5 AU as a function of planets mass (Figure 18), we find that these rates also increase slowly as compared to the predictions of a power law model with $\alpha = 0$.  This implies a negative $\alpha$ value.   
 
\begin{figure}
\includegraphics[width=0.5\textwidth]{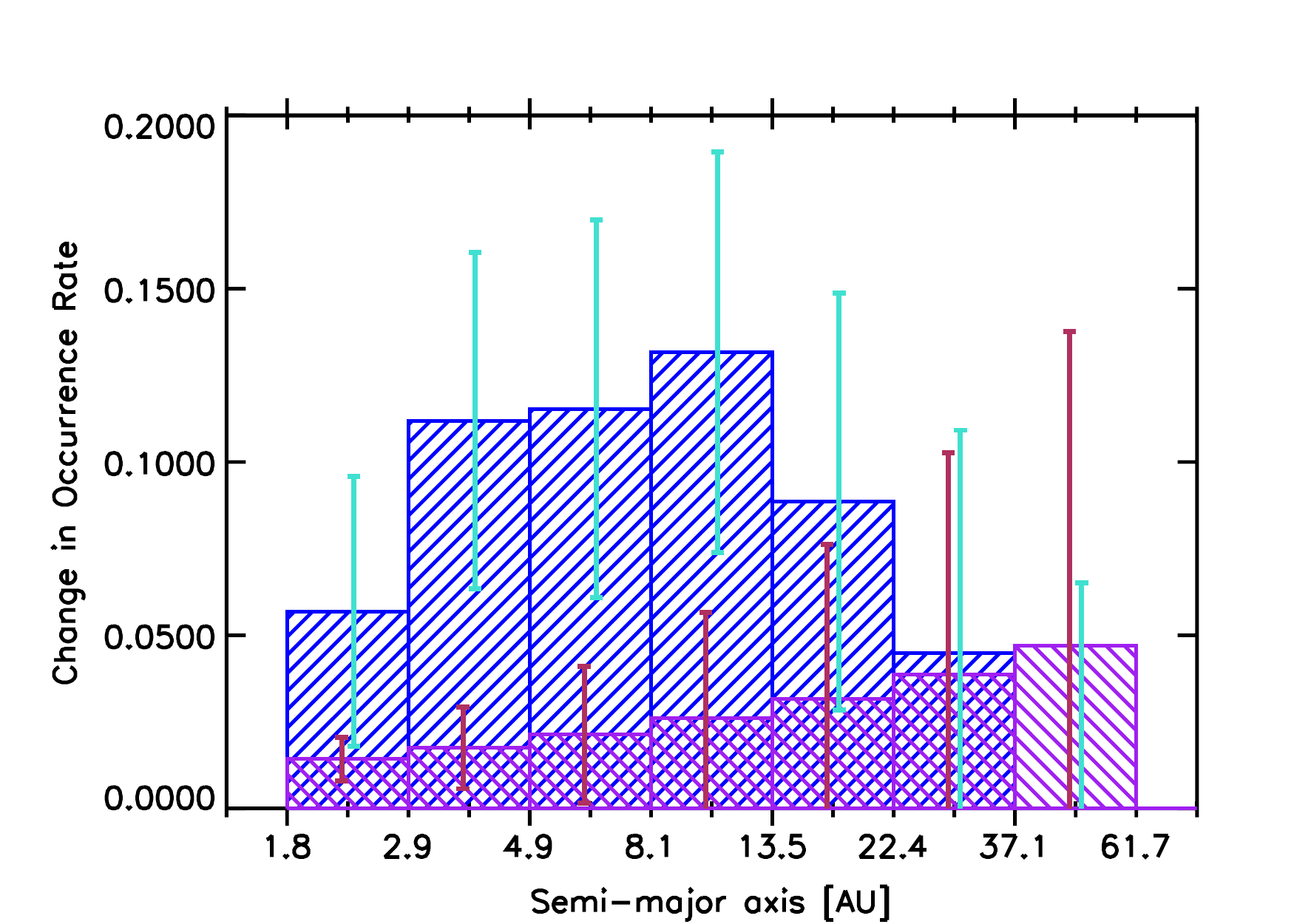} 
\caption{This plot shows the change in occurrence rate between adjoining semi-major axis steps as a function of the upper semi-major axis integration limit.  The results for the Cumming et al power law distribution are plotted in purple, while the results from this survey are plotted in blue. For the fits for our survey we include all planets in these systems outside 1 AU, not just outer companions as in the rest of our analysis.  This allows us to study the relative distribution of planets in these systems across a broad range of semi-major axes.  The sensitivity limit of the Cumming et al survey is $\sim$3 AU.   For our survey, we are $\sim$50$\%$ complete between 1 - 20 $M_{\rm Jup}$ and 5 - 100 AU.  We note that the slight upward trend of the purple histogram bins corresponds to a $\beta$ value that is 2.6$\sigma$ away from zero.}
\end{figure}

\begin{figure}
\includegraphics[width=0.5\textwidth]{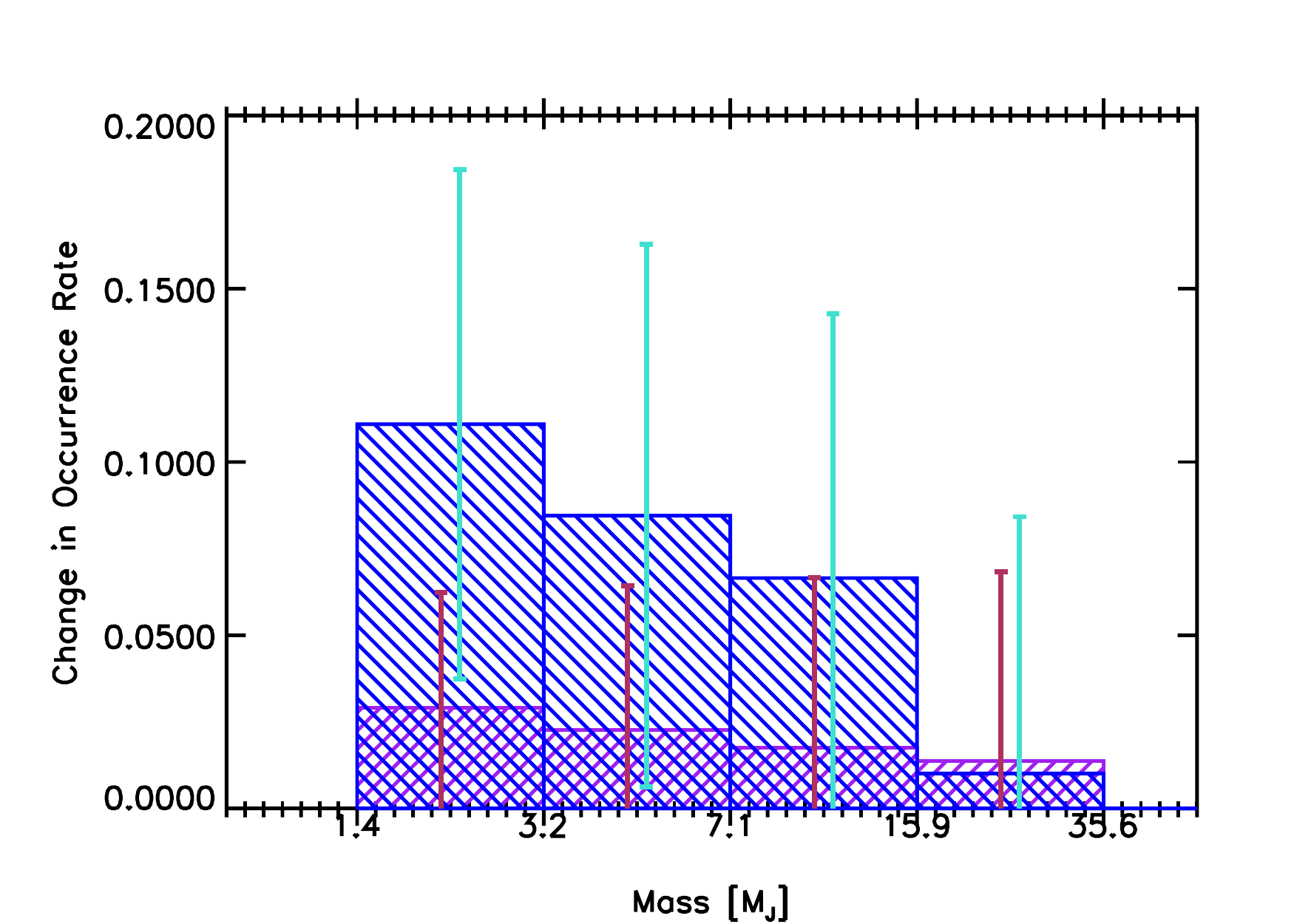} 
\caption{This plot shows the change in occurrence rate between adjoining mass steps as a function of the upper mass integration limit.  The results from the Cumming et al power law distribution are plotted in purple, while the results from this survey are plotted in blue.  We note that Cumming et al only includes planets with masses below 10 $M_{\rm Jup}$ in their survey, whereas we include companions with masses up to 20 $M_{\rm Jup}$.  The occurrence rates for larger masses shown in this plot are therefore an extrapolation based on our best-fit power law models.  The slight downward trend in the purple histograms corresponds to an $\alpha$ value that is 1.6$\sigma$ away from zero.}
\end{figure}

We next compare our constraints on the mass and semi-major axis distribution of long-period companions to predictions based on studies of short-period planets around FGK stars.  Since values of $\alpha$ and $\beta$ are broadly consistent among these studies (e.g. Bowler et al 2010), the results from \citet{Cumming2008} will be taken as representative:  $\alpha = -0.31 \pm 0.2$ and $\beta = 0.26 \pm 0.1$.  These values were derived for planet masses between $0.3 - 10$ $M_{\rm Jup}$ and periods less than 2000 days (approximately 3 AU).  We would like to know whether or not the population of companions beyond 5 AU is consistent with predictions based on the power law coefficients from this study.  We answer this question by repeating our previous calculation using the Cumming et al power law, where we determine the change in the integrated occurrence rate per log mass and semi-major axis steps over the parameter range of interest.  We calculate the uncertainties on these changes in occurrence rate by assuming Gaussian distributions for $\alpha$ and $\beta$ and using a Monte Carlo method to get a distribution of occurrence rates for each semi-major axis and mass integration range.  We then determine the uncertainties on the changes in occurrence rates by adding the uncertainties on the occurrence rates in quadrature.  We note that due to correlations between $\alpha$ and $\beta$ these uncertainties are slightly overestimated.  We then compare these results to those obtained by fitting to our sample of long-period planets in Figures 17 and 18.

As shown in Figure 17, the Cumming et al. power law predicts an increase in the frequency of planets as a function of increasing semi-major axis, whereas our fits suggest a declining frequency for gas giant companions beyond the conservative 3 - 10 AU range.  This implied disagreement between the integrated occurrence rates for our sample as compared to the extrapolated occurrence rates of Cumming et al is not surprising, as \citet{Cumming2008} only fits gas giant planets interior to 3 AU.  We speculate that this difference may indicate either a peak in the frequency of gas giant planets in the 3-10 AU range, or a difference between the population of outer giant planet companions in these systems and the overall giant planet population.  In contrast to this result, Figure 18 indicates that the mass distribution of the long-period companions in our study is consistent with the negative $\alpha$ value (i.e. increasing frequency with decreasing planet mass) reported by Cumming et al. for the population of planets interior to 3 AU.  

We next consider how the frequency of companions in these systems varies as a function of other parameters, including the inner planet mass, semi-major axis, and stellar mass.  We select an integration range of $1 - 20$ $M_{\rm Jup}$ and 5 - 20 AU for these companions; this range is large enough to include all known companions detected by our survey, while still remaining small enough to ensure that we do not extrapolate too far beyond the region in which we are sensitive to companions.  We find that within this integration range, the total occurrence rate for massive, long-period companions is $52.4^{+4.5}_{-4.7}\%$.

\citet{Johnson2010_2} showed that planet occurrence rates and system architecture vary as a function of stellar mass.  The A and M star systems are the high and low extremes of the sample's stellar mass range.  To address the concern that including A and M star systems would influence our final results, we ran the entire grid search and MCMC analyses again excluding the 29 A and M star systems in the sample.  The occurrence rate for this FGK-only sample is $54.6^{+4.8}_{-4.8}\%$.  We therefore conclude that the occurrence rates for the sample with and without the A and M stars are consistent with each other at the $0.4\sigma$ level.   

Following the total occurrence rate calculation, we calculated the occurrence rate of massive, long-period companions as a function of inner-planet semi-major axis.  We divided the total sample up into three bins - systems with planets interior to 0.1 AU (hot gas giants), systems with planets between 0.1 and 1 AU (warm gas giants), and systems with planets between 1 and 5 AU (cold gas giants).  For each bin, we repeated our fits to derive new values of $C$, $\alpha$, and $\beta$, which we integrated over a range of $1 - 20$ $M_{\rm Jup}$ and 5 - 20 AU.  Our results are presented in Figure 19.  The hot gas giant companion frequency is 2.4$\sigma$ higher than that of the warm gas giants, and 2.3$\sigma$ higher than that of the cold gas giants.  This suggests that gas giants with orbital semi-major axes interior to 0.1 AU may have a higher companion fraction than their long-period counterparts, albeit with the caveat that this short-period bin is dominated by our transiting hot Jupiter sample.  These planets typically have fewer radial velocity measurements than planets detected using the radial velocity technique, which could result in an underestimate of the stellar jitter for these stars.  

If this enhanced companion fraction for short-period planets is confirmed by future studies, it would suggest that three body interactions may be an important mechanism for hot Jupiter migration.  Alternatively, this trend might also result from differences in the properties of the protoplanetary disks in these systems.  If we suppose that each disk that successfully generates gas giant planets produces them at some characteristic radius (e.g. the ice line - see \citet{Bitsch2013}) separated by some time span, and these planets subsequently migrate inwards via type II migration. Gas giants that migrate early in the disk's lifetime will reach the inner magnetospheric cavity of the disk, and due to eccentricity excitation mechanisms (Rice et al (2008)), will rapidly accrete onto the host star over a timescale that is short compared to the lifetime of the disk. As the disk ages however, photoevaporation will grow the radius of the inner disk cavity. Accordingly, for those gas giants that arrive later in the lifetime of the disk, the inner disk edge will have been eaten away to the point that the eccentricity excitation mechanisms are no longer effective at shepherding the planets into the host stars, allowing migration to halt. We note that there is a very narrow window of time where the aforementioned processes allow for a successful formation of a hot Jupiter (which may self-consistently explain their inherent rarity - see \citet{Rice2008}).  We would thus expect hot Jupiters to form primarily around stars that hosted disks that were especially efficient at giant planet formation, thus increasing the chances of having a planet reach the inner disk edge during the small window of time where hot Jupiter formation is possible.  These highly efficient disks would also be expected to produce more than one gas giant planet, which leads to the expectation that hot Jupiters would be more likely to have companions.

\begin{figure}
\includegraphics[width=0.5\textwidth]{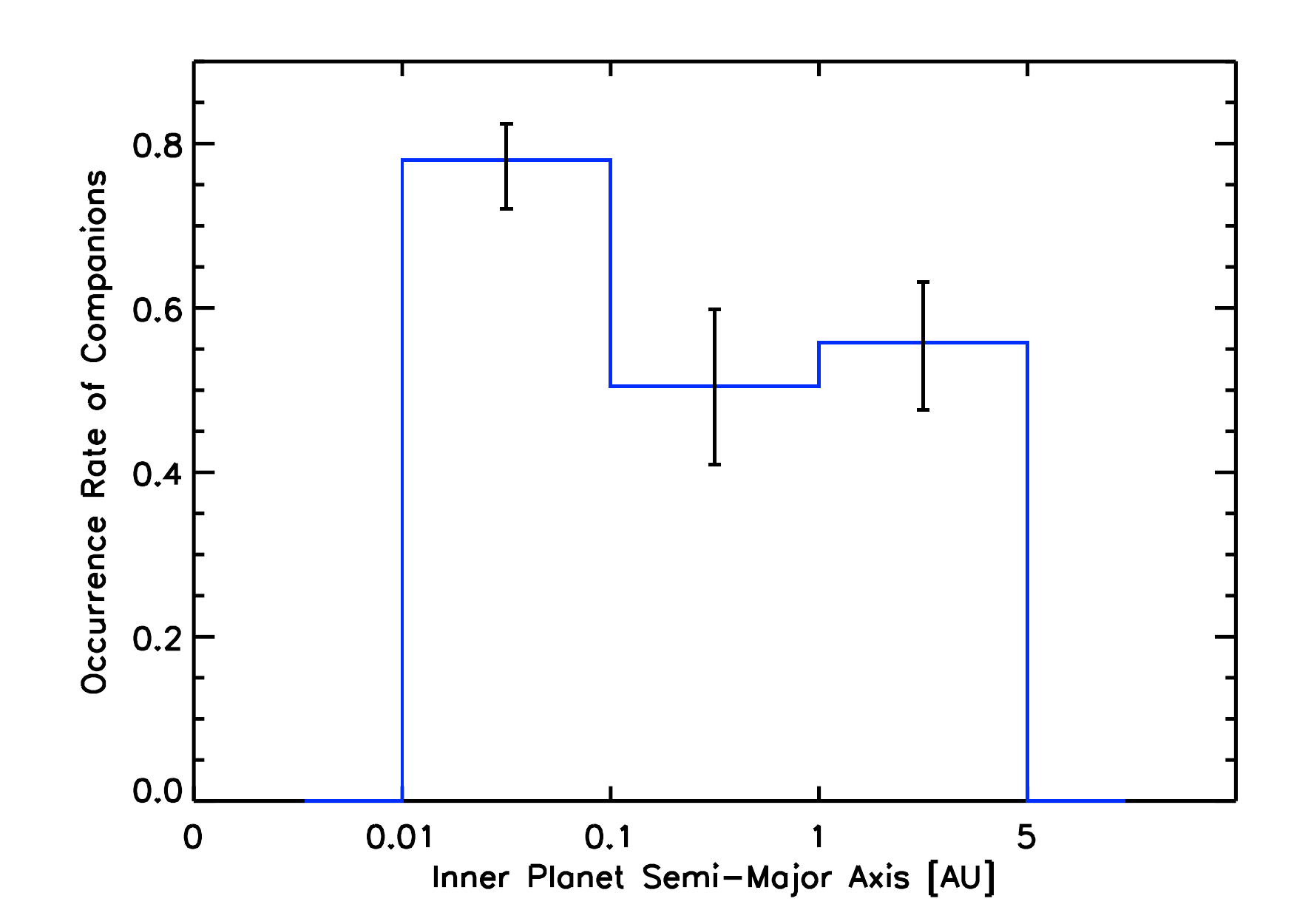} 
\caption{Occurrence rate as a function of inner planet semi-major axis.  The values for each histogram starting at the leftmost bin are:  $75.1^{+4.4}_{-5.9}\%$, $48.8^{+9.4}_{-9.5}\%$, and $53.7^{+7.3}_{-8.2}\%$.}
\end{figure}

We also calculated the occurrence rate of companions as a function of inner planet mass.  We divided the sample up into three bins, corresponding to planets with masses between 0.05 and 0.5 $M_{\rm Jup}$, $0.5 - 5$ $M_{\rm Jup}$, and $5 - 15$ $M_{\rm Jup}$.  Our results are plotted in Figure 20.  We find that intermediate mass planets may be more likely to have a massive, long-period companion, although all three bins are consistent at the 2$\sigma$ level.  We note that our ability to discern trends in companion rate as a function of planet mass is limited by the relatively small sample sizes in the lowest and highest mass bins, which result in correspondingly large uncertainties on their companion rates.

\begin{figure}
\includegraphics[width=0.5\textwidth]{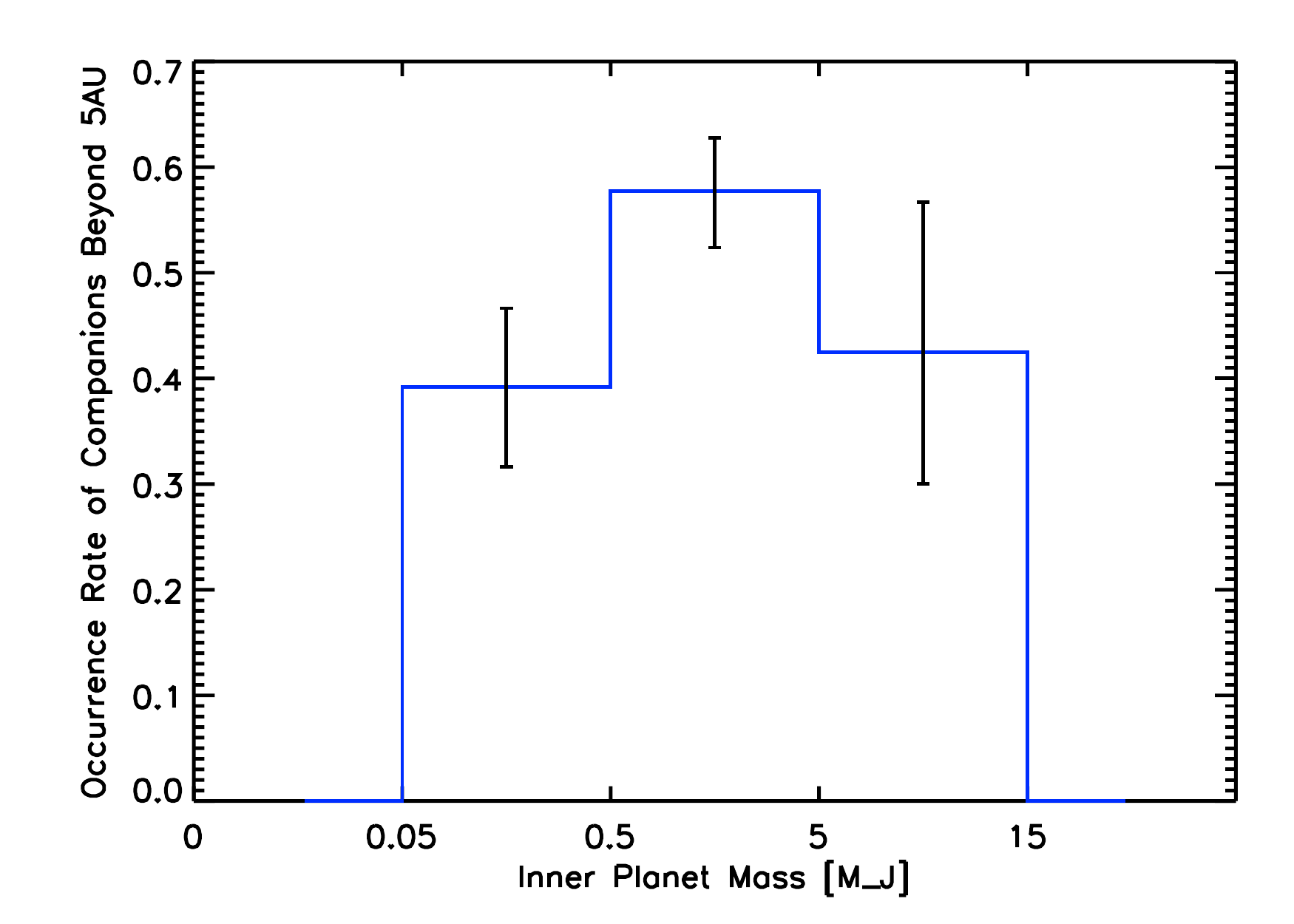} 
\caption{Occurrence rate as a function of inner planet mass.  The values for each histogram starting at the leftmost bin are:  $39.2^{+7.4}_{-7.6}\%$, $57.7^{+5.1}_{-5.3}\%$, and $42.5^{+14.2}_{-12.5}\%$.}
\end{figure}

Finally, we calculated the occurrence rate of companions outside of 5 AU as a function of stellar mass.  Once again, we divided the sample up into three bins - systems with stellar masses from 0.08 - 0.8 $M_{\odot}$ (M and K stars), 0.8 - 1.4 $M_{\odot}$ (G and F stars), and 1.4 - 2.1 $M_{\odot}$ (A stars).  Our results are plotted in Figure 21.  We find that the occurrence rates for each stellar mass bin are consistent with each other at the $0.2\sigma$ level.  Earlier studies indicated that the occurrence rate for gas giant planets interior to 3 AU is higher around A stars than F and G stars \citep{Johnson2010_2}; our results for companions beyond 5 AU suggest that these differences may be reduced at large orbital separations, albeit with large uncertainties due to the small number of A stars included in our sample.  We note that while mass estimates for the evolved A stars have been debated in the literature \citep{Schlauffman2013, Johnson2013, Johnson2013_2, Lloyd2013, Lloyd2011}, this has a minimal impact on our conclusions in this study as we find that these evolved stars have the same frequency of companions as the main sequence FGKM stars in our sample.  

\begin{figure}
\includegraphics[width=0.5\textwidth]{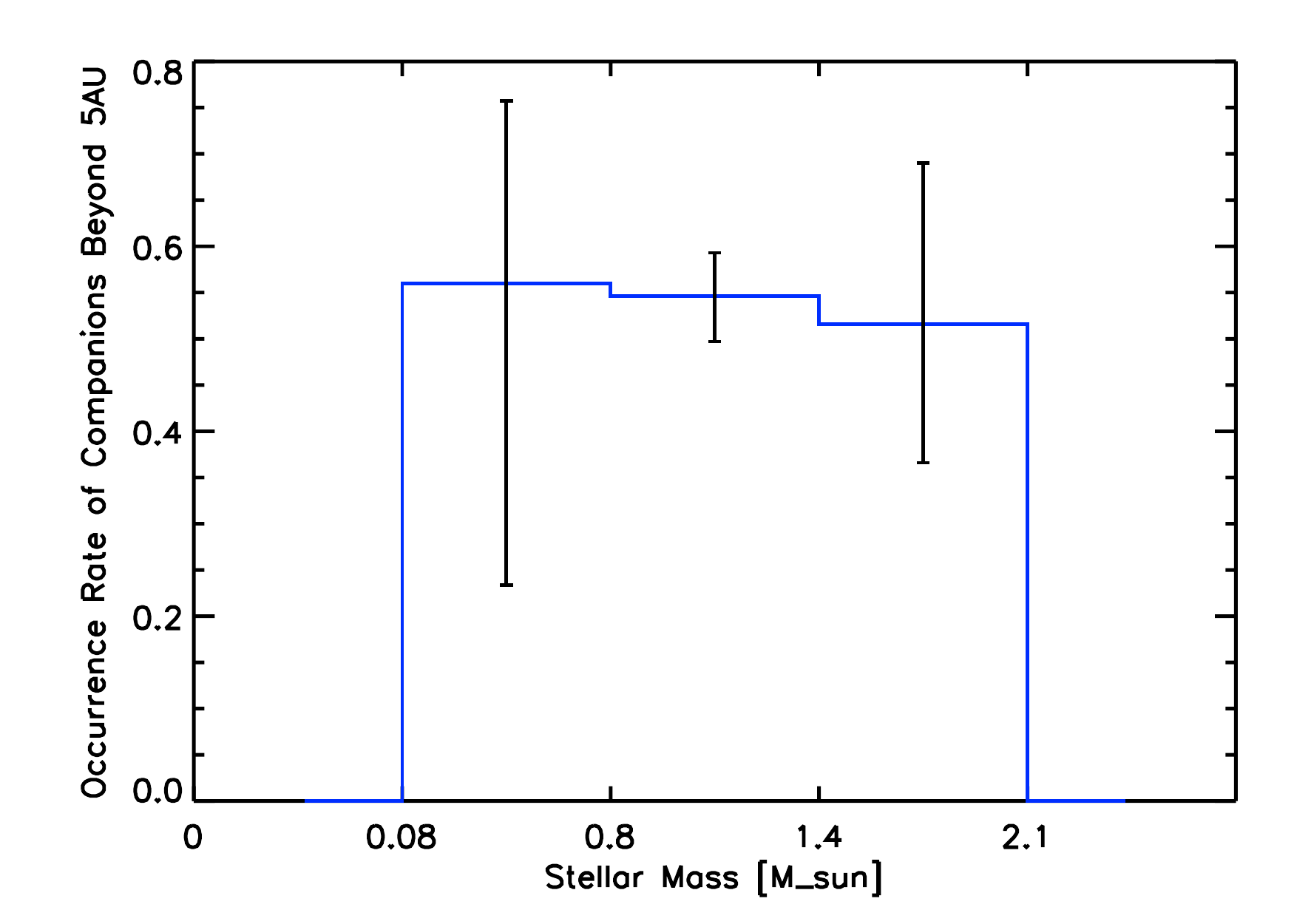} 
\caption{Occurrence rate of massive outer companions as a function of stellar mass.  The values for each histogram starting at the leftmost bin are:  $56.0^{+19.7}_{-32.6}\%$, $54.6^{+4.7}_{-4.9}\%$, and $51.6^{+17.4}_{-15.0}\%$.}
\end{figure}

\subsection{Eccentricity Distribution}

In addition to the results described above, we also seek to quantify how the eccentricity distribution of exoplanets in single planet systems might differ from that of exoplanets in two planet systems or systems with an outer body, as indicated by a radial velocity trend.  We quantify these differences by fitting the set of inner planet eccentricities for each sample using the beta distribution \citep{Kipping2013}:

\begin{equation}
P_{\beta}(e;a,b) = \frac{\Gamma (a+b)}{\Gamma (a) \Gamma(b)}e^{a-1}(1 - e)^{b-1}.
\end{equation}

\noindent We account for the uncertainties in the measured eccentricities for each planet by repeating our beta distribution fit 10,000 times, where each time we draw a random eccentricity from the MCMC posterior probability distribution for each individual planet.  The resulting distributions of best-fit $a$ and $b$ values therefore reflect both the measured eccentricities and their uncertainties.  Figure 22 plots the distribution of best-fit eccentricities for the two groups of planets.  We excluded planets interior to 0.1 AU whose eccentricities might be circularized due to tidal forces from the primary star from this plot as well as the beta distribution fits.  Figure 23 compares the two-dimensional posterior probability distributions in $a$ and $b$ for each of the two groups, taking into account the uncertainties on each planet eccentricity.  We find that the two-planet systems appear to have systematically higher eccentricities than their single planet counterparts, with a significance greater than 3$\sigma$.    

This result appears to contradict previous studies, which found that multi-planet systems have lower eccentricities \citep{Chatterjee2008, Howard2013, Limbach2014, Wright2009}.  This difference may be explained if the separation between inner and outer planets is larger for cases where the inner planet has a large orbital eccentricity.  Previous surveys were typically only sensitive to a 1 $M_{\rm Jup}$ planet out to 3 - 5 AU, suggesting that many of the multi-planet systems detected by our survey would have been misclassified as single planet systems.  

The most detailed study of this correlation to date was presented in \citet{Limbach2014}.  This study used 403 cataloged RV exoplanets from exoplanet.org \citep{Han2014} to determine a relationship between eccentricity and system multiplicity.  127 of these planets were members of known multi-planet systems, with up to six planets in each system.  When the authors calculated the mean eccentricity as a function of the number of planets in each system, they found that systems with more planets had lower eccentricities.   We note that the difference between our new study and this one may be due to the fact that the majority of their planets have relatively short orbital periods.  For systems with three or more planets, this means that the spacing between planets is typically small enough to require less eccentric orbits in order to ensure that the system remains stable over the lifetime of the system.  Furthermore, their analysis did not take into account the uncertainties on individual exoplanet eccentricities, which can be substantial.  \citet{Howard2013} reaches a similar conclusion in their simpler analysis of published RV planets.   This study compared eccentricity distributions of single giant planets to giant planets in multi-planet systems, and found that eccentricities of planets in multi-planet systems are lower on average.  

Because \citet{Limbach2014} did not carry out their own fits to the radial velocity data, they did not consistently allow for the possibility of long-term radial velocity accelerations due to unresolved outer companions.  Previous studies by \citet{Fischer2001} and \citet{Rodigas2009} demonstrate that undetected outer planets can systematically bias eccentricity estimates for the inner planet to larger values.  This is also a problem for systems where the signal to noise of the planet detection is low or the data are sparsely sampled \citep{Shen2008}.  Although we use a smaller sample of planets for our study than \citet{Limbach2014}, our systems all have high signal to noise detections and long radial velocity baselines, which we use to fit and remove long-term accelerations that might otherwise bias our eccentricity estimates.  

In contrast to these other studies, Dong et al (2014) found that warm Jupiters with companions have higher eccentricities than single warm Jupiters.  However, we note that this study relied on a relatively small sample of planets (9 systems with e $>$ 0.4 and 17 with e $<$ 0.2), and the authors did not report uncertainties on their estimated occurrence rates for either sample.  In this study the authors also point out that in order to migrate a warm Jupiter inwards via dynamical interactions with an outer body, the perturber in question must be close enough to overcome GR precession of the inner planet.  We use this constraint, presented in their Equation 4, to test this formation scenario for the warm Jupiter population in our sample.  Of the 42 warm Jupiter systems in our sample, 15 have resolved companions and 4 have statistically significant linear trends.  We find that for the resolved companions, 13 out of the 15 companions satisfy the criterion for high-eccentricity migration (namely that warm Jupiters must reach a critical periastron distance of 0.1 AU within a Kozai-Lidov oscillation).  We take the best fit masses and semi-major axes for the companions causing the trends from their probability distributions, and use these values to calculate the upper limit on the separation ratio between the warm Jupiter and the companion.  We find that zero out of the four systems satisfy the criterion for high-e migration.  Combining the resolved and trend systems, 13 out of 19 warm Jupiter systems with companions satisfy the criterion.  However, we note that the criterion presented in Dong et al (2014) is necessary but insufficient for high-eccentricity migration.  While our observations in principle do not rule out Kozai-Lidov migration for the warm Jupiter population, in order to decide if migration is relevant the character of the angular-momentum exchange cycle must be understood.  In order to do this to lowest order, the mass and semi-major axis of the perturbing orbit, as well as the mutual inclination, must be known.

\begin{figure}
\includegraphics[width=0.5\textwidth]{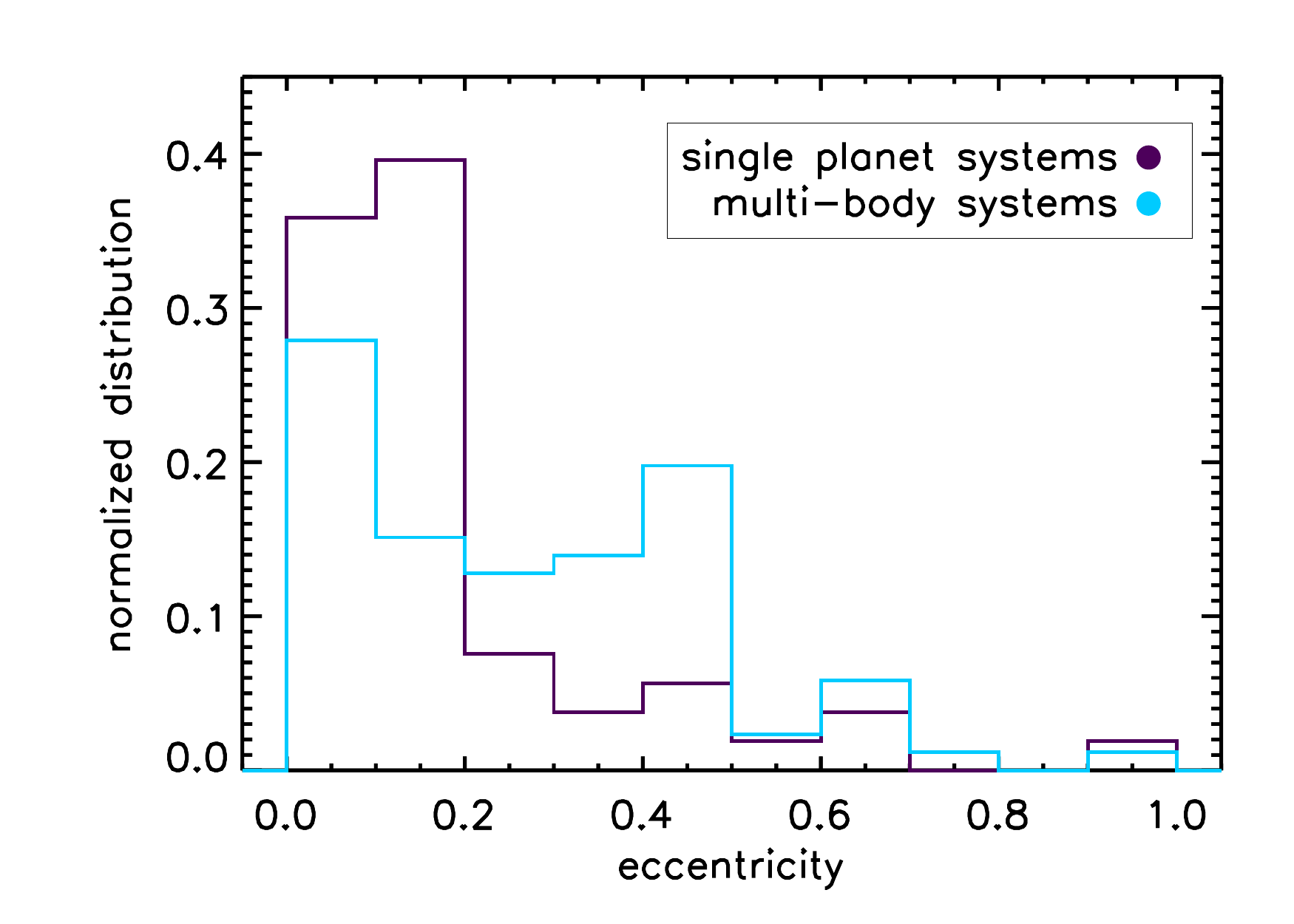}
\caption{Eccentricity distributions of the planets in the full sample.  The purple line shows this histogram for all single planets without outer planets or RV trends, while the blue histogram shows the distribution for planets in two planet systems and single planets with trends.}
\end{figure}

\begin{figure}
\includegraphics[width=0.5\textwidth]{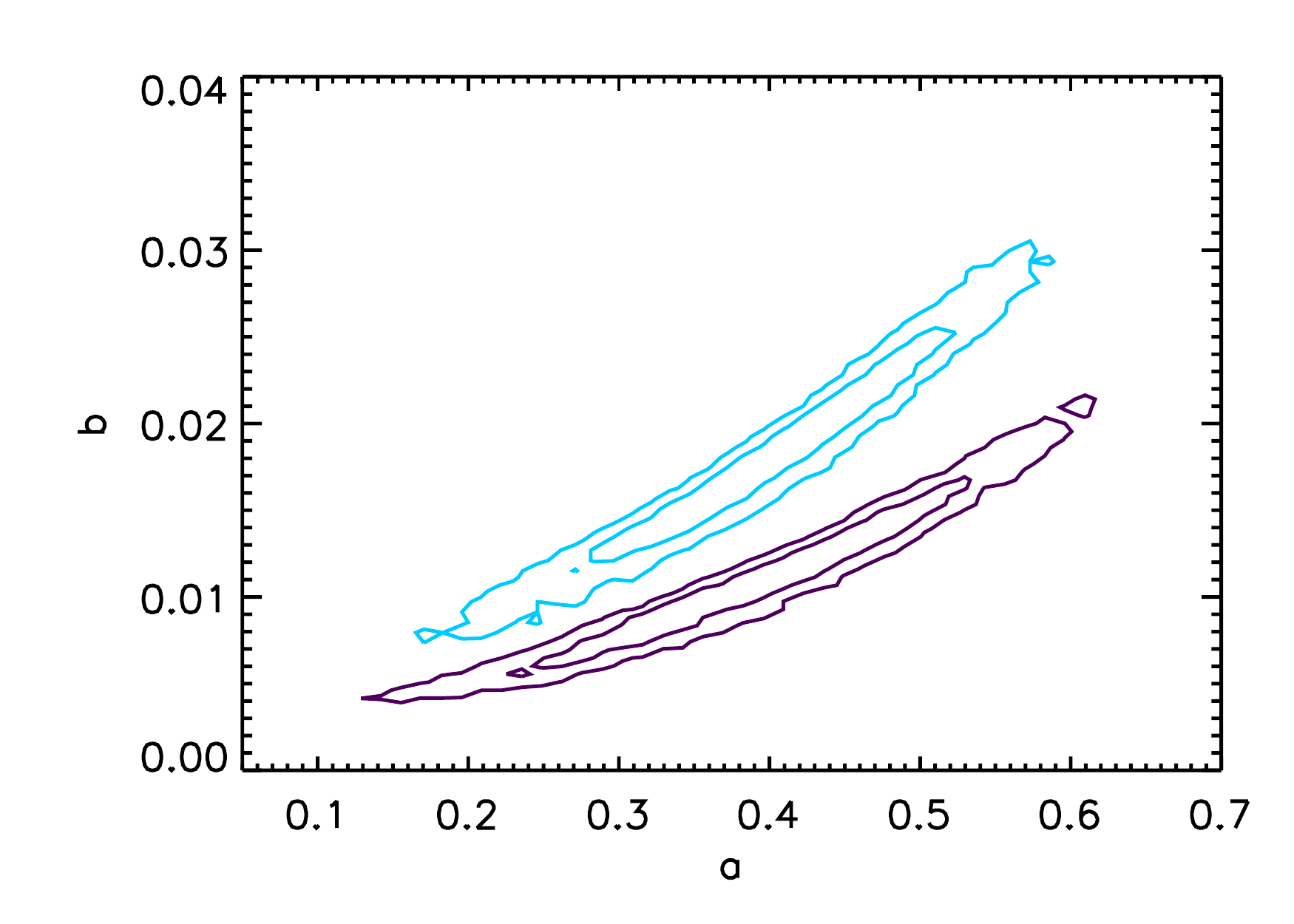} 
\caption{Two dimensional likelihood distributions of $a$ and $b$.  The purple contours represent the $1\sigma$ and $2\sigma$ contours of the two planet systems and single planets with positive trend detections.  The blue contours represent the $1\sigma$ and $2\sigma$ contours of the single planet systems with no outer bodies.}
\end{figure}

\section{Conclusions}

We conducted a Doppler survey at Keck combined with NIRC2 K-band AO imaging to search for massive, long period companions to a sample of 123 known one and two planet systems detected using the radial velocity method.  These companions manifest as long term radial velocity trends in systems where the RV baseline is not long enough to resolve a full orbit.  We extended archival RV baselines by up to 12 years for the stars in our sample, and found that 25 systems had statistically significant radial velocity trends, six of which displayed significant curvature (HD 68988, HD 50499, HD 72659, HD 92788, HD 75898, and HD 158038).  We found that trends detected in HD 1461 and HD 97658 correlxated with the Ca II H$\&$K line strengths, indicating that these trends were likely due to stellar activity and not due to a wide-separation companion.  These systems were removed from further analysis.  We also checked each system for stellar companions, and found that HD 164509, HD 126614, and HD 195109 had stellar companions that could account for the linear RV accelerations.  These systems were also removed from further analysis.  

For the remaining 20 trend systems, we placed lower limits on companion masses and semi-major axes from the RV trends, and upper limits from the AO contrast curves of the corresponding systems.  We quantified the sensitivity of our survey and found that on average we were able to detect a 1 $M_{\rm Jup}$ planet out to 20 AU, and a Saturn mass planet out to 8 AU with $50\%$ completeness.  We fit the companion probability distributions with a double power law in mass and semi-major axis, and integrated this power law to determine the giant planet companion occurrence rate.

We found the total occurrence rate of companions over a mass range of 1 - 20 $M_{\rm Jup}$ and semi-major axis range of 5-20 AU to be $52.4^{+4.5}_{-4.7}\%$, and obtained a comparable occurrence rate when the A and M star systems were removed from the calculation.  The distribution of these long-period companions is best matched by models with a declining frequency as a function of increasing semi-major axis, and appears to be inconsistent with an extrapolation from fits to the population of gas giant planets interior to 3 AU described in Cumming et al. (2008).  This suggests that either the radial distribution of gas giants peaks between 3 - 10 AU, or that the distribution of outer gas giant companions differs from that of the overall gas giant population.  

When calculating the occurrence rate as a function of inner planet semi-major axis, we found that the hot gas giants were more likely to have a massive outer companion as compared to their cold gas giant counterparts.  This result suggests that dynamical interactions between planets may be an important migration mechanism for gas giant planets.

When we compared the eccentricity distributions of single planets in this sample with no outer bodies to planets in two-planet systems and single planets with a positive trend detection, we found that in multi-body systems, the eccentricity distribution was significantly higher than that of single planet systems with no outer bodies.  The higher average eccentricities in these systems suggest that dynamical interactions between gas giant planets play a significant role in the evolution of these systems.  

If we wish to better understand the role that dynamical evolution plays in these systems, there are several possible approaches to consider.  First, continued RV monitoring would help to better constrain companion orbits and masses.  Second, deep imaging of the trend systems could probe down to brown dwarf masses and determine whether any of the observed trends could be caused by stellar instead of planetary mass companions.  If any brown dwarf companions are detected via direct imaging, the existence of complementary radial velocity data would allow us to dynamically measure their masses, which would provide a valuable test of stellar evolution models in the low mass regime \citep{Crepp20122}.  Finally, long term RV monitoring of systems with lower mass planets and/or systems with three or more short period planets detected by transit surveys such as Kepler could allow us to determine if the companion occurrence rate of these systems differs from that of their gas giant counterparts.  A significant limitation of this last suggestion is the need to detect low mass planetary systems orbiting bright, nearby stars - most Kepler stars are time consuming to observe with RVs, but K2, and later TESS, should provide a good sample of low mass planets orbiting nearby stars.

\section*{}

We thank David Hogg and Ben Montet for helpful conversations.  This work was supported by NASA grant NNX14AD24G, and was based on observations at the W. M. Keck Observatory granted by the University of Hawaii, the University of California, the California Institute of Technology, Yale University, and NASA. We thank the observers who contributed to the measurements reported here and acknowledge the efforts of the Keck Observatory staff. We extend special thanks to those of Hawaiian ancestry on whose sacred mountain of Mauna Kea we are privileged to be guests.


\begin{thebibliography}{}

\bibitem[Albrecht et al(2012)]{Albrecht2012}
Albrecht, S., Winn, J. N., Johnson, J. A., et al. 2012b, ApJ, 757, 18

\bibitem[Albrecht et al(2013)]{Albrecht2013}
Albrecht, S. et al 2013, ApJ, 771, 1

\bibitem[Alibert et al(2005)]{Alibert2005}
Alibert, Y. et al 2005, A$\&$A, 434, 1

\bibitem[Anglada-Escude et al(2012)]{Anglada2012}
Anglada-Escude, G. et al 2012, ApJ, 746, 1

\bibitem[Apps et al(2010)]{Apps2010}
Apps, K. et al 2010, PASP, 122, 888

\bibitem[Baraffe et al(1998)]{baraffe98}
Baraffe, I., Chabrier, G., Allard, F., $\&$ Hauschildt, P. H. 1998, A$\&$A, 337, 403

\bibitem[Barbieri et al(2009)]{Barbieri2009}
Barbieri, M. et al 2009, A$\&$A, 503, 2

\bibitem[Batygin(2012)]{Batygin2012}
Batygin, K. 2012, Nature, 491, 418

\bibitem[Batygin $\&$ Adams(2013)]{Batygin2013}
Batygin, K., $\&$ Adams, F. C. 2013, ApJ, 778, 169

\bibitem[Bechter et al.(2014)]{Bechter2014}
Bechter, E. B. et al 2014, ApJ, 788, 1

\bibitem[Becker et al(2015)]{Becker2015}
Becker, J. C. et al 2015, arXiv:150802411B

\bibitem[Bitsch et al(2013)]{Bitsch2013}
Bitsch, B. et al 2013, A$\&$A, 555

\bibitem[Boisse et al(2012)]{Boisse2012}
Boisse, I. et al 2012, A$\&$A, 545

\bibitem[Bonfils et al(2013)]{Bonfils2013}
Bonfils, X. et al 2013, A$\&$A, 549

\bibitem[Bouchy et al(2005)]{Bouchy2005}
Bouchy, F. et al 2005, A$\&$A, 444, 1

\bibitem[Bourrier $\&$ Hebrard(2014)]{Bourrier2014}
Bourrier, V. $\&$ Hebrard, Guillaume 2014, A$\&$A, 569

\bibitem[Bowler et al(2010)]{Bowler2010}
Bowler, B. P. et al 2010, ApJ, 709, 1

\bibitem[Butler et al(2006)]{Butler2006}
Butler, R. P. et al 2006, PASP, 118, 850

\bibitem[Butler et al(2006)]{Vogt2006}
Butler, R. P. et al 2006, ApJ, 646, 1

\bibitem[Butler et al(1996)]{Butler96}
Butler, R. P. et al 1996, PASP, 108

\bibitem[Chatterjee et al(2008)]{Chatterjee2008}
 Chatterjee, S. et al 2008, ApJ 686, 580 
 

\bibitem[Crepp et al(2012)]{Crepp2012}
Crepp, J. R. et al 2012, ApJ, 761, 1

\bibitem[Crepp et al(2012)]{Crepp20122}
Crepp, J. R. et al 2012, ApJ, 751, 97
 
 \bibitem[Cumming et al(2008)]{Cumming2008}
Cumming, A., Butler, R. P., Marcy, G., W., et al. 2008, PASP, 120, 531
 
 \bibitem[Da Silva et al(2007)]{Silva2007}
Da Silva, R. et al 2007, 473, 1

\bibitem[Dawson(2014)]{Dawson2014}
Dawson, R. I. 2014, ApJL, 790, 2, L31

\bibitem[Dawson $\&$ Murray-Clay(2013)]{Dawson2013}
Dawson, R. I. $\&$ Murray-Clay, R. A. 2013, ApJL, 767, 2

\bibitem[Delfosse et al(2000)]{Delfosse2000}
Delfosse, X. et al 2000, A$\&$A, 364

\bibitem[Desidera $\&$ Barbieri(2007)]{Desidera2007}
Desidera, S. $\&$ Barbieri, M. 2007, A$\&$A, 462, 1

\bibitem[Diaz et al(2012)]{Diaz2012}
Diaz, R. F. et al 2012, A$\&$A, 538

\bibitem[Dong et al(2014)]{Dong2014}
Dong, S. et al 2014, ApJL, 781, 1

\bibitem[Dragomir et al(2013)]{Dragomir2013}
Dragomir, D. et al 2013, ApJL, 772, 1

\bibitem[Endl et al(2006)]{Endl2006}
Endl, M. et al 2006, AJ, 131, 6

\bibitem[Endl et al(2008)]{Endl2007}
Endl, M. et al 2008, ApJ, 673, 2

\bibitem[Fabrycky $\&$ Tremaine(2007)]{Fabrycky2007}
Fabrycky, D. $\&$ Tremaine, S. 2007, ApJ, 669, 1298

\bibitem[Fischer et al(2006)]{Fischer2006}
Fischer, D. A. et al 2006, ApJ, 637, 2

\bibitem[Fischer et al(2007)]{Fischer2007}
Fischer, D. A. et al 2007, ApJ, 669, 2

\bibitem[Fischer et al(2001)]{Fischer2001}
Fischer, D. A. et al 2001, ApJ, 551, 1107

\bibitem[Fischer et al(2003)]{Fischer2003}
Fischer, D. A. et al 2003, ApJ, 590, 2

\bibitem[Forveille et al(2009)]{Forveille2009}
Forveille, T. et al 2009, A$\&$A, 493, 2

\bibitem[Fressin et al(2013)]{Fressin2013}
Fressin, F. et al 2013, ApJ, 766, 81

\bibitem[Fulton et al(2015)]{Fulton2015}
Fulton, B. J. et al 2015, ApJ, 805, 2

\bibitem[Giguere et al(2012)]{Giguere2012}
Giguere, M. J. et al 2012, ApJ, 744, 4

\bibitem[Gilliland et al(2011)]{Gilliland2011}
Filliland, R. L. et al 2011, ApJ, 726, 1

\bibitem[Goldreich $\&$ Tremaine(1980)]{Goldreich1980}
Goldreich, P., $\&$ Tremaine, S. 1980, ApJ, 241, 425

\bibitem[Haghighipour et al(2012)]{Haghighipour2012}
Haghighipour, N. et al 2012, ApJ, 756, 1

\bibitem[Haghighipour et al(2010)]{Haghighipour2010}
Haghighipour, N. et al 2010, ApJ, 715, 1

\bibitem[Han et al(2014)]{Han2014}
Han, E. et al 2014, PASP, 126, 827

\bibitem[Harakawa et al(2010)]{Harakawa2010}
Harakawa, H. et al 2010, ApJ, 715, 1

\bibitem[Hebrard et al(2011)]{Hebrard2011}
Hebrard, G., Ehrenreich, D., Bouchy, F., et al. 2011, A$\&$A, 527, L11

\bibitem[Howard et al(2011)]{Howard20112}
Howard, A. W. et al 2011, ApJ, 726, 2

\bibitem[Howard et al(2011)]{Howard2011}
Howard, A. W. et al 2011, ApJ, 730, 1

\bibitem[Howard(2013)]{Howard2013}
Howard, A. W. 2013, Science, 340, 572

\bibitem[Howard et al(2009)]{Howard2009}
Howard, A. W. et al 2009, ApJ, 696, 1


\bibitem[Howard et al(2009)]{Howard2009}
Howard, A. et al 2009, ApJ, 749, 134

\bibitem[Howard et al(2010)]{Howard2010}
Howard, A. W. et al 2010, ApJ, 721, 2

\bibitem[Howard et al(2012)]{Howard2012}
Howard, A. W. et al. 2012, ApJS, 201, 15

\bibitem[Howard et al(2014)]{Howard2014}
Howard, A. W. et al 2014, ApJ, 794, 1

\bibitem[Huber et al(2013)]{Huber2013}
Huber, D. et al 2013, Science, 342, 6156

\bibitem[Husser et al(2013)]{husser13}
Husser, T.-O., Wende-von Berg, S., Dreizler, S., et al. 2013, A$\&$A, 553, A6

\bibitem[Isaacson $\&$ Fischer(2010)]{Isaacson2010}
Isaacson, H., $\&$ Fischer, D. 2010, ApJ, 725, 875

\bibitem[Johnson $\&$ Wright(2013)]{Johnson2013}
Johnson, J. A. $\&$ Wright, J. T. 2013, arXiv:1307.3441J

\bibitem[Johnson et al(2010)]{Johnson2010}
Johnson, J. A. et al 2010, PASP, 122, 892

\bibitem[Johnson et al(2011)]{Johnson2011}
Johnson, J.A. 2011, ApJS, 197, 2

\bibitem[Johnson et al(2013)]{Johnson2013_2}
Johnson, J. A. et al 2013, ApJ, 763, 1

\bibitem[Johnson et al(2010)]{Johnson2010}
Johnson, J. et al 2010, PASP, 122, 149

\bibitem[Johnson et al(2010)]{Johnson20102}
Johnson, J. A. et al 2010, PASP, 122, 888

\bibitem[Johnson et al(2010)]{Johnson2010_2}
Johnson, J.A. et al 2010, PASP, 122, 894

\bibitem[Johnson et al(2010)]{Johnson20103}
Johnson, J. A. et al 2010, ApJL, 721, 2

\bibitem[Johnson et al(2007)]{Johnson20072}
Johnson, J. A. et al 2007, ApJ, 665, 1

\bibitem[Johnson et al(2007)]{Johnson2007}
Johnson, J. A. et al 2007, ApJ, 670, 1

\bibitem[Johnson et al(2011)]{Johnson20112}
Johnson, J. A. et al 2011, ApJ, 141, 1

\bibitem[Johnson et al(2006)]{Johnson2006}
Johnson, J. A. et al 2006, ApJ, 647, 1


\bibitem[Jones et al(2010)]{Jones2010}
Jones, H. R. et al 2010, MNRAS, 403, 4

\bibitem[Juric $\&$ Tremaine(2008)]{Juric2008}
Juric, M., $\&$ Tremaine, S. 2008, ApJ, 686, 603

\bibitem[Kane et al(2012)]{Kane2012}
Kane, S. R. et al 2012, MNRAS, 425, 1

\bibitem[Kane et al(2015)]{Kane2015}
Kane, S. R. et al 2015, submitted to ApJ, arXiv:  1504.04066v1

\bibitem[Kass $\&$ Raftery(1995)]{Kass1995}
Kass, R. E. $\&$ Raftery, A. E. 1995, J. Am. Statist. Assoc., 90, 430

\bibitem[Kipping(2013)]{Kipping2013}
Kipping, D. M. 2013, MNRAS, 434, L51

\bibitem[Knutson et al(2014)]{Knutson2014}
Knutson, H. A. et al 2014, ApJ, 785, 126

\bibitem[Limbach $\&$ Turner(2014)]{Limbach2014}
Limbach, M. A. $\&$ Turner, E. L. 2014, arXiv:1404.2552L

\bibitem[Lin $\&$ Papaloizou(1986)]{Lin1986}
Lin, D. N. C., $\&$ Papaloizou, J. C. B. 1986, ApJ, 309, 846

\bibitem[e.g., Lin et al(1996)]{Lin1996}
Lin, D. N. C., Bodenheimer, P., $\&$ Richardson, D. C. 1996, Nature, 380, 606

\bibitem[Liu et al(2002)]{Liu2002}
Liu, M. C., Fischer, D. A., Graham, J. R., et al. 2002, ApJ, 571, 519

\bibitem[Lloyd(2011)]{Lloyd2011}
Lloyd, J. P. 2011, ApJL, 739, 2

\bibitem[Lloyd(2013)]{Lloyd2013}
Lloyd, J. P. 2013, ApJL, 774, 1

\bibitem[Malmberg et al(2007)]{Malmberg2007}
Malmberg, D., Davies, M. B., $\&$ Chambers, J. E. 2007, MNRAS, 377, L1

\bibitem[Marcy $\&$ Butler(2000)]{Marcy2000}
Marcy, G. W. $\&$ Butler, R. P. 2000, ApJ, 536, 1

\bibitem[Marcy $\&$ Butler(1992)]{Marcy92}
Marcy, G. W. $\&$ Butler, R. P. 1992, PASP, 104, 674

\bibitem[Mayor et al(2004)]{Mayor2004}
Mayor, M. et al 2004, A$\&$A, 415

\bibitem[Melo et al(2007)]{Melo2007}
Melo, C. et al 2007, A$\&$A, 467, 2

\bibitem[Meschiari et al(2011)]{Meschiari2011}
Meschiari, S. et al 2011, ApJ, 727, 2

\bibitem[Metchev $\&$ Hillenbrand(2009)]{Metchev2009}
Metchev, S. A. $\&$ Hillenbrand, L. A. 2009, ApJS, 181, 1

\bibitem[Montet et al(2014)]{Montet2014}
Montet, B. T. et al 2014, ApJ, 781, 1

\bibitem[Mortier et al(2013)]{Mortier2013}
Mortier, A. et al 2013, A$\&$A, 556

\bibitem[Morton $\&$ Winn(2014)]{Morton2014}
Morton, T. D. $\&$ Winn, J. N. 2014, ApJ, 796, 1

\bibitem[Moutou et al(2009)]{Moutou2009}
Moutou, C. et al 2009, A$\&$A, 498, 1

\bibitem[Moutou et al(2011)]{Moutou2011}
Moutou, C. et al 2011, A$\&$A, 527

\bibitem[Mugrauer et al(2005)]{Mugrauer2005}
Mugrauer, M. et al 2005, A$\&$A, 440, 3

\bibitem[Mugrauer et al(2007)]{Mugrauer2007}
Mugrauer, M. et al 2007, Proceedings IAU Symposium, No. 240

\bibitem[Nagasawa et al(2008)]{Nagasawa2008}
Nagasawa, M., Ida, S., $\&$ Bessho, T. 2008, ApJ, 678, 498

\bibitem[Naoz et al(2012)]{Naoz2012}
Naoz, S. M. et al 2012, ApJ, 754, 36

\bibitem[Ngo et al(2015)]{Ngo2014}
Ngo, H. et al 2015, ApJ, 800, 138

\bibitem[Peek et al(2009)]{Peek2009}
Peek, K. M. G. et al 2009, PASP, 121, 880

\bibitem[Pepe et al(2011)]{Pepe2011}
Pepe, F. et al 2011, A$\&$A, 534

\bibitem[Perrier et al(2003)]{Perrier2003}
Perrier, C. et al 2003, A$\&$A, 410

\bibitem[Petigura et al(2013)]{Petigura2013}
Petigura, E. et al 2013, PNAS, 110, 48

\bibitem[Pilyavsky et al(2011)]{Pilyavsky2011}
Pilyavsky, G. et al 2011, ApJ, 743, 2

\bibitem[Pollack et al(1996)]{Pollack1996}
Pollack, J. B. et al 1996, Icarus, 124, 1

\bibitem[Rafikov(2006)]{Rafikov2006}
Rafikov, R. R. 2006, ApJ, 648, 1

\bibitem[Raghavan et al(2006)]{Raghavan2006}
Raghavan, D. et al 2006, ApJ, 646, 1

\bibitem[Rasio $\&$ Ford(1996)]{Rasio1996}
Rasio, F. A. $\&$ Ford, E. B. 1996, Science, 274

\bibitem[Rice et al(2008)]{Rice2008}
Rice, W. K. M. et al 2008, MNRAS, 384, 3

\bibitem[Rivera et al(2010)]{Rivera2010}
Rivera, E. J. et al 2010, ApJ, 708, 2

\bibitem[Robinson et al(2007)]{Robinson2007}
Robinson, S. E. et al 2007, ApJ, 670, 2

\bibitem[Rodigas $\&$ Hinz(2009)]{Rodigas2009}
Rodigas, T. J. $\&$ Hinz, P. M. 2009, ApJ, 702, 1

\bibitem[Roell et al(2012)]{Roell2012}
Roell, T. et al 2012, A$\&$A, 542

\bibitem[Santos et al(2010)]{Santos2010}
Santos, N. C. et al 2010, A$\&$A, 511, 54

\bibitem[Schlaufman $\&$ Winn(2013)]{Schlauffman2013}
Schlaufman, K. C. $\&$ Winn, J. N. 2013, ApJ, 772, 2

\bibitem[Segransan et al(2011)]{Segransan2011}
Segransan, D. et al 2011, A$\&$A, 535

\bibitem[Shen $\&$ Turner(2008)]{Shen2008}
Shen, Y. $\&$ Turner, E. L. 2008, ApJ, 685, 1

\bibitem[Spalding $\&$ Batygin(2014)]{Spalding2014}
Spalding, C. $\&$ Batygin, K. 2014, ApJ, 790, 1

\bibitem[Steffen et al(2012)]{Steffen2012}
Steffen, J. H. et al 2012, Proc Natl Acad Sci, 109, 21

\bibitem[Storch et al(2014)]{Storch2014}
Storch, N. I. et al 2014, Science, 345, 6202

\bibitem[Takeda et al(2007)]{Takeda2007}
Takeda, G. et al 2007, ApJS, 168, 2

\bibitem[Tanaka et al(2002)]{Tanaka2002}
Tanaka, H., Takeuchi, T., $\&$ Ward, W. R. 2002, ApJ, 565, 1257

\bibitem[Torres(1999)]{Torres1999}
Torres, G. 1999, PASP, 111, 169

\bibitem[Torres(2008)]{Torres2008}
Torres, G. et al 2008, ApJ, 677, 2

\bibitem[Udry et al(2002)]{Udry2002}
Udry, S. et al 2002, A$\&$A, 390

\bibitem[Valenti $\&$ Fischer(2005)]{Valenti2005}
Valenti, J. A. $\&$ Fischer, D. A. 2005, ApJ, 159, 1

\bibitem[Valenti et al(2009)]{Valenti2009}
Valenti, J. A. et al 2009, 702, 2

\bibitem[Valenti et al(1995)]{Valenti95}
Valenti, J. A. et al 1995, PASP, 107

\bibitem[Vogt et al(2005)]{Vogt2005}
Vogt, S. S. et al 2005, ApJ, 632, 1

\bibitem[Vogt et al(2000)]{Vogt2000}
Vogt, S. et al 2000, ApJ, 536, 2

\bibitem[Vogt et al(2002)]{Vogt2002}
Vogt, S. et al 2002, A[J, 568, 1

\bibitem[Vogt et al(1994)]{Vogt94}
Vogt, S. S. et al 1994, Proc. SPIE, 2198, 362

\bibitem[Wang et al(2012)]{Wang2012}
Wang, X. et al 2012, ApJ, 761, 1

\bibitem[Winn et al(2010)]{Winn2010}
Winn, J. N., Fabrycky, D., Albrecht, S., $\&$ Johnson, J. A. 2010a, ApJL,
718, L145

\bibitem[Wittenmyer et al(2009)]{Wittenmyer2009}
Wittenmyer, R. A. et al 2009, ApJS, 182, 1

\bibitem[Wright et al(2011)]{Wright2011}
Wright, J. T. et al. 2011, PASP, 123, 41

\bibitem[Wright et al(2007)]{Wright2007}
Wright, J. T. et al 2007, ApJ, 657, 1

\bibitem[Wright et al(2004)]{Wright2004}
Wright, J. et al 2004, ApJS, 152, 261

\bibitem[Wright et al(2008)]{Wright2008}
Wright, J. T. et al 2008, ApJ, 683, 1

\bibitem[Wright et al(2009)]{Wright2009}
Wright, J. T. et al 2009, ApJ, 693, 2

\bibitem[Wright et al(2004)]{Wright2004}
Wright, J. T. et al 2004, ApJS, 152, 2

\bibitem[Wu $\&$ Lithwick(2010)]{Wu2010}
Wu, Y. $\&$ Lithwick, Y. 2010, ApJ, 735, 109




\end{thebibliography}
\end{document}